\documentclass[a4paper,11pt]{article}
\pdfoutput=1 
\usepackage{verbatim}

\usepackage{jheppub}
\usepackage{amsmath,amsfonts,amssymb,amsthm,graphicx,color}
\usepackage[T1]{fontenc} 
\usepackage{enumerate}
\usepackage{tikz, tikz-3dplot, pgfplots}
\usepackage[utf8]{inputenc}
\usepackage{slashed}
\usepackage{subcaption}

\usepackage{changepage}
\usepackage{float}
\usepackage{tikz}
\usetikzlibrary{decorations.text}
\usepackage{enumitem}
\setlist{itemsep=0pt}
\usetikzlibrary{math}

\usepackage{tikz}
\usepackage{pgfplots}
\usetikzlibrary{decorations.text,intersections, pgfplots.fillbetween}
\usetikzlibrary{math}
 \usetikzlibrary{snakes,3d,shapes.geometric,shadows.blur}
\usepackage{tikz-3dplot}

\newcommand{\comm}[1]{} 

\def\({\left(}
\def\){\right)}
\def\[{\left[}
\def\]{\right]}

\def\One{{\hbox{ 1\kern-.8mm l}}}

\def\barray{\begin{array}}
\def\earray{\end{array}}
\def\be{\begin{equation}}
\def\ee{\end{equation}}
\def\bea{\begin{eqnarray}}
\def\eea{\end{eqnarray}}
\def\bal{\begin{align}}
\def\eal{\end{align}}

\def\-{\,-\,}
\def\={\,=\,}
\def\+{\,+\,}
\def\equi{\,\equiv\,}

\numberwithin{equation}{section} 

\definecolor{cardinal}{rgb}{0.6,0,0}
\definecolor{darkgreen}{rgb}{0,0.4,0}
\definecolor{golden}{rgb}{0.92, 0.7, 0}
\definecolor{midnight}{rgb}{0, 0, 0.5}
\definecolor{darkblue}{rgb}{0, 0, 0.7}
\definecolor{purple}{rgb}{0.5, 0, 0.5}

\def\IR{\mathbb{R}}

\def\cA{{\cal A}}

\def\cF{{\cal F}}
\def\cG{{\cal G}}

\def\cL{{\cal L}}
\def\cM{{\cal M}}
\def\cN{{\cal N}}

\def\cO{{\cal O}}

\def\cR{{\cal R}}
\def\cS{{\cal S}}
\def\cT{{\cal T}}

\def\cZ{{\cal Z}}

\makeatletter
\newcommand\footnoteref[1]{\protected@xdef\@thefnmark{\ref{#1}}\@footnotemark}
\makeatother

\usetikzlibrary{positioning,calc,fadings,decorations.pathreplacing}

\tikzset{
 diffuse color/.initial = black,                       
}

\tikzset{
 linear opacity/.initial=0.5,                          
 linear stroke/.style = {                              
   preaction={                                         
     draw=\pgfkeysvalueof{/tikz/diffuse color},        
     line width = (2.0-#1)*\pgflinewidth,              
     opacity=\pgfkeysvalueof{/tikz/linear opacity},white}},  
 diffuse gradient/.style={                             
   draw = none,                                        
   linear opacity=#1,                                  
   linear stroke/.list={0.0,#1,...,1.0}},              
 diffuse gradient/.default=1,                          
}

\tikzset{
 non-linear stroke/.style = {                          
   preaction={                                         
     draw=\pgfkeysvalueof{/tikz/diffuse color},        
     line width = (2.0-#1)*\pgflinewidth,              
     opacity=#1,white}},                                     
 diffuse falloff/.style={                              
   draw = none,                                        
   non-linear stroke/.list={0.0,#1,...,1.0}},          
 diffuse falloff/.default=1,                           
}
 
\tikzset{%
  >=latex, 
  inner sep=0pt,%
  outer sep=2pt,%
  mark coordinate/.style={inner sep=0pt,outer sep=0pt,minimum size=3pt,
    fill=black,circle}%
}

\title{\boldmath Non-BPS Floating Branes and Bubbling Geometries
}

\author{Pierre Heidmann} 
\affiliation{Department of Physics and Astronomy, Johns Hopkins University, 3400 North Charles Street, Baltimore, MD 21218, USA}

\emailAdd{pheidma1@jhu.edu}

\abstract{We derive a non-BPS linear ansatz using the charged Weyl formalism in string and M-theory backgrounds. Generic solutions are static and axially-symmetric with an arbitrary number of non-BPS sources corresponding to various brane, momentum and KKm charges. Regular sources are either four-charge non-extremal black holes or smooth non-BPS bubbles. We construct several families such as chains of non-extremal black holes or smooth non-BPS bubbling geometries and study their physics.  The smooth horizonless geometries can have the same mass and charges as non-extremal black holes. Furthermore, we find examples that scale towards the four-charge BPS black hole when the non-BPS parameters are taken to be small, but the horizon is smoothly resolved by adding a small amount of non-extremality.}

\preprint{}
\begin{document}

\maketitle
\flushbottom

\newpage

\section{Introduction}
\label{sec:Intro}

The construction of solitons with non-trivial topologies and matter fields in supergravity theories has generally been limited to the frontier of supersymmetry. Supersymmetry provides BPS conditions that force the sources to have an electromagnetic potential that is equal to the gravitational potential. Intuitively, this allows linearization of the equations of motion and generation of backgrounds with multiple sources that do not interact strongly. In this context, BPS equations led to ``floating brane'' ansatz where BPS brane sources ``float'' relative to each other, preventing gravitational collapse by electromagnetic flux. 

It is an interesting and important  question to ask whether such ansatz can survive the non-BPS regime where the electromagnetic forces are not canceled by the gravitational force. This would require non-trivial novel mechanisms to allow for non-collapsing sources. In this paper, we will answer in the affirmative by deriving a non-BPS floating brane ansatz in the context of M-theory on T$^6\times$S$^1$ and other dual frameworks. The construction is based on a recent method, the ``charged'' Weyl formalism \cite{Bah:2020pdz,Bah:2021owp,Bah:2021rki}, which is simple enough to induce a linear system but sufficiently rich to allow non-trivial non-BPS structures.

The classification of BPS solitons in supergravity have generated a great deal of interest in studying semi-classical descriptions of coherent quantum gravity states and in exploring AdS/CFT dualities from a bulk point of view \cite{Lin:2004nb,Kanitscheider:2006zf,Avery:2010qw,Giusto:2013bda,Chakrabarty:2015foa,Giusto:2019qig,Bena:2018bbd}. Within the microstate geometry program for instance (see \cite{Warner:2019jll,Heidmann:2019gvg,Bena:2013dka,Bena:2007kg} for a  review), large classifications of BPS solitons, which exhibit the same supersymmetries, mass, spin, and charges as BPS black holes, but whose horizon is resolved by a geometric transition of BPS brane sources, have been allowed with the help of linear ansatz. Examples of this are the $\tfrac{1}{2}$-BPS microstate geometries of the two-charge black hole \cite{Lunin:2001jy,Lunin:2002qf,Lunin:2002iz}, or the $\tfrac{1}{4}$-BPS microstate geometries of the three-charge black hole such as multicenter bubbling solutions \cite{Bena:2004de,Bena:2005va,Berglund:2005vb,Bena:2010gg,Vasilakis:2011ki,Heidmann:2017cxt,Bena:2017fvm} or superstrata \cite{Bena:2015bea,Bena:2017xbt,Bena:2016agb,Ceplak:2018pws,Heidmann:2019zws,Heidmann:2019xrd}.

Moving inside the non-BPS regime requires to deal with Einstein-Maxwell equations for which finding a linear system can be tedious. The Weyl formalism was originally formulated to linearize Einstein equations for four-dimensional general relativity (GR) in vacuum by restricting to axisymmetric and static solutions \cite{Weyl:book}. The generalization to GR in $D$ dimensions has been done by Emparan and Reall \cite{Emparan:2001wk}. All solutions are given in terms of $D-3$ harmonic functions whose sources are segments called rods on the symmetry axis. The physical sources correspond to regular coordinate singularities, either along the time direction inducing a black hole horizon, or along a compact spatial direction inducing a vacuum bubble. Therefore, chains of black holes and bubbles \cite{Elvang:2002br} or smooth chains of vacuum bubbles \cite{Bah:2021owp} can be generated.

There are several reasons for adding matter fields to vacuum Weyl constructions. The first is stability. Indeed, a Kaluza-Klein (KK) bubble in vacuum suffers from an instability that forces its decay by ``eating up'' the whole spacetime \cite{Witten:1981gj}. However, the addition of suitable electromagnetic flux stabilizes the bubble at a fixed radius \cite{Stotyn:2011tv,StabilityPaper}. Second, it has been shown in \cite{Gibbons:2013tqa} that electromagnetic flux is a necessary ingredient to have a regular structure that can be as compact as a black hole horizon or that can develop an AdS throat. Nevertheless, gauge fields add non-linearity for which the extraction of a linear branch that does not reduce to the BPS equations is far from guaranteed.

 In \cite{Bah:2020pdz,Bah:2021owp,Bah:2021rki}, the Weyl formalism has been generalized to a ``charged'' Weyl formalism for Einstein-Maxwell theories. Solutions are still determined by $D-3$ harmonic functions. They are sourced by rods that can be now charged under the gauge field(s) and still correspond to either black hole horizon or smooth bubbles. In this system, the sources are set to have the same charge-to-mass ratio. Moreover, as these are non-BPS systems, this ratio can be generically different from $1$. In \cite{Bah:2021owp,Bah:2021rki}, a ``bottom-up'' approach has been adopted to highlight the physics of the new non-BPS solitons, such as chains of smooth charged bubbles in six dimensions or ``bubble bag ends''. In the present paper, we adopt a top-down perspective by applying the same formalism to string theory backgrounds to derive and explore a non-BPS linear ansatz.

\subsection{Summary of the results}

Our initial construction takes place in M-theory on T$^6\times$S$^1$. We consider that the gauge field is sourced by three sets of M2 branes that wrap three orthogonal 2-tori inside T$^6$. Moreover, we allow for a magnetic KKm vector associated to the S$^1$.\footnote{We do not consider M5 brane charges on the T$^6$ to avoid additional non-linearity from Chern-Simons interactions.}  By applying the charged Weyl formalism, we derive what we call a ``non-BPS floating brane ansatz'' for the M2-M2-M2-KKm system in M-theory. The solutions are determined in terms of eight harmonic functions. Four are related to the deformations of the T$^6$ while the other four are associated with the gauge potentials. In this paper, we focus on solutions that are asymptotic to $\IR^{1,3}\times$T$^6\times$S$^1$ or $\IR^{1,4}\times$T$^6$. 

The asymptotically-$\IR^{1,3}$ solutions are static, axisymmetric and induced by an arbitrary number of non-BPS sources on the symmetry axis. A physical source corresponds to a regular coordinate degeneracy of either the time or one of the T$^6\times$S$^1$ directions. If the time degenerates, it corresponds to the horizon of a static non-extremal M2-M2-M2-KKm black holes \cite{Cvetic:1995kv,Chong:2004na,Chow:2014cca}. If one of the T$^6$ directions or the S$^1$ shrinks, it defines the locus of a M2-KKm bubble or a KKm bubble respectively. The sources are separated by strings with negative tension, or struts, which disappear if they are connected \cite{Bah:2021owp}. We explicitly construct solutions that are chains of non-extremal static black holes, separated by struts or by KKm bubbles. In addition, we construct non-BPS  bubbling geometries of M2-KKm bubbles on a line. They can have the same charges and mass as static non-extremal M2-M2-M2-KKm black holes, but they are horizonless and smooth. Moreover, we show that they have a BPS limit where they converge to the BPS black hole. Therefore, when they are slightly non-BPS, they are almost indistinguishable from the BPS black hole, develop an AdS$_2$ throat, and resolve its horizon into non-BPS bubbles by adding a small amount of non-extremality. This is the first example of such a resolution in the non-BPS regime.

We also consider the non-BPS floating brane ansatz in different duality frames such as the D1-D5-P-KKm frame in type IIB. We show that there are types of bubbles that are smooth only in a specific duality frame, and we construct bubbling geometries in the D1-D5 framework.

Solutions that are asymptotic to five-dimensional flat space plus compactification circles can be obtained by considering the S$^1$ to be the Hopf fiber of a four-dimensional base. The non-BPS sources correspond to either three-charge non-extremal black holes \cite{Cvetic:1996xz,Cvetic:1997uw,Cvetic:1998xh}, smooth M2 bubbles or vacuum bubbles. We explicitly construct and study black hole chains and smooth bubbling geometries that are asymptotically-flat in five dimensions.

In section \ref{sec:Ansatz}, we derive the non-BPS floating brane ansatz by solving Einstein-Maxwell equations with the charged Weyl formalism. We discuss the different regimes and the reduction in five and four dimensions. In section \ref{sec:Classification}, we classify the asymptotically-flat solutions and the regular sources. In the sections \ref{sec:Sol4d} and \ref{sec:Sol5d}, we construct explicit families of non-BPS four-dimensional and five-dimensional solutions, and we discuss their physics.

\section{Non-BPS floating branes in M-theory}
\label{sec:Ansatz}

We consider M-theory solutions on T$^6=$T$^2\times$T$^2\times$T$^2$ for which each two-torus can be wrapped by M2 branes. Moreover, we allow one of the other four spatial directions to be an extra S$^1$ that can carry KKm charges. The dynamics in the bosonic sector of M-theory is given by 
\begin{equation}
\left(16 \pi G_{11}\right) S_{M}=\int R \,\star \mathbb{I} \-\frac{1}{2}  \,F_{4} \wedge \star F_{4}\+\frac{1}{6} \,A_{3} \wedge F_{4} \wedge F_{4},
\label{eq:Action11D}
\end{equation}
where $G_{11}$ is the eleven-dimensional Newton constant and $F_4 = dA_3$ is the four-form field strength of the three-form gauge field. The Einstein-Maxwell equations are
\begin{equation}
\begin{aligned}
R_{\mu \nu} \-\frac{1}{12}\left(F_{4\, \mu \alpha \beta \gamma} \,F_{4 \,\nu}{ }^{\alpha \beta \gamma} \- \frac{1}{12} \,g_{\mu \nu} \,F_{4}^{\,\,\alpha \beta \gamma \delta}F_{4\,\alpha \beta \gamma \delta}\right) &\=0, \\
d \star F_{4} \- \frac{1}{2} \,F_{4} \wedge F_{4} & \=0 .
\end{aligned}
\label{eq:EOMMtheory}
\end{equation}
We parametrize the T$^6$ by $\{y_a\}_{a=1,..,6}$, and each direction is $2\pi R_{y_a}$ periodic. The remaining four-dimensional spatial directions will be decomposed into a three-dimensional basis in cylindrical Weyl coordinates $(\rho,z,\phi)$ and a U(1) fiber, $y_7$. This fiber will correspond either to a compact circle with $2\pi R_{y_7}$ periodicity for asymptotically $\IR^{1,3}\times$S$^1\times$T$^6$ solutions, or to an angle with $2\pi$ periodicity for  solutions that asymptote to $\IR^{1,4}\times$T$^6$.

\subsection{Equations of motion}

We derive the equations of motion, obtained from \eqref{eq:EOMMtheory}, for solutions that are \emph{static}, \emph{axisymmetric}, and sourced by a Kaluza-Klein magnetic vector and three stacks of M2 branes wrapping the two-tori. Without restriction, we consider the following ansatz that fits the equations:
\begin{align}
ds_{11}^2 = &- \frac{dt^2}{\left(W_0 Z_1 Z_2 Z_3 \right)^{\frac{2}{3}}} + \left(\frac{Z_1 Z_2 Z_3}{W_0^2}\right)^{\frac{1}{3}} \left[\frac{1}{Z_0} \left(dy_7 +H_0 d\phi\right)^2 + Z_0 \left( e^{2\nu} \left(d\rho^2 + dz^2 \right) +\rho^2 d\phi^2\right)\right] \nonumber\\
&\hspace{-0.5cm} + \left(W_0Z_1 Z_2 Z_3\right)^{\frac{1}{3}} \left[\frac{1}{Z_1} \left(\frac{dy_1^2}{W_1} + W_1 \,dy_2^2 \right) + \frac{1}{Z_2} \left(\frac{dy_3^2}{W_2} + W_2 \,dy_4^2 \right) + \frac{1}{Z_3} \left(\frac{dy_5^2}{W_3} + W_3 \,dy_6^2 \right)\right], \nonumber\\
F_4 = &d\left[ T_1 \,dt \wedge dy_1 \wedge dy_2 \+T_2 \,dt \wedge dy_3 \wedge dy_4 \+T_3\,dt \wedge dy_5 \wedge dy_6 \right]\,. \label{eq:nonBPSfloating}
\end{align}
There are three electric gauge potentials $T_I$ induced by the stacks of M2 branes and a magnetic gauge potential $H_0$ for the KKm vector. We have introduced four warp factors, $Z_\Lambda$, which couple naturally with each gauge potential. In addition, we have four independent warp factors, $W_\Lambda$, which are associated with torus deformations. Specifically, $W_0$ corresponds to a K\"ahler structure deformation of the T$^6$ since $\det($T$^6)=W_0^2$, while the other $W_I$ define complex structure deformations of the internal T$^2$.  Finally, $e^{2\nu}$ determines the nature of the three-dimensional base.  All functions depend only on $\rho$ and $z$ by symmetry. More generically, we use capital greek letters for the label $\Lambda=0,1,2,3$ and capital latin letter for the label $I=1,2,3$. We introduce the cylindrical Laplacian operator of a flat three-dimensional base for axisymmetric functions:
\begin{equation}
\Delta \equi \frac{1}{\rho}\,\partial_\rho \left( \rho \,\partial_\rho \right) \+ \partial_z^2\,.
\end{equation}
The equations of motion \eqref{eq:EOMMtheory} can be decomposed into three almost-linear sectors:
\begin{align}
&\text{\underline{Torus sector:}} \quad \Delta \log W_\Lambda  \= 0\,,\nonumber\\
&\text{\underline{Maxwell sector:}} \nonumber \\
& \Delta \log Z_I \= - {Z_I}^2 \,\left[ (\partial_\rho T_I)^2 + (\partial_z T_I)^2 \right] ,\qquad \partial_\rho \left( \rho\,{Z_I}^2\,\partial_\rho T_I \right)\+\partial_z \left(  \rho\,{Z_I}^2\,\partial_\rho T_I \right)  \=0\,,\label{eq:EOMWeyl}\\
& \Delta \log Z_0 \= - \frac{1}{\rho^2\,{Z_0}^2} \,\left[ (\partial_\rho H_0)^2 + (\partial_z H_0)^2 \right] ,\qquad \partial_\rho \left(\frac{1}{ \rho\,{Z_0}^2}\,\partial_\rho H_0 \right)\+\partial_z \left(  \frac{1}{\rho\,{Z_0}^2}\,\partial_z H_0 \right)  \=0\,,\nonumber\\
&\text{\underline{Base sector:}}  \quad \frac{1}{\rho}\,\partial_z \nu  \=  \cS_z \left(W_\Lambda,Z_\Lambda,T_I,H_0\right)\,, \qquad \frac{1}{\rho}\,\partial_\rho\nu \=  \cS_\rho \left(W_\Lambda,Z_\Lambda,T_I,H_0\right) \nonumber\,,
\end{align}
where $\cS$ are source functions that depend non-trivially on $ \left(W_\Lambda,Z_\Lambda,T_I,H_0\right)$ \eqref{App:nueq}. The equations can be treated  linearly except the Maxwell sector which is a set of non-linear coupled differential equations.

\subsection{Non-BPS floating brane ansatz}
\label{sec:WeylAns}

In \cite{Bah:2020pdz}, a procedure has been found to extract linear closed-form solutions from the above equations without reducing to BPS equations. In the present context, charged Weyl solutions are determined by \emph{eight harmonic functions}
\begin{equation}
\Delta \log W_\Lambda\=\Delta L_\Lambda \= 0\,.
\end{equation}
The pairs of scalars $(Z_I, T_I)$ and $(Z_0,H_0)$ are given by\footnotemark
\begin{equation}
Z_\Lambda =  \frac{\sinh(a_\Lambda \,L_\Lambda +b_\Lambda)}{a_\Lambda}  \,,\qquad \star_3 d\left(H_0\,d\phi\right) = dL_0\,, \qquad T_I =  - \frac{1}{Z_I} \,\sqrt{1+a_I^2 Z_I^2} \,,
\label{eq:WGF&H}
\end{equation}
where we have defined four pairs of gauge field parameters $(a_\Lambda,b_\Lambda)$ with $a_\Lambda >0$. 
The equations for $\nu$ are 
\begin{equation}
\begin{split}
\frac{2}{\rho}\,\partial_z \nu \= &  \sum_{\Lambda=0}^3 \left[ \partial_\rho \log W_\Lambda \,\partial_z \log W_\Lambda + a_\Lambda^2\, \partial_\rho L_\Lambda \,\partial_z L_\Lambda \right], \\
\frac{4}{\rho}\,\partial_\rho \nu \=&   \sum_{\Lambda=0}^3 \left[ \left(\partial_\rho \log W_\Lambda\right)^2 -\left( \partial_z \log W_\Lambda\right)^2+ a_\Lambda^2\, \left( \left(\partial_\rho L_\Lambda\right)^2 - \left(\partial_z L_\Lambda\right)^2 \right)\right]\,.\label{eq:nuEq}
\end{split}
\end{equation}
The integrability is guaranteed by the harmonicity of the functions on the right-hand sides. These equations can be integrated in a case-by-case manner depending on the choice of sources for the harmonic functions.
\footnotetext{ More generically, the scalars are given by
\begin{equation}
Z_\Lambda = \cG^{(\Lambda)}_\ell\left(L_\Lambda\right) \,,\qquad \star_3 d\left(H_0\,d\phi\right) = dL_0\,, \qquad dT_I =  \frac{dL_I}{ \cG^{(I)}_\ell(L_I)^2} \,,
\label{eq:WGF&Hfoot}
\end{equation}
where $\cG^{(\Lambda)}_\ell$ can be freely chosen among the following five generating functions of one variable:
\begin{align}
\cG^{(\Lambda)}_1 (x) &= \frac{\sinh(a_\Lambda x+b_\Lambda)}{a_\Lambda} \,,\qquad \cG^{(\Lambda)}_2 (x) = i\,\frac{\cosh(a_\Lambda x+b_\Lambda)}{a_\Lambda}\,,\qquad \cG^{(\Lambda)}_3 (x) = x +b_\Lambda\,,\nonumber \\
\cG^{(\Lambda)}_4 (x) &= \frac{\sin(a_\Lambda x+b_\Lambda)}{a_\Lambda} \,,\qquad \cG^{(\Lambda)}_5(x) =\frac{\cos(a_\Lambda x+b_\Lambda)}{a_\Lambda}\,,\qquad a_\Lambda \in \mathbb{R}_+, \quad b_\Lambda\in \IR \,.
\label{eq:DefFi}
\end{align}
However, we did not find regular solutions for the branches $\ell=2,4,5$ (see \cite{Bah:2020pdz,Bah:2021owp,Bah:2021rki} for more details). Therefore, we consider only the branch $\ell=1$ from which the function $\ell=3$ can be obtained with \eqref{eq:BPSlimit}.}

It is important to point out that $W_\Lambda$ are ``vacuum'' warp factors, as they are not coupled to any gauge potentials, whereas $Z_\Lambda$ are generated and induced by the gauge potentials. However, one can take a neutral limit by sending all gauge potentials to zero while keeping $Z_\Lambda$ non-trivial, which then satisfy the vacuum equation $\Delta \log Z_\Lambda =0$. The neutral limit is given by
\begin{equation}
(b_\Lambda, a_\Lambda) \to \infty \qquad\text{with}\quad e^{b_\Lambda}/a_\Lambda \quad\text{and}\quad a_\Lambda L_\Lambda\quad \text{held fixed.}
\label{eq:neutrallimit}
\end{equation}
It shows that the present ansatz is generically in the non-BPS regime of the M2-M2-M2-KKm system.

Moreover, at the other side of the parameter space,
\begin{equation}
(b_\Lambda, a_\Lambda) \to 0 \qquad\text{with}\quad b_\Lambda/a_\Lambda \quad \text{held fixed,}
\label{eq:BPSlimit}
\end{equation}
the solutions converge to the linear branch $Z_\Lambda = L_\Lambda+b_\Lambda/a_\Lambda$. As we will see in the next section, this leads to the BPS equations for the M2-M2-M2-KKm system. Therefore, the ansatz admits a BPS limit considering the above transformation.

It is remarkable that one can deviate significantly from the BPS regime while maintaining the linearity of the Einstein-Maxwell equations. Moreover, it only changes the warp factors from a linear function (BPS regime) to a $\sinh$ function (non-BPS regime) in terms of harmonic functions. As we will see later, the price to pay for having a linear ansatz will be to use non-BPS sources that have a fixed charge-to-mass ratio, but not necessarily 1.

\subsection{BPS regime}

We consider that the two-tori and six-torus are rigid, $W_\Lambda=1$. Moreover, we take the limit \eqref{eq:BPSlimit} for the scalars $(Z_\Lambda, H_0,T_I)$ \eqref{eq:WGF&H}, which leads to $L_\Lambda \= Z_\Lambda - b_\Lambda/a_\Lambda$. Therefore, the solutions can be better written such that $Z_\Lambda$ are harmonic functions and the gauge potentials are determined by
\begin{equation}
\Delta Z_\Lambda \= 0\,,\qquad \star_3 d (H_0\,d\phi) \= dZ_0\,,\qquad T_I \= - \frac{1}{Z_I}\,.
\end{equation}
We recognize the BPS equations of motion one can obtain from supersymmetric static M2-M2-M2-KKm solutions. With such a choice, $\nu=0$ \eqref{eq:nuEq}, and the base is flat $\IR^3$ as expected. Physical solutions are necessarily sourced by point particles on the $z$-axis:
\begin{equation}
Z_\Lambda \= \sum_i^n \frac{q_i^{(\Lambda)}}{\sqrt{\rho^2+(z-z_i)^2}}\+ c_\Lambda\,, \qquad H_0 \= \sum_i^n \frac{q_i^{(0)}\,(z-z_i)}{\sqrt{\rho^2+(z-z_i)^2}}\+ \bar{c}_0\,.
\end{equation}
The sources at $(\rho=0,z=z_i)$ are either static extremal four-charge black holes if $q_i^{(\Lambda)} \neq 0$ or smooth Gibbons-Hawking centers if $q_i^{(I)}=0$. 

Therefore, the linear non-BPS floating-brane ansatz contains the BPS floating brane ansatz for the static M2-M2-M2-KKm system.


\subsection{Non-BPS regime}
\label{sec:GenWeylSol}

Generic non-BPS solutions are constructed for non-zero $(a_\Lambda,b_\Lambda)$ in \eqref{eq:WGF&H}. One can check that sourcing the harmonic functions by point sources lead to naked singularities. We therefore consider that the harmonic functions are sourced by \emph{rods}, that is segment sources.

\begin{figure}[h]
\centering
\includegraphics[width=0.23\textwidth]{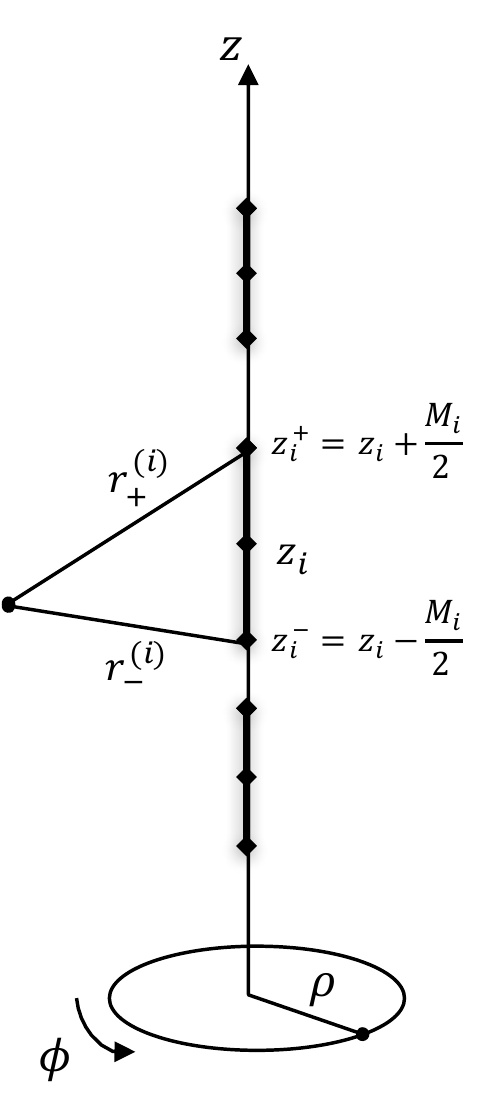}
\caption{Schematic description of axisymmetric rod sources.}
\label{fig:RodsSources}
\end{figure}

In this section, we establish our formalism and highlight the basic ingredients that can be added linearly through the ansatz. We save the construction of the solutions for later.

 We consider $n$ distinct rods of length $M_i>0$ along the $z$-axis centered around $z=z_i$. Without loss of generality, we can order them as $z_i < z_j$ for $i<j$. Our conventions are illustrated in Fig.\ref{fig:RodsSources}. The coordinates of the rod endpoints on the $z$-axis and the distance to them are given by
\begin{equation}
z^\pm_i \equi z_i \pm \frac{M_i}{2}\,, \qquad r_\pm^{(i)} \equiv \sqrt{\rho^2 + \left(z-z^\pm_i\right)^2}\,.
\label{eq:coordinatesEndpoints}
\end{equation}
Moreover, we introduce the following generating functions
\begin{align}
& R_\pm^{(i)} \equiv r_+^{(i)}+r_-^{(i)}\pm M_i\,, \label{eq:Rpmdef}\\
&E_{\pm \pm}^{(i,j)} \equi r_\pm^{(i)} r_\pm^{(j)} + \left(z-z_i^{\pm}  \right)\left(z-z_j^{\pm}  \right) +\rho^2\,, \qquad \nu_{ij} \equi \log \frac{E_{+-}^{(i,j)}E_{-+}^{(i,j)}}{E_{++}^{(i,j)}E_{--}^{(i,j)}}\,.\nonumber
\end{align}
They satisfy the equations
\begin{align}
&\Delta\left(\log \frac{R_+^{(i)}}{R_-^{(i)}}  \right) \=0 \,,\qquad \frac{1}{\rho} \partial_z \nu_{ij} \= \partial_{(\rho} \left(\log \frac{R_+^{(i)}}{R_-^{(i)}}  \right)  \partial_{z)} \left(\log \frac{R_+^{(j)}}{R_-^{(j)}}  \right)\,,\nonumber\\
&\frac{1}{\rho} \partial_\rho \nu_{ij} \= \partial_{\rho} \left(\log \frac{R_+^{(i)}}{R_-^{(i)}}  \right)  \partial_{\rho} \left(\log \frac{R_+^{(j)}}{R_-^{(j)}}  \right) \-  \partial_{z} \left(\log \frac{R_+^{(i)}}{R_-^{(i)}}  \right)  \partial_{z} \left(\log \frac{R_+^{(j)}}{R_-^{(j)}}  \right)\,.
\end{align}
The functions $\log \frac{R_+^{(i)}}{R_-^{(i)}} $ can be used as the building blocks for the eight harmonic functions:
\begin{equation}
\log W_\Lambda = \sum_{i=1}^n G^{(\Lambda)}_i \,\log  \frac{R_+^{(i)}}{R_-^{(i)}}\,,\qquad L_\Lambda = \sum_{i=1}^n P^{(\Lambda)}_i \,\log  \frac{R_+^{(i)}}{R_-^{(i)}}\,,
\label{eq:HarmFuncGen}
\end{equation}
where the constants, $G^{(\Lambda)}_i$ and $P^{(\Lambda)}_i$, define the \emph{weights} of the $i^\text{th}$ rod. The warp factors and gauge potentials follow from \eqref{eq:WGF&H},
\begin{equation}
Z_\Lambda \= \frac{\sinh \left( a_\Lambda L_\Lambda +b_\Lambda \right)}{a_\Lambda}\,,\qquad T_I \= -\frac{1}{Z_I}\,\sqrt{1+a_I^2\,Z_I^2}\,,\qquad H_0 \= \sum_{i=1}^n P^{(0)}_i \left(r_-^{(i)}-r_+^{(i)} \right)\,,
\label{eq:HarmFuncGen2}
\end{equation} 
and $\nu$ is induced by the generic functions $\nu_{ij}$:
\begin{equation}
\nu \= \frac{1}{4} \sum_{i,j=1}^n \alpha_{ij}\,\nu_{ij}\,, \qquad \alpha_{ij} \equi  \sum_{\Lambda =0}^3  \left[ G_i^{(\Lambda)}G_j^{(\Lambda)}+ a_\Lambda^2\,P_i^{(\Lambda)}P_j^{(\Lambda)} \right]\,.
\end{equation}

As we will see precisely later, the weights of the torus warp factors,  $G_i^{(\Lambda)}$, are not involved in any of the conserved charges of the solutions as they are deformations of the T$^6$.  However, the four $P_i^{(\Lambda)}$ are associated to the four charges carried by the $i^\text{th}$ rod, $(Q^1_{\text{M2},i},Q^2_{\text{M2},i},Q^3_{\text{M2},i},Q_{\text{KKm},i})$, \footnote{We have normalized the charges in units of volume: 
\begin{equation}
\begin{split}
Q^I_{\text{M2},i} &= \frac{1}{4\pi\,\text{Vol}(T_I^4\times S^1)}\,\int_{z=z_i^+}^{z_i^-}  \int_{\phi=0}^{2\pi} \int_{T_I^4\times S^1}\, \star F_4 \bigl|_{\rho=0} = M_i \,P^{(I)}_i,\\
 Q_{\text{KKm},i} &=- \frac{1}{4\pi}\int_{z=z_i^+}^{z_i^-}  \int_{\phi=0}^{2\pi}  \,d(H_0\,d\phi)\bigl|_{\rho=0} =  M_i\, P_i^{(0)},
\end{split}
\end{equation}
where $T_I^4$ is the four-torus orthogonal to the $I^\text{th}$ stack of M2 branes and $S^1$ is the $y_7$-fiber. }

\begin{equation}
Q^I_{\text{M2},i}  \= M_i \,P^{(I)}_i,\qquad Q_{\text{KKm},i}  \=  M_i\, P_i^{(0)}.
\label{eq:individualMcharges}
\end{equation}

Moreover, we will have to add a semi-infinite rod to construct asymptotically-$\IR^{1,4}\times$T$^6$ solutions as we will see in detail in the section \ref{sec:5dNonBPSGen}.  It can be obtained from the above expressions by sending one of the endpoints to infinity. 

Finally, the non-BPS solutions have a very large phase space determined by the geometry of the rod sources, the eight weights associated with each and the four pairs of gauge field parameters $(a_\Lambda,b_\Lambda)$. However, the parameters will be constrained to have physical solutions as we will discuss in section \ref{sec:Classification}.

\subsection{Profile in five and four dimensions}

In this section, we describe the ansatz in five and four dimensions after reduction along the T$^6$ and the S$^1$. We will keep the least number of degrees of freedom that encompasses the ansatz. Note that we do not expect to obtain the Lagrangian of the STU model as it is usually the case for BPS multicenter solutions  \cite{Bena:2004de,Bena:2005va,Berglund:2005vb,Bena:2010gg,Vasilakis:2011ki,Heidmann:2017cxt,Bena:2017fvm} and some of their non-BPS extensions \cite{Bena:2009qv,DallAgata:2010srl,Bossard:2014yta,Bena:2015drs}. Indeed, we are not compactifying along a rigid T$^6$ since it has non-trivial deformations due to the $W_\Lambda$ \eqref{eq:nonBPSfloating}. 

We will therefore apply a series of KK reductions from M-theory using the generic rules of \cite{Lu:1995yn,Cremmer:1997ct}. The reduction rules and the derivation are detailed in the appendix \ref{App:5d&4dframe}.

\subsubsection{Reduction to five dimensions} 
\label{sec:5dreduction}

We first perform a series of KK reductions to five dimensions by compactifying along $\{y_i\}_{i=1,..,6}$. For that purpose, we suitably define seven scalars $(X^I,\Phi_\Lambda)$ and three one-form gauge fields such that the M-theory metric and gauge field \eqref{eq:nonBPSfloating} is now given as
\begin{equation}
d s_{11}^{2} =e^{-\frac{2}{3} \Phi_0}  \, d s_{5}^{2} \+ e^{\frac{1}{3}\Phi_0}\, \sum_{I=1}^{3} X^I \left( e^{-\Phi_I} dy_{2I-1}^2+ e^{\Phi_I} dy_{2I}^2 \right)\,,\quad F_4  = \sum_{I=1}^{3} F^{(I)} \wedge dy_{2I-1} \wedge dy_{2I} ,
\label{eq:KKredGen}
\end{equation}
where $F^{(I)}=dA^{(I)}$ are two-form field strengths. The reduced five-dimensional action is (see the appendix \ref{App:5d&4dframe})
\begin{align}
(16 \pi G_5) S_5 \= &\int \left( \cL_{STU}^{5D}\- \frac{1}{2}\,\sum_{\Lambda=0}^3 \star d\Phi_\Lambda \wedge \Phi_\Lambda \right),
\label{eq:KKredAction5d}
\end{align}
where $G_5=\frac{G_{11}}{(2\pi)^6 \prod_{a=1}^6 R_{y_a}}$ is the five-dimensional Newton's constant, $\star$ is the Hodge star operator in five dimensions, and $\cL_{STU}^{5D}$ is the STU Lagrangian of five-dimensional $\cN=2$ supergravity:
\begin{equation}
\cL_{STU}^{5D} = R\,\star \mathbb{I} - Q_{IJ} \left( \star F^{(I)} \wedge F^{(J)} + \star d X^I \wedge d X^J \right) + \frac{|\epsilon_{IJK}|}{2} \,A^{(I)} \wedge F^{(J)} \wedge F^{(K)}\,,
\end{equation}
where repeated indices are summed over, $\epsilon_{IJK}$ is the rank-$3$ Levi-Civita tensor, and we have defined the scalar kinetic term $Q_{IJ}=\frac{1}{2}\, \text{diag} \left((X^1)^{-2} ,(X^2)^{-2} ,(X^3)^{-2} \right)$. 

In this framework, the ansatz \eqref{eq:nonBPSfloating} defines solutions of the above action such that
\begin{align}
ds_5^2 &\= - \frac{dt^2}{\left(Z_1 Z_2 Z_3 \right)^{\frac{2}{3}}} \+ \left(Z_1 Z_2 Z_3 \right)^{\frac{1}{3}} \left[\frac{1}{Z_0} \left(dy_7 +H_0 \,d\phi\right)^2 \+ Z_0 \left( e^{2\nu} \left(d\rho^2 + dz^2 \right) +\rho^2 d\phi^2\right)\right], \nonumber \\
X^I &\= \frac{|\epsilon_{IJK}|}{2} \left(\frac{Z_J Z_K}{Z_I^2} \right)^\frac{1}{3},\qquad \Phi_\Lambda \= \log W_\Lambda \,,\qquad F^{(I)} \= dA^{(I)} \= dT_I \wedge dt\,.
\label{eq:5dframework}
\end{align}
Remarkably, the torus warp factors in M-theory, $W_\Lambda$, decouple entirely and the five-dimensional metric is independent of these factors.  Therefore, from a five-dimensional point of view, these functions are simple harmonic scalar decorations on an STU background. 

For solutions that are asymptotic to five-dimensional flat space, the ADM mass is obtained from the following asymptotic expansion\footnote{We use the conventions of \cite{Myers:1986un}.}
\begin{equation}
\cM_5 \= \frac{3\pi}{8 G_5}\,\underset{r\to \infty}{\lim} \, r^2\,\left( 1-\left(Z_1 Z_2 Z_3 \right)^{-\frac{2}{3}}\right)\,,
\label{eq:ADMmass5d}
\end{equation}
where $r$ is the five-dimensional radial coordinate, related to the cylindrical coordinates such as $(\rho,z)=\frac{1}{2}(r^2\sin 2\theta ,r^2\cos 2\theta)$. Note that the sources in $W_\Lambda$ does not indeed induce mass.

\subsubsection{Reduction to four dimensions}
\label{sec:4dreduction}

We further reduce to four dimensions after compactification along $y_7$. One can simply use known results about the reduction of the STU model from five to four dimensions (see for instance \cite{Goldstein:2008fq,DallAgata:2010srl,Bah:2021jno}). Therefore, the reduced four-dimensional action is given by 
\begin{align}
(16 \pi G_4) S_4 \= &\int \left( \cL_{STU}^{4D} \- \frac{1}{2}\,\sum_{\Lambda=0}^3 \star d\Phi_\Lambda \wedge \Phi_\Lambda \right),
\label{eq:KKredAction4d}
\end{align}
where $G_4 \= \frac{G_5}{2\pi R_{y_7}}$ is the four-dimensional Newton's constant, $\star$ is now the Hodge star in four dimensions, and $\cL_{STU}^{4D}$ is the STU Lagrangian of four-dimensional $\cN=2$ supergravity. This Lagrangian is better written in terms of three independent complex scalars $z^I$ and four two-form field strengths, $\bar{F}^\Lambda= d\bar{A}^\Lambda$,
\begin{equation}
\cL_{STU}^{4D} \= R \,\star \mathbb{I} - 2 g_{IJ} \,\star dz^I \wedge d\bar{z}^J \- \frac{1}{2} \mathcal{I}_{\Lambda \Sigma} \,\star \bar{F}^\Lambda \wedge \bar{F}^\Sigma + \frac{1}{2} \cR_{\Lambda \Sigma}\, \bar{F}^\Lambda \wedge \bar{F}^\Sigma.
\end{equation}
We refer the reader to the appendix \ref{App:4dframe} for the scalar and gauge field couplings, $g_{IJ}$, $\mathcal{I}_{\Lambda \Sigma}$ and $\cR_{\Lambda \Sigma}$ \eqref{eq:gscalcouplingApp}, \eqref{eq:IscalcouplingApp} and \eqref{eq:RscalcouplingApp}. The four-dimensional metric, scalar and gauge fields are derived from five-dimensional fields such as
\begin{equation}
ds_5^2 = e^{2\Phi} \,ds_4^2 + e^{-4\Phi} (dy_7-\bar{A}^0)^2,\quad
A^{(I)} = \bar{A}^I \+ C^I (dy_7 - \bar{A}^0),\quad z^I = C^I + i \,e^{-2\Phi} X^I\,.
\end{equation}
In this context, our ansatz \eqref{eq:nonBPSfloating} consists of a metric, three purely imaginary complex scalars, four real scalars and four gauge fields given by
\begin{align}
ds_4^2 &\= -\frac{dt^2}{\sqrt{Z_0 Z_1 Z_2 Z_3}} \+ \sqrt{Z_0 Z_1 Z_2 Z_3} \left( e^{2\nu} \left(d\rho^2 + dz^2 \right) +\rho^2 d\phi^2\right), \nonumber \\
z^I &\= i\,\frac{|\epsilon_{IJK}|}{2} \sqrt{\frac{Z_J Z_K}{Z_0\,Z_I} },\quad \Phi_\Lambda \= \log W_\Lambda  \,,\quad \bar{F}^{0}  \= -dH_0 \wedge d\phi \,,\quad \bar{F}^{I}  \= dT_I \wedge dt\,.
\label{eq:4dSol}
\end{align}
For solutions that are asymptotic to four-dimensional flat space, the ADM mass is obtained from the following asymptotic expansion
\begin{equation}
\cM_4 \= \frac{1}{2 G_4}\,\underset{r\to \infty}{\lim} \, r\,\left( 1-\left(Z_0 Z_1 Z_2 Z_3 \right)^{-\frac{1}{2}}\right)\,,
\label{eq:ADMmass4d}
\end{equation}
where $r$ is the radial coordinate in four dimensions, $(\rho,z)=(r\cos \theta,r\sin \theta)$.

\section{Classification of asymptotically-flat solutions}
\label{sec:Classification}

In this section, we construct families of non-BPS solutions introduced in section \ref{sec:GenWeylSol}, and discuss their regularity. We first discuss solutions that are asymptotic to a four-dimensional flat space plus compactification circles and then build solutions asymptotic to five-dimensional flat space. 
For the self-consistency of this section, we remind the ansatz of metric and gauge field in M-theory \eqref{eq:nonBPSfloating}:
\begin{align} ds_{11}^2 = &- \frac{dt^2}{\left(W_0Z_1 Z_2 Z_3 \right)^{\frac{2}{3}}} + \left(\frac{Z_1 Z_2 Z_3}{W_0^2}\right)^{\frac{1}{3}} \left[\frac{1}{Z_0} \left(dy_7 +H_0 d\phi\right)^2 + Z_0 \left( e^{2\nu} \left(d\rho^2 + dz^2 \right) +\rho^2 d\phi^2\right)\right] \nonumber\\
&\hspace{-0.5cm} + \left(W_0 Z_1 Z_2 Z_3\right)^{\frac{1}{3}} \left[\frac{1}{Z_1} \left(\frac{dy_1^2}{W_1} + W_1 \,dy_2^2 \right) + \frac{1}{Z_2} \left(\frac{dy_3^2}{W_2} + W_2 \,dy_4^2 \right) + \frac{1}{Z_3} \left(\frac{dy_5^2}{W_3} + W_3 \,dy_6^2 \right)\right], \nonumber\\
F_4 = &d\left[ T_1 \,dt \wedge dy_1 \wedge dy_2 \+T_2 \,dt \wedge dy_3 \wedge dy_4 \+T_3\,dt \wedge dy_5 \wedge dy_6 \right]\,. \label{eq:nonBPSfloatingRemind}
\end{align}

\subsection{Four-dimensional solutions}
\label{sec:4dNonBPSGen}

We consider that $y_7$ is an extra S$^1$ of $2\pi R_{y_7}$ periodicity.
Generic non-BPS solutions are given by \eqref{eq:WGF&H}, and are sourced by $n$ rods as depicted in Fig.\ref{fig:RodsSources}. The warp factors and gauge potentials are given by eight weights at each rod, $(P_i^{(\Lambda)},G_i^{(\Lambda)})$, and four pairs of gauge field parameters, $(a_\Lambda,b_\Lambda)$, such that
\begin{align}
&Z_\Lambda= \frac{1}{2a_\Lambda} \left[e^{b_\Lambda} \prod_{i=1}^n \left(  \frac{R_+^{(i)}}{R_-^{(i)}}\right)^{a_\Lambda P^{(\Lambda)}_i}-e^{-b_\Lambda} \prod_{i=1}^n \left(  \frac{R_-^{(i)}}{R_+^{(i)}}\right)^{a_\Lambda P^{(\Lambda)}_i} \right]\,,\qquad W_\Lambda = \prod_{i=1}^n \left(  \frac{R_+^{(i)}}{R_-^{(i)}}\right)^{G^{(\Lambda)}_i} \,,\nonumber\\
&H_0 \=\sum_{i=1}^n P^{(0)}_i \left(r_-^{(i)}-r_+^{(i)} \right)\,,\quad T_I\=-\frac{1}{Z_I}\,\sqrt{1+a_I^2\,Z_I^2},\quad e^{2\nu} \=\prod_{i,j=1}^n\, \left(  \frac{E_{+-}^{(i,j)}E_{-+}^{(i,j)}}{E_{++}^{(i,j)}E_{--}^{(i,j)}}\right)^{\frac{1}{2}\,\alpha_{ij}} \,,
\label{eq:WarpfactorsRods}
\end{align}
where we remind the main functions and variables
\begin{align}
& z^\pm_i \equi z_i \pm \frac{M_i}{2}\,, \qquad r_\pm^{(i)} \equiv \sqrt{\rho^2 + \left(z-z^\pm_i\right)^2}\,,\qquad  R_\pm^{(i)} \equiv r_+^{(i)}+r_-^{(i)}\pm M_i\,, \label{eq:RpmdefRemind}\\
&E_{\pm \pm}^{(i,j)} \equi r_\pm^{(i)} r_\pm^{(j)} + \left(z-z_i^{\pm}  \right)\left(z-z_j^{\pm}  \right) +\rho^2\,, \qquad \nu_{ij} \equi \log \frac{E_{+-}^{(i,j)}E_{-+}^{(i,j)}}{E_{++}^{(i,j)}E_{--}^{(i,j)}}\,,\nonumber
\end{align}
and the exponents
\begin{equation} \alpha_{ij} \equi  \sum_{\Lambda =0}^3  \left[ G_i^{(\Lambda)}G_j^{(\Lambda)}+ a_\Lambda^2\,P_i^{(\Lambda)}P_j^{(\Lambda)} \right]\,.
\label{eq:Defalpha}
\end{equation}

We have defined a family of non-BPS solutions given by $10 n + 8$ parameters. We now have to discuss the regularity of the solutions that constrains the parameter space. 

\subsubsection{Regularity constraints}
\label{sec:regM4d}

Potential constraints come from coordinate singularities on the $z$-axis, regularity of the spacetime elsewhere, and conditions on the asymptotic. The details of the analysis can be found in the appendix \ref{App:Reg4d}.

\begin{itemize}
\item[•] \textbf{Condition on the asymptotics:}

We consider the asymptotic spherical coordinates $(\rho,z)=(r\sin \theta, r\cos\theta)$. At large $r$, we have
\begin{equation}
Z_\Lambda \,\sim\, \frac{\sinh b_\Lambda}{a_\Lambda}, \qquad W_\Lambda \,\sim\, 1\,,\qquad e^{2\nu} \sim 1\,.
\end{equation} 
Therefore, the four-dimensional metric of the solutions \eqref{eq:4dSol} is asymptotically flat if
\begin{equation}
 \prod_{\Lambda=0}^3  \frac{\sinh b_\Lambda}{a_\Lambda} \= 1\,.
\label{eq:condonasymp}
\end{equation}
\item[•] \textbf{Regularity out of the $z$-axis:}

 From \eqref{eq:RpmdefRemind}, it is clear that the ratio $R_+^{(i)}/R_-^{(i)}$ diverges at the $i^\text{th}$ rod. To avoid extra zeroes out of the $z$-axis, one needs $Z_\Lambda > 0$ \eqref{eq:WarpfactorsRods}, which requires, if $b_\Lambda \neq +\infty$, 
\begin{equation}
b_\Lambda \geq 0 \,,\quad a_\Lambda \,P_i^{(\Lambda)}\, \geq\, 0 \,,\qquad \Lambda=0,...,3, \,\, i=1,...,n.
\label{eq:RegPositivity}
\end{equation}
With this condition, the solutions are regular out of the $z$-axis. 
\item[•] \textbf{Regularity at the rods on the $z$-axis:}

We consider the local spherical coordinates centered around the $i^\text{th}$ rod, \be
\rho= \sqrt{r_{i}\left(r_{i}+M_{i}\right)} \sin \theta_i,\, \qquad z\= \left(r_{i}+\frac{M_{i}}{2}\right) \cos \theta_{i}+z_{i}. 
\label{eq:localcoordrod}
\ee
 At the rod, $r_i \to 0$, we have
\begin{equation}
\frac{R_+^{(i)}}{R_-^{(i)}} \propto \frac{1}{r_i}\,,\qquad  \frac{E_{+-}^{(i,i)}E_{-+}^{(i,i)}}{E_{++}^{(i,i)}E_{--}^{(i,i)}} \propto r_i^2,
\end{equation}
while all other quantities are well-behaved. From \eqref{eq:WarpfactorsRods}, we have\footnote{Since $a_\Lambda P_i^{(\Lambda)}\geq 0$ is required if $b_\Lambda \neq \infty$, $Z_\Lambda$ can vanish at the rod only if one takes $b_\Lambda = \infty$, that is the neutral limit \eqref{eq:neutrallimit} for the pair $(Z_\Lambda,T_\Lambda)$. In that case, $Z_\Lambda \,\propto\, r_i^{- a_\Lambda P_i^{(\Lambda)}}$ where $a_\Lambda P_i^{(\Lambda)}$ can be taken negative.}
\begin{equation}
Z_\Lambda \,\propto\, r_i^{- a_\Lambda P_i^{(\Lambda)} }\,, \qquad W_\Lambda \,\propto\, r_i^{-G_i^{(\Lambda)}}\,,\qquad e^{2\nu} \,\propto \, r_i^{\alpha_{ii}}\,,
\end{equation}
and therefore, the metric \eqref{eq:nonBPSfloatingRemind} is singular at the rod, except for specific choices of weights that will characterize the physical rods of the non-BPS floating brane ansatz in M-theory. It will be convenient to use the following aspect ratios, $d_i$,
\begin{equation}
\begin{split}
d_1 &\equi 1\,,\qquad d_i \equi  \prod_{j=1}^{i-1} \prod_{k=i}^n \left(\dfrac{(z_k^- - z_j^+)(z_k^+ - z_j^-)}{(z_k^+ - z_j^+)(z_k^- - z_j^-)}  \right)^{\alpha_{jk}}\quad \text{when } i=2,\ldots n\,.
\end{split}
\label{eq:dialphaDefcharged}
\end{equation}
The regularity at each rod requires that the sources are in one of these categories (see appendix \ref{App:AttheRod} for more details):
\begin{itemize}
\item[-] \underline{\emph{Black rods:}} the $y_a$-fibers and the $\phi$-circle have a finite size at the $i^\text{th}$ rod if
\begin{equation}
G^{(\Lambda)}_i \= 0 \,,\qquad P^{(\Lambda)}_i \= \frac{1}{2 \, a_\Lambda}\,.
\label{eq:blackrodWeight}
\end{equation}
One can check that $\alpha_{ii}=1$ and the timelike Killing vector $\partial_t$ shrinks at the rod which therefore defines the locus of a horizon. The topology of the horizon is either T$^7\times$S$^2$ or T$^6\times$S$^3$ depending on the close environment around the rod (see Fig.\ref{fig:RodCategories} and \ref{fig:TouchingRods}).\footnote{\label{footnote1}The horizon topology depends on the close environment of the black rod. If the rod is connected on one side as in Fig.\ref{fig:TouchingRods}, it corresponds to a T$^6\times$S$^3$ horizon. Otherwise, it is a T$^7\times$S$^2$ horizon  as in Fig.\ref{fig:RodCategories}.} 

Moreover, the surface gravity and horizon area associated to the black hole give\footnote{The surface gravity can be read from the local geometry such as $ds^2 \propto d\rho_i^2 - \rho_i^2 \kappa_i^2 dt^2 + \ldots$ with $\rho_i\to 0$, and is associated to the temperature of the black hole, $\cT_i = \kappa_i/(2\pi )$. Moreover, note that if the rod is connected to another one, for instance $z_{i-1}^+=z_i^-$, the surface gravity is still finite since $(z_{i-1}^+-z_i^-)/d_i=\text{cst}$.}
\be
\begin{aligned}
\kappa_{i} & \equiv \frac{2}{d_{i} M_{i}}  \prod_{\Lambda=0}^4 \sqrt{\frac{a_{\Lambda}}{e^{b_{\Lambda}}}}\, \prod_{j \neq i}\left(\frac{z_{j}^{+}-z_{i}^{-}}{z_{j}^{-}-z_{i}^{-}}\right)^{\operatorname{sign}(i-j) \alpha_{i j}},\\
A_{i} &=\frac{(2 \pi)^8\,M_{i} }{\kappa_{i}}\,\prod_{a=1}^7 R_{y_a}.
\end{aligned}
\label{eq:Area&SurfaceGrav}
\ee
The rod on its own corresponds to the static limit of a four-charge non-extremal black hole \cite{Cvetic:1995kv,Chong:2004na,Chow:2014cca}.\footnote{Since all $P_i^{(\Lambda)}$ are non-zero, the rod carries non-zero M2-M2-M2-KKm charges given by \eqref{eq:individualMcharges}.} However, in the present ansatz, one can stack linearly multiple black holes on a line and their interactions can be studied.
\item[-]  \underline{\emph{Bubble rods:}} One of the T$^6$ directions, for instance $y_1$ or $y_2$, shrinks at the $i^\text{th}$ rod while all other fibers are finite by taking
\begin{equation}
\pm G_i^{(1)} =-G_i^{(0)} = a_0 P_i^{(0)} = a_1 P_i^{(1)} =\frac{1}{2} , \quad G_i^{(2)}=G_i^{(3)}=P_i^{(2)}=P_i^{(3)}= 0,
\label{eq:bubblerodWeight1}
\end{equation}
where the ``$+$'' corresponds to a shrinking $y_1$-circle and the ``$-$'' corresponds to a shrinking $y_2$-circle. The weights for which another T$^6$ direction degenerates can be obtained by permuting the $I=1,2,3$ indexes. Similarly, the $y_7$ circle shrinks while all other fibers remain finite by considering
\begin{equation}
a_0\,P_i^{(0)}\= 1 , \qquad G_i^{(\Lambda)} \= P_i^{(I)} \= 0\,.
\label{eq:bubblerodWeight2}
\end{equation}
For each possibility, the $(r_i, y_a)$-subspace corresponds to an origin of an $\IR^2$ defining the locus of a bubble with a T$^6\times$S$^2$ or T$^5\times$S$^3$ (see Fig.\ref{fig:RodCategories} and \ref{fig:TouchingRods}).\footnote{\label{footnote2}As for a black rod, the local topology of the bubble depends if the bubble rod is disconnected (Fig.\ref{fig:RodCategories}) or connected (Fig.\ref{fig:TouchingRods}).} It corresponds to a smooth coordinate degeneracy if an algebraic equation, which we can denote as a \emph{bubble equation}, is satisfied. For a rod where the $y_1$ or $y_2$ circle degenerates smoothly, the bubble equation relates the asymptotic radius to the internal parameters as follows\footnote{If the rod is connected to another one, e.g. $z_{i-1}^+=z_i^-$, the constraint is always finite since $(z_{i-1}^+-z_i^-)/d_i=\text{cst}$.}
\be
k_{i}\,R_{y_{a}}  \= d_{i} M_{i}\,\sqrt{\frac{e^{b_{1}+b_0}}{a_{1}a_{0}}}\, \prod_{j \neq i}\left(\frac{z_{j}^{+}-z_{i}^{-}}{z_{j}^{-}-z_{i}^{-}}\right)^{\operatorname{sign}(j-i) \alpha_{i j}}, \quad  a=1 \text{ or }2,
\label{eq:RegCondBu}
\ee
where $k_i$ is an orbifold parameter $k_i\in \mathbb{N}$ that corresponds to the order of the conical deficit on the $\IR^2$. The constraint corresponding to the degeneracy of another T$^6$ direction or the $y_7$-circle is obtained by permuting $\frac{e^{b_1}}{a_1} \to \frac{e^{b_I}}{a_I}$ or $\frac{e^{b_1}}{a_1} \to \frac{e^{b_0}}{a_0}$ respectively. Moreover, one can check that the gauge field is regular at the rod (see Appendix \ref{App:AttheRod}).

The smooth bubble where one of the T$^6$ directions degenerates has nonzero $P_i^{(0)}$ and $P_i^{(I)}$. Therefore, it carries only a M2 brane charge and a KKm charge, given by \eqref{eq:individualMcharges}. Similarly, the smooth bubble that shrinks the $y_7$ direction carries only a KKm charge. Moreover, these charges can be freely reduced to zero without changing the topology of the bubble by taking the neutral limit \eqref{eq:neutrallimit}, and the resulting objects will be smooth vacuum bubbles in M-theory.

Therefore, \emph{seven types of physical rod sources} correspond to a smooth coordinate degeneracy of one the T$^6\times$S$^1$ directions defined by an origin of an $\IR^2$ where a smooth bubble is sitting.
\end{itemize}
\begin{figure}[h]
\centering
\includegraphics[width=0.8\textwidth]{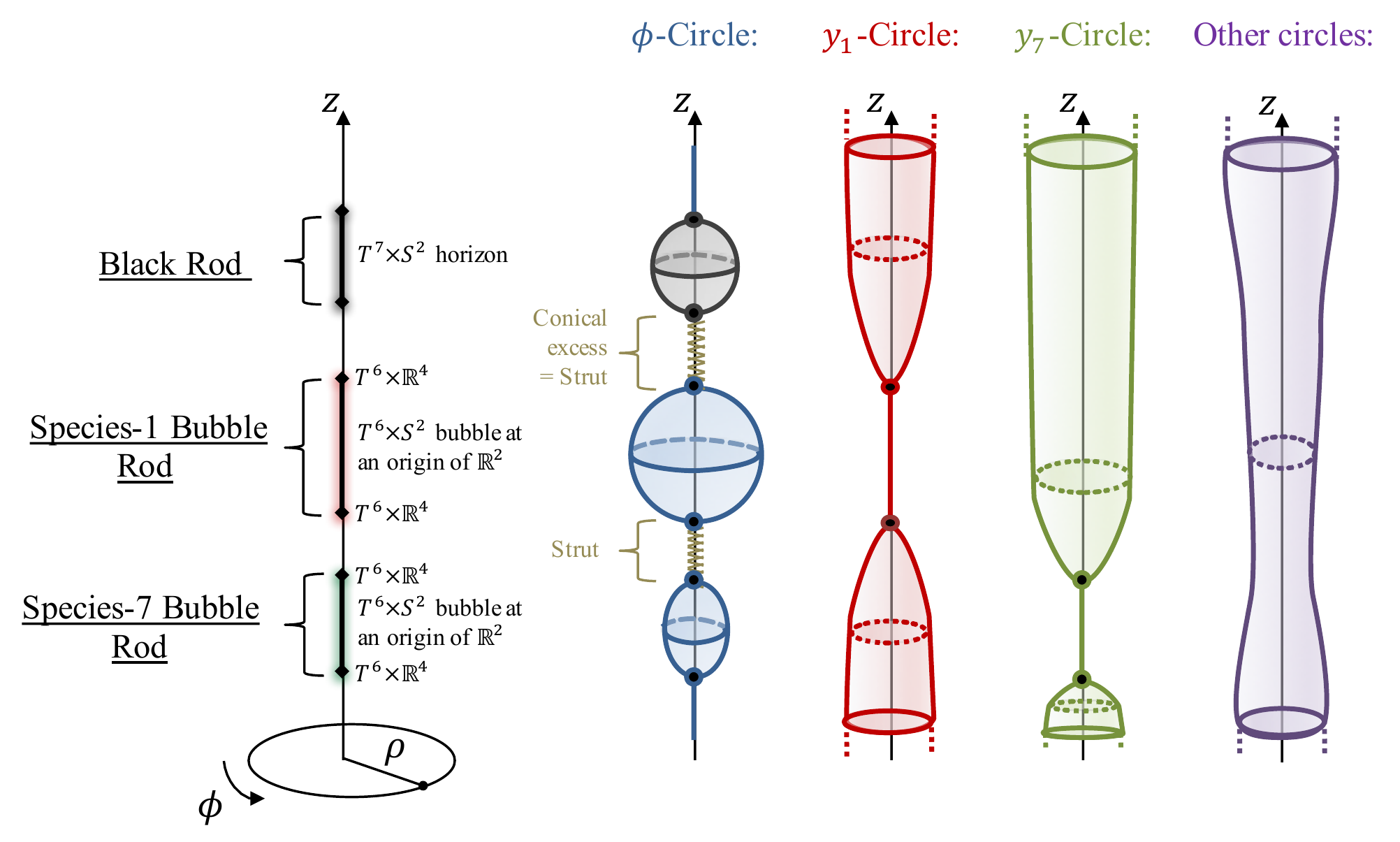}
\caption{Schematic description of three physical rods and the behavior of the circles on the $z$-axis when the rod are disconnected. The black rod sources a static M2-M2-M2-KKm black hole with a T$^7\times$S$^2$ horizon. The species-1 bubble rod corresponds to the degeneracy of the $y_1$-circle inducing a T$^6\times$S$^2$ M2-KKm bubble The species-7 bubble rod induces the degeneracy of the $y_7$-circle and a T$^6\times$S$^2$ KKm bubble. Each section in between the sources has a conical excess, defining a strut.}
\label{fig:RodCategories}
\end{figure}

\begin{figure}[h]
\centering
\includegraphics[width=0.8\textwidth]{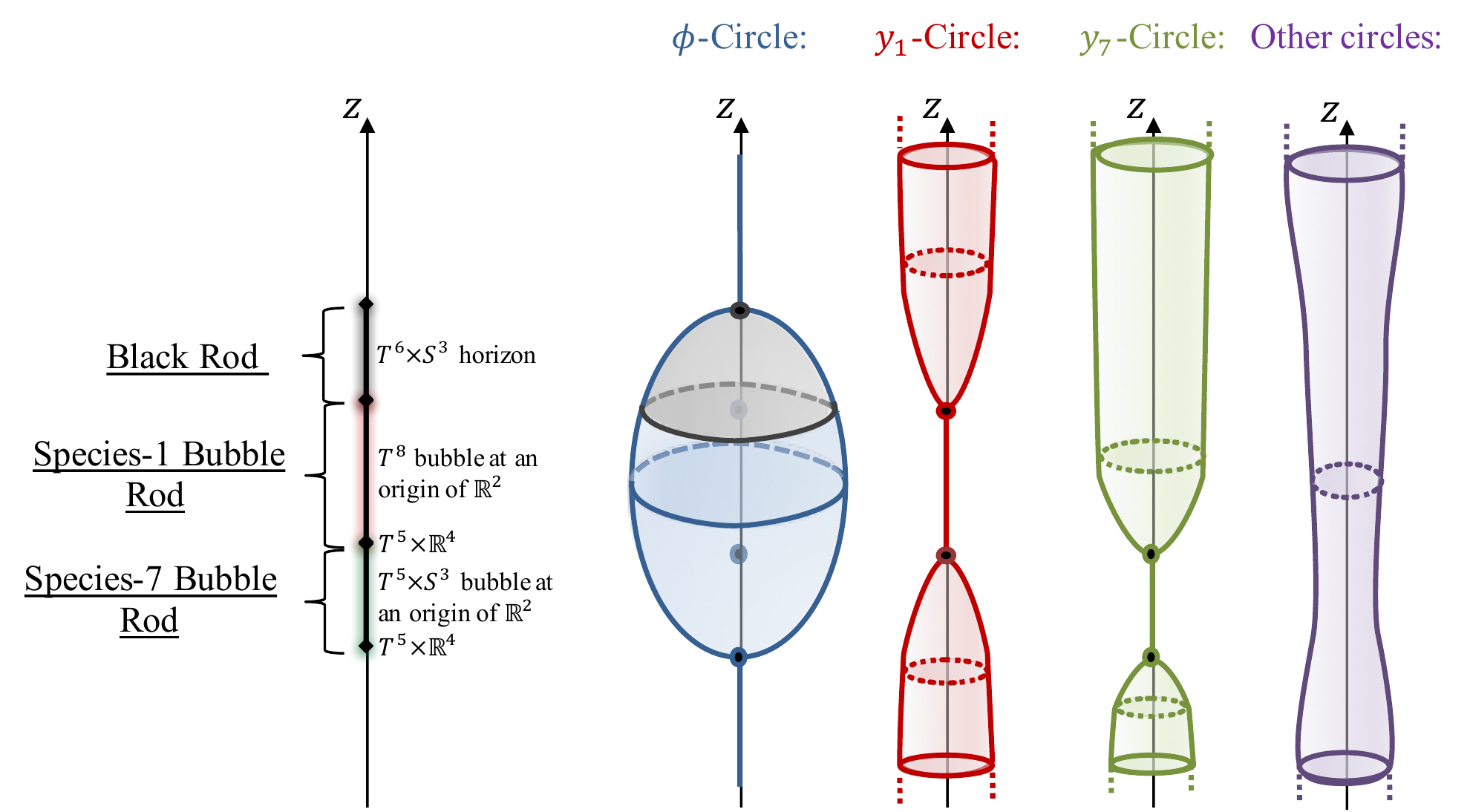}
\caption{Schematic description of connected rods by making the rods of Fig.\ref{fig:RodCategories} touch. The $\phi$-circle does not shrink along the rod configuration, and the solution is strut-free.}
\label{fig:TouchingRods}
\end{figure}

There are 8 categories of sources that can be used to generate physical asymptotically-$\IR^{1,3}$ solutions in M-theory: a non-extremal M2-M2-M2-KKm black hole, 6 non-BPS M2-KKm bubbles and a non-BPS KKm bubble. Remarkably, we have for any choices:
\begin{equation}
\begin{split}
\alpha_{jk}& \= \begin{cases} 
1 \qquad &\text{if the }j^\text{th}\text{ and }k^\text{th}\text{ rods are of the same nature,}  \\
\frac{1}{2} \qquad &\text{otherwise,}
\end{cases} 
\end{split}
\label{eq:dialphaDefcharged2}
\end{equation}
which means that the three-dimensional base does not depend on the specific nature of the sources.

Finally, one can think of other types of sources that are singular in M-theory but regular in other duality frameworks. This will be clarified when discussing the ansatz in other string theory frameworks in the section \ref{sec:dualFrame}.

The effective mass induced by the $i^\text{th}$ rod, $\cM_i$, can be derived from \eqref{eq:ADMmass4d} by isolating the contribution of the rod. We find
\begin{equation}
\cM_i \= \frac{1}{4 G_4} \left(a_0 \coth b_0\, Q_{\text{KKm},i}+ \sum_{I=1}^3 a_I \coth b_I\, Q_{\text{M2},i}^I\right)\,,
\end{equation}
where the brane charges carried by the rod are given in \eqref{eq:individualMcharges}. Therefore, the charge-to-mass ratios induced by the branes are determined by $a_\Lambda \coth b_\Lambda$, and are the same for all the rods. It is a consequence for having a linear ansatz. If it still allows to remarkably deviate from the BPS regime, one cannot have two rods with opposite charges for instance.

\newpage
\item[•] \textbf{Regularity on the $z$-axis out of the rods:}

At $\rho=0$ and $z \notin [ z_i^-,z_i^+]$, the functions $R_+^{(i)}/R_-^{(i)}$ and $E_{\pm \pm }^{(i,j)}$ are finite and non-zero. Therefore, the warp factors are well-defined and the $\phi$-circle degenerates as the cylindrical coordinate singularity. The analysis reduces to the regularity of the three-dimensional base, $ds_3^2 =  e^{2\nu} \left(d\rho^2 + dz^2 \right) +\rho^2 d\phi^2$, and the warp factor $e^{2\nu}$ can induce conical singularities. More precisely, we have
\begin{equation}
e^{2 \nu} \sim \begin{cases}
~~ 1  &\text{if } z < z_{1}^-  \text{ and } z > z_{n}^+,\\
~~ d_i^2\qquad  &\text{if } z \in \, ]z_{i-1}^+, z_i^- [\,,\quad  i=2,..,n,
\end{cases}
\end{equation}
where we remind that $d_i$ are the aspect ratios \eqref{eq:dialphaDefcharged}. Therefore, the base metric above and below the rod configuration, $z < z_{1}^-$ and $z > z_{n}^+$, has no conical singularity and is smooth there. However, in between the $(i-1)^\text{th}$ and $i^\text{th}$ rods, we have $ds_3^2 \sim d_i^2 \left(d\rho^2 + dz^2 +\frac{\rho^2}{d_i^2} d\phi^2 \right)$. Since $d_i <1$, the segment has a conical excess and describes a strut, i.e. a string with a negative tension (see Fig.\ref{fig:RodCategories}). The strut is a singularity that accounts for the lack of repulsion between the $(i-1)^\text{th}$ and $i^\text{th}$ sources. It can be associated with a curvature singularity and an energy as done in \cite{Costa:2000kf} and reviewed in \cite{Bah:2021owp}.

The existence of a strut between two rods requires that they are disconnected. Therefore, by connecting the rods, which implies that they are of different nature, the solutions are free of struts and conical excesses \cite{Elvang:2002br,Bah:2021owp,Bah:2021rki}. These solutions consist in connecting bubbles and possibly non-extremal black holes on a line (see Fig.\ref{fig:TouchingRods}). For such configurations, the sources do not require a singular mechanism to be prevented from collapsing and are kept apart by topological bubbles \cite{Bah:2021owp}. This provides a new paradigm in the microstate geometry program for constructing smooth bubbling geometries without the aid of supersymmetry. The standard lore of supersymmetry is that gravitational attraction is counterbalanced by electromagnetic repulsion. In the present ansatz, the solutions are generically non-BPS with arbitrarily low electromagnetic potentials, and the lack of repulsion is balanced by the inherent pressure of squeezed bubbles \cite{Bah:2021owp}. 
\end{itemize}

In conclusion, the non-BPS floating brane ansatz contains a large phase space of axisymmetric regular static solutions that are asymptotic to four-dimensional flat space plus compactification circles. Typical solutions consist of non-extremal four-charge black holes and 7 types of smooth bubbles on a line, which are kept apart by struts if disconnected. In addition, smooth bubbling geometries can be constructed by considering chains of connected bubbles of different species. We will construct explicit solutions in section \ref{sec:Sol4d}.

\subsection{Five-dimensional solutions}
\label{sec:5dNonBPSGen}

To construct five-dimensional non-BPS solutions, the four-dimensional base, $(\rho,z,\phi,y_7)$, must be considered as a four-dimensional infinite space. That is, we take $y_7$ to be an angle of periodicity $2\pi$. As detailed in the appendix \ref{App:Reg5d}, one must also consider the neutral limit \eqref{eq:neutrallimit} for the pair $(Z_0,H_0)$ and source $Z_0$ by a semi-infinite rod.

\begin{figure}[h]
\centering
\includegraphics[width=0.22\textwidth]{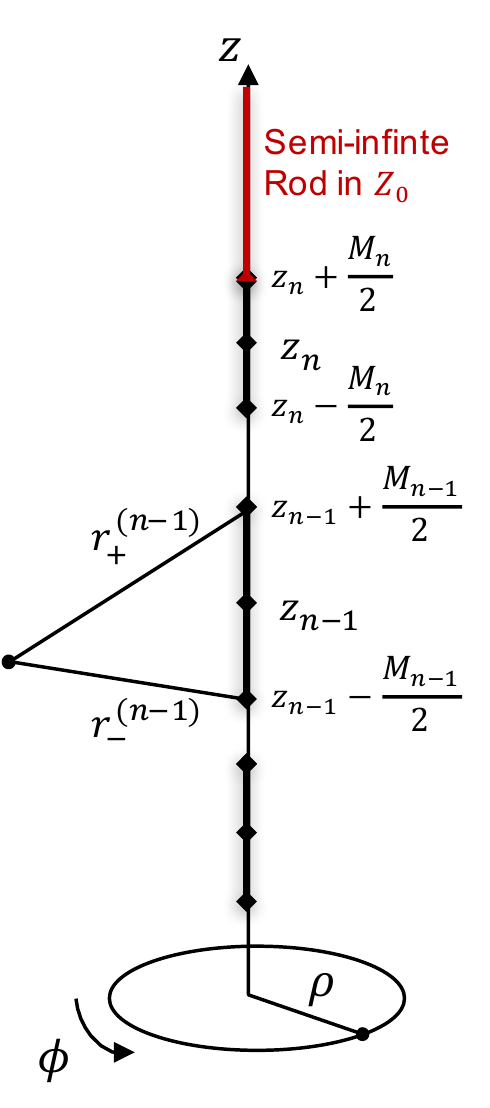}
\caption{Schematic description of axisymmetric rod sources with a semi-infinite rod in $Z_0$.}
\label{fig:RodsSources5d}
\end{figure}

Generic five-dimensional solutions are therefore given by \eqref{eq:nonBPSfloatingRemind}, sourced by $n$ finite rod of length $M_i$ and a $(n+1)^\text{th}$ semi-infinite rod (see Fig.\ref{fig:RodsSources5d}). We will consider that the $n^\text{th}$ and $(n+1)^\text{th}$ rods are connected to avoid a conical excess between them, $z_n^+=z_{n+1}^-$. The warp factors and gauge potentials are similar to the ones given for four-dimensional solutions \eqref{eq:WarpfactorsRods} with the neutral limit \eqref{eq:neutrallimit} for $(Z_0,H_0)$ and additional terms corresponding to the semi-infinite rod, that is
\begin{align}
&Z_I= \frac{1}{2a_I} \left[e^{b_I} \prod_{i=1}^n \left(  \frac{R_+^{(i)}}{R_-^{(i)}}\right)^{a_I P^{(I)}_i}-e^{-b_I} \prod_{i=1}^n \left(  \frac{R_-^{(i)}}{R_+^{(i)}}\right)^{a_I P^{(I)}_i} \right]\,,\quad T_I\=-\frac{1}{Z_I}\,\sqrt{1+a_I^2\,Z_I^2} \,,\nonumber\\
& Z_0 \=   \left(r_+^{(n)}-(z-z_n^+) \right)^{-G_{n+1}}\,\prod_{i=1}^n \left(  \frac{R_+^{(i)}}{R_-^{(i)}}\right)^{G_i} \,,\quad H_0 \=0\,,\quad W_\Lambda = \prod_{i=1}^n \left(  \frac{R_+^{(i)}}{R_-^{(i)}}\right)^{G^{(\Lambda)}_i}  \, , \label{eq:WarpfactorsRods5d} 
\end{align}
\begin{align}
&e^{2\nu} \= \frac{1}{2} \left(\frac{r_+^{(n)}-(z-z_n^+)}{r_+^{(n)}} \right)^{{G_{n+1}}^2}\,\prod_{i=1}^n \left(\frac{E_{++}^{(n,i)}\,R_-^{(i)}}{E_{+-}^{(n,i)}\,R_+^{(i)}} \right)^{G_{n+1}G_i }\,\prod_{i,j=1}^n\, \left(  \frac{E_{+-}^{(i,j)}E_{-+}^{(i,j)}}{E_{++}^{(i,j)}E_{--}^{(i,j)}}\right)^{\frac{1}{2}\,\alpha_{ij}} \,, \nonumber
\end{align}
where the main functions are given in \eqref{eq:RpmdefRemind} and the exponents, $\alpha_{ij}$, are now given by\footnote{The exponents are slightly different from \eqref{eq:Defalpha} due to the neutral limit \eqref{eq:neutrallimit} taken for the pair $(Z_0,H_0)$. More precisely, we have taken $(a_0,b_0) \to \infty$ with $G_i = a_0 P^{(0)}_i$ finite in \eqref{eq:HarmFuncGen} and \eqref{eq:HarmFuncGen2}.}
\begin{align}
\alpha_{ij} &\equi  G_i G_j\+ \sum_{\Lambda =0}^3  G_i^{(\Lambda)}G_j^{(\Lambda)} \+  \sum_{I=1}^3 a_I^2\,P_i^{(I)}P_j^{(I)}\,. \label{eq:alpha5d}
\end{align}

Note that the solutions have no KKm charges along $y_7$  because $H_0=0$. We have therefore defined a family of non-BPS M2-M2-M2 solutions given by $10 n + 8$ parameters. 

\subsubsection{Regularity constraints}
\label{sec:regM5d}

We will be brief in the regularity analysis since it is similar to the discussion for four-dimensional solutions. We refer to the appendix \ref{App:Reg5d} for more details.

The five-dimensional metric \eqref{eq:5dframework} is asymptotically flat if 
\begin{equation}
 \prod_{I=1}^3  \frac{\sinh b_I}{a_I} \= 1\,.
\label{eq:condonasymp5d}
\end{equation}
The solution is regular everywhere out of the finite rods, especially at the semi-infinite rod where the $y_7$ degenerates smoothly as an origin of a $\IR^2$ if and only if
\begin{equation}
G_{n+1} \= 1\,.
\label{eq:CondGn+1}
\end{equation}
Moreover, the solution is regular at the finite $i^\text{th}$ rod if its weights fall into one of these eight categories (see appendix \ref{App:AttheRod2} for more details):
\begin{itemize}
\item[-] \underline{\emph{Black rods:}} 
We consider that
\begin{equation}
G^{(\Lambda)}_i \=0\,,\qquad G_i\= a_I P^{(I)}_i\= \frac{1}{2},
\label{eq:blackrodWeight5d}
\end{equation}
and the rod corresponds to a horizon of a black hole where the timelike Killing vector $\partial_t$ shrinks. The surface gravity, temperature and area of the black hole are given by
\begin{align}
\kappa_i &\equi \frac{2 \sqrt{z_n^+-z_i^+}}{d_i\,M_i}\, \sqrt{\frac{a_1 a_2 a_3}{e^{b_1+b_2+b_3}}}\,\prod_{j=1}^i  \left( \frac{z_n^+ - z_j^-}{z_n^+ - z_j^+}\right)^{ G_j} \,\prod_{j\neq i} \left(\frac{z_j^+ - z_i^-}{z_j^- -z_i^-} \right)^{\text{sign}(i-j)\, \alpha_{ij}}\,, \nonumber\\
\cT_i &\= \frac{\kappa_i}{2\pi}\,,\qquad A_i \= \frac{(2\pi)^8\,M_i}{\kappa_i}\,\prod_{a=1}^6 R_{y_a}\,,
\label{eq:surfaceGrav5d}
\end{align}
where $d_i$ is still given by \eqref{eq:dialphaDefcharged}. Because the three $P_i^{(I)}$ are finite, the black hole carries generically three M2 charges given by \eqref{eq:individualMcharges}. Each charge can be taken to be zero by considering the neutral limit \eqref{eq:neutrallimit}, which corresponds to $a_I =  \sinh b_I\to \infty$. 

More precisely, the rod corresponds to the horizon of a three-charge static non-extremal black hole for which the single-center solutions have been derived in \cite{Cvetic:1996xz,Cvetic:1997uw,Cvetic:1998xh}. 

\item[-]  \underline{\emph{Bubble rods:}} The $y_1$ or $y_2$ circle shrinks at the $i^\text{th}$ rod while all other fibers are finite if we consider
\begin{equation}
 \pm G_i^{(1)} = -G_i^{(0)} =G_i = a_1 P_i^{(1)} =\frac{1}{2} , \quad G_i^{(2)}=G_i^{(3)}=P_i^{(2)}=P_i^{(3)}= 0,
\label{eq:bubblerodWeight15d}
\end{equation}
where the ``$+$'' corresponds to a shrinking $y_1$-circle and the ``$-$'' corresponds to a shrinking $y_2$-circle. One can obtain the weights for another degenerating T$^6$ direction by permuting the $I=1,2,3$ indexes. Similarly, on can make the $y_7$ circle degenerate by considering
\begin{equation}
G_i \= 1 , \qquad G_i^{(\Lambda)} \= P_i^{(I)} \= 0\,.
\label{eq:bubblerodWeight25d}
\end{equation}
For each possibility, the local geometry corresponds to a bubble sitting at a smooth origin of a $\IR^2$ if a bubble equation is satisfied. For a rod that makes the circle $y_1$ or $y_2$ degenerate smoothly, the bubble equation constrains the internal parameters as a function of the extra-dimension radius such as
\be
k_i \,R_{y_a} \=  \frac{d_{i} M_{i}}{\sqrt{z_n^+-z_i^+}}\,\sqrt{\frac{e^{b_{1}}}{a_{1}}}\, \prod_{j=1}^i  \left( \frac{z_n^+ - z_j^+}{z_n^+ - z_j^-}\right)^{ G_j} \, \prod_{j \neq i}\left(\frac{z_{j}^{+}-z_{i}^{-}}{z_{j}^{-}-z_{i}^{-}}\right)^{\operatorname{sign}(j-i) \alpha_{i j}}, \quad  a=1 \text{ or }2,
\label{eq:RegCondBu5d}
\ee
where $k_i$ is an orbifold parameter, $k_i\in \mathbb{N}$, that corresponds to the order of the conical deficit on the $\IR^2$. The bubble equation corresponding to another degenerating T$^6$ direction or the $y_7$-circle is obtained by permuting $\frac{e^{b_1}}{a_1} \to \frac{e^{b_I}}{a_I}$ or $\frac{e^{b_1}}{a_1} \to 1$ respectively. 

The smooth bubble where one of the T$^6$ direction degenerates carries only a M2 brane charge, given by \eqref{eq:individualMcharges}, because only one $P^{(I)}_i$ is nonzero \eqref{eq:bubblerodWeight15d}. Similarly, the smooth bubble where the $y_7$ direction shrinks is purely topological and carries no charge. We have therefore six types of M2 bubbles and one type of vacuum bubble that can smoothly source the solutions.
\end{itemize}

The mass induced by each rod, $\cM_i$, can be derived from \eqref{eq:ADMmass5d} and related to the brane charges \eqref{eq:individualMcharges} such as
\begin{equation}
\cM_i \= \frac{\pi}{2 G_5} \,\sum_{I=1}^3 a_I \coth b_I \,Q_{\text{M2},i}^I\,.
\end{equation}
The charge-to-mass ratios for each stack of M2 branes are fixed by $a_I \coth b_I$ for all rods.

As for four-dimensional solutions, a strut is separating each pair of disconnected rods. More precisely, if the $(i-1)^\text{th}$ and $i^\text{th}$ rods are not connected, the $\phi$-circle degenerates in between them with a conical excess of order $d_i^{-1}>1$ \eqref{eq:dialphaDefcharged}. This singularity can be bypassed by considering connected rods. For such configurations, the struts disappear and the non-BPS sources are held apart by pure topology.

To conclude, the family of solutions, given by \eqref{eq:WarpfactorsRods5d}, contains a large phase space of regular non-BPS axisymmetric static solutions that are asymptotic to five-dimensional flat space plus compactification circles. Typical solutions consist of non-BPS non-extremal three-charge black holes and 7 types of smooth bubbles on a line, which are held apart by struts if disconnected. Moreover, smooth bubbling geometries can be constructed by considering chains of connected bubbles of different species. We will construct explicit solutions in section \ref{sec:Sol5d}.

\subsection{Other duality frames}
\label{sec:dualFrame}

So far we have limited the discussion to the M-theory frame, but the construction can be dualized to different string theory frames through dimensional reduction and T-dualities. Each frame will have new types of regular rod sources that are singular in M-theory. To illustrate this property, we will focus on dualization in the D1-D5-P-KKm frame, but we refer the interested reader to the appendix \ref{App:dualFrame} where we have dualized the solutions in six different frames, from the D2-D2-D2-D6 to the P-M5-M2-KKm frame. In addition, the study of the D1-D5-P-KKm frame will provide a link to the author's previous constructions of smooth bubbling geometries \cite{Bah:2020pdz,Bah:2021owp,Bah:2021rki}.

After reduction along $y_5$ and a series of T-dualities along $y_1$, $y_2$ and $y_6$, the M2-M2-M2-KKm solutions \eqref{eq:nonBPSfloatingRemind} transform to D1-D5-P-KKm solutions where the common direction for the momentum P charge, the D1 and D5 branes is $y_6$. The metric, the R-R gauge fields $C^{(p)}$, the NS-NS two-form gauge field $B_2$ and the dilaton in type IIB are given by 
\begin{align}
ds_{10}^2 \= &- \frac{dt^2}{Z_3 \sqrt{W_3W_0 Z_1 Z_2}} \+ \sqrt{\frac{ Z_1 Z_2}{W_3 W_0}}  \left[\frac{1}{Z_0} \left(dy_7 +H_0 \,d\phi\right)^2 \+ Z_0\left( e^{2\nu} \left(d\rho^2 + dz^2 \right) +\rho^2 d\phi^2\right)\right] \nonumber\\
& +\sqrt{\frac{Z_1}{Z_2}}\left[\sqrt{\frac{W_3}{W_0}} \,\left(W_1\,dy_1^2+ \frac{dy_2^2}{W_1}\right) +\sqrt{\frac{W_0}{W_3}} \left(\frac{dy_3^2}{W_2} + W_2 \,dy_4^2 \right) \right] \label{eq:D1D5PKKm} \\
& + \frac{Z_3}{\sqrt{W_3 W_0 Z_1 Z_2}}\,\left(dy_6+T_3\, dt \right)^2\,, \qquad \Phi \= \frac{1}{2} \log \frac{Z_1}{W_0 W_3 Z_2}\,,\qquad B_2 \= 0\,, \nonumber \\
C^{(0)} \=&  0 \,,\qquad C^{(2)}\=  H_2 \,d\phi \wedge dy_7 \- T_1 \,dt \wedge dy_6\,,\qquad C^{(4)} \= 0\,,\nonumber 
\end{align}
where $H_2$ is the electromagnetic dual of $T_2$, and is given by the same expression as $H_0$ in \eqref{eq:WarpfactorsRods} but with $P_i^{(0)} \to P_i^{(2)}$ (see appendix \ref{App:dualFrame}).

One can recognize the type IIB frame of topological stars and five-dimensional charged Weyl solutions of \cite{Bah:2020pdz} by taking
\begin{equation}
W_0 \= W_1\= W_2\= W_3 \= 1 \,,\qquad Z_0 \=Z_1 \=Z_2 \=  Z \,,\qquad T_3 \= 0 \,,\qquad Z_3 \= W\,,
\end{equation}
Taking $T_3=0$ while keeping a non-trivial $Z_3=W$ is possible by considering the neutral limit for the pair $(Z_3,T_3)$ \eqref{eq:neutrallimit}, that is no P-charges. If we now consider $Z_0$ to be different from $Z_1=Z_2$, we retrieve the framework of \cite{Bah:2021owp,Bah:2021rki} where six-dimensional smooth bubbling solutions, as bubble bag ends \cite{Bah:2021rki}, have been constructed.

The warp factors and gauge potentials are governed by the linear system of equations obtained from the non-BPS floating brane ansatz of section \ref{sec:WeylAns}.

For the non-BPS solutions, the physical rod sources are different from the eighth found in sections \ref{sec:regM4d} and \ref{sec:regM5d}. For instance,  the $y_6$-circle degenerates at the $i^\text{th}$ rod for a solutions given by \eqref{eq:WarpfactorsRods} defining the locus of a bubble if we consider its associated weights to be 
\begin{equation}
G^{(\Lambda)}_i \= 0 \,,\qquad a_0 P^{(0)}_i \= a_1 P^{(1)}_i \= a_2 P^{(2)}_i \= -a_3 P^{(3)}_i \=\frac{1}{2}\,.
\label{eq:bubblerodWeight4}
\end{equation}
As a black rod \eqref{eq:blackrodWeight}, they do not require to turn on the weights of the torus warp factors $W_\Lambda$, and therefore the T$^4$ is  rigid as for BPS solutions. The only difference is that $a_3 P_i^{(3)}=-\frac{1}{2}$. However, taking $a_3 P_i^{(3)}$ negative is compatible with the regularity condition \eqref{eq:RegPositivity} only if one takes the neutral limit, $(a_3,b_3)\to \infty$ \eqref{eq:neutrallimit}. It requires to take the gauge potential for the P charge to be zero and the solutions have no momentum charges as in \cite{Bah:2020pdz,Bah:2021owp,Bah:2021rki}. A rod given by \eqref{eq:bubblerodWeight4} corresponds to a smooth D1-D5-KKm bubble in type IIB.\footnote{Smoothness requires a bubble equation of a similar type to those obtained previously.} However, it is singular in the M-theory frame \eqref{eq:nonBPSfloatingRemind}. Therefore, there are specific species of bubble rods that correspond to smooth loci in a unique string theory frame.

\section{Explicit four-dimensional solutions}
\label{sec:Sol4d}

In this section, we derive explicit solutions that are asymptotic to $\IR^{1,3}$ with extra compact dimensions using the non-BPS floating brane ansatz in different string theory frames. There is a very large number of configurations that one can think about. First, we will construct chain of non-extremal four-charge black holes, either separated by struts or smooth bubbles, and discuss their physics. Second, we will construct smooth bubbling geometries in the manner of \cite{Bah:2021owp,Bah:2021rki}, that have the same mass and charges as non-extremal four-charge black holes.

\subsection{Chain of static non-extremal black holes}

We first study solutions for which the main ingredients are non-extremal four-charge black holes on a line.

\subsubsection{Black holes separated by struts}

\begin{figure}[h]
\centering
\includegraphics[width=0.38\textwidth]{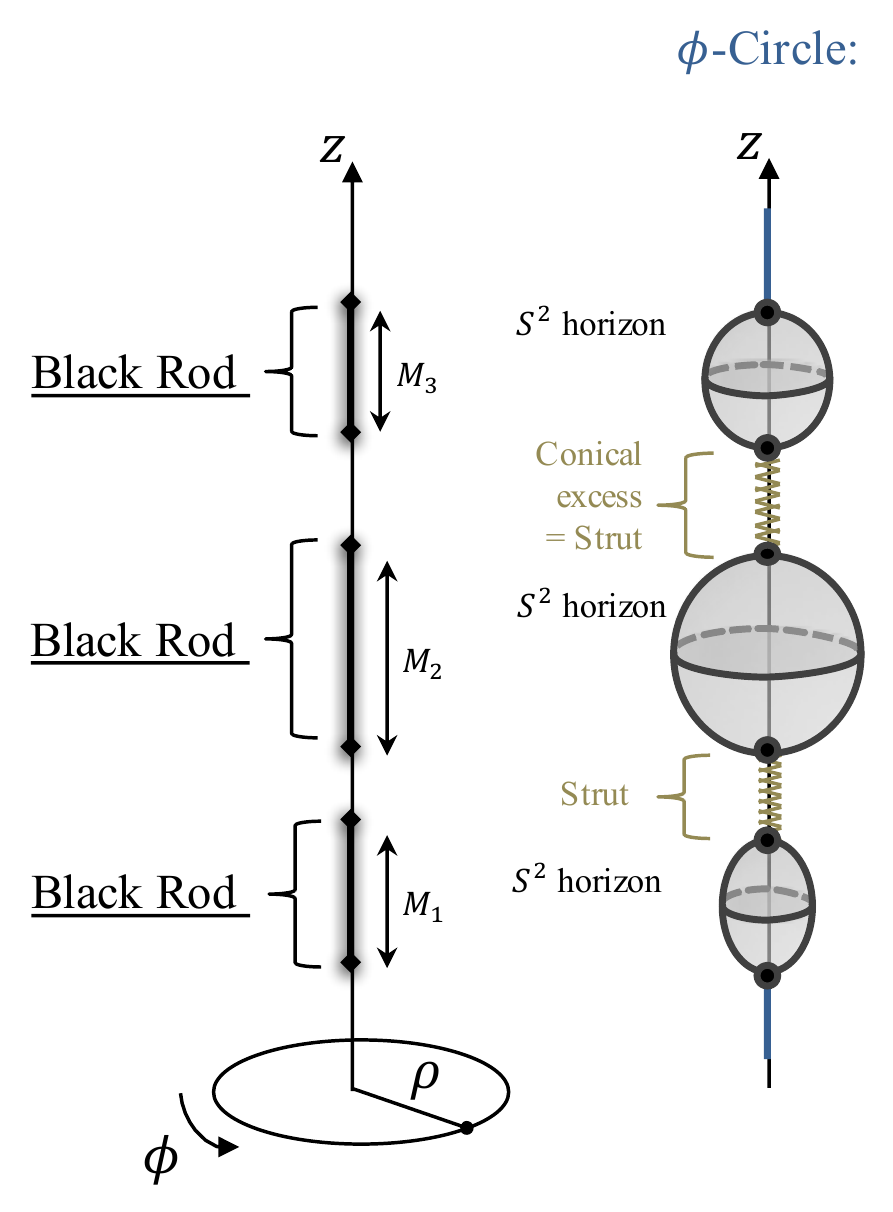}
\caption{Description of a chain of static four-charge non-extremal black holes in the four-dimensional framework \eqref{eq:4dprofilechainofBH} and the behavior of the $\phi$-circle on the $z$-axis.}
\label{fig:ChainofBHStruts}
\end{figure}

We consider $n$ finite rods of length $M_i$ on the $z$-axis with weights corresponding to black rods given by \eqref{eq:blackrodWeight} (see Fig.\ref{fig:ChainofBHStruts}). It is convenient to divide the warp factors, $Z_\Lambda$ \eqref{eq:WarpfactorsRods}, into two parts: one will encode changes in topology and the other will correspond to a well-behaved flux decoration such as
\begin{equation}
Z_\Lambda \= \frac{\cZ_\Lambda}{\sqrt{U_t}}\,,\qquad U_t \equi \prod_{i=1}^n
\left(1-\frac{2 M_i}{R_+^{(i)}} \right)\,,\qquad \cZ_\Lambda \equi \frac{e^{b_\Lambda}-e^{-b_\Lambda}\,U_t}{2 a_\Lambda}
\end{equation}
The remaining functions \eqref{eq:WarpfactorsRods} give
\begin{align}
& W_\Lambda = 1,\quad e^{2\nu} \=\prod_{i,j=1}^n\, \sqrt{ \frac{E_{+-}^{(i,j)}E_{-+}^{(i,j)}}{E_{++}^{(i,j)}E_{--}^{(i,j)}}},\quad  H_0 \= \frac{1}{2 a_0}\sum_{i=1}^n \left(r_-^{(i)}-r_+^{(i)} \right),\quad T_I\=-\sqrt{a_I^2+\frac{U_t}{\cZ_I^2}}, \nonumber 
\end{align}
where the generating functions $R_\pm^{(i)}$ and $E_{\pm \pm}^{(i,j)}$ are given in \eqref{eq:RpmdefRemind}. The metric and gauge potential in M-theory are given in \eqref{eq:nonBPSfloatingRemind}, while the four-dimensional reduction along T$^6\times$S$^1$ is given by \eqref{eq:4dSol}:
\begin{align}
ds_4^2 &\= -\frac{U_t\,dt^2}{\sqrt{\cZ_0 \cZ_1 \cZ_2 \cZ_3}} \+ \frac{\sqrt{\cZ_0 \cZ_1 \cZ_2 \cZ_3}}{U_t} \left( e^{2\nu} \left(d\rho^2 + dz^2 \right) +\rho^2 d\phi^2\right), \nonumber \\
z^I &\= i\,\frac{|\epsilon_{IJK}|}{2} \sqrt{\frac{\cZ_J \cZ_K}{\cZ_0\,\cZ_I} },\quad \Phi_\Lambda \=0 \,,\quad \bar{F}^{0}  \= -dH_0 \wedge d\phi \,,\quad \bar{F}^{I}  \= dT_I \wedge dt\,.
\label{eq:4dprofilechainofBH}
\end{align}
The function $U_t$ behaves as a product of ``Schwarzschild factors'' that vanishes at each rod and makes the timelike Killing vector shrink. It induces the topology of the solutions and forms the horizons. The $\cZ_\Lambda$ depend only on the gauge field parameters, and are always positive and finite. In the neutral limit \eqref{eq:neutrallimit}, all $\cZ_\lambda$ converge to 1 while $U_t$ remains unchanged. We retrieved the solutions of Schwarzschild black holes on a line \cite{Israel1964,Gibbons:1979nf}.

At large distance $r$, with $(\rho,z)=(r\sin\theta,r\cos\theta)$, the geometry is asymptotically flat if \eqref{eq:condonasymp} is satisfied, and one can read the ADM mass from \eqref{eq:ADMmass4d}:
\begin{equation}
\cM \= \frac{\sum_{\Lambda=0}^3 \coth b_\Lambda}{8 G_4}\,\sum_{i=1}^n M_i\,.
\label{eq:MassBHchain}
\end{equation}
Moreover, the solutions carry four charges that are M2-M2-M2-KKm charges in M-theory. They are given by the sum of the rod charges carried by each rod \eqref{eq:individualMcharges}: 
\begin{equation}
Q^I_{\text{M2}} \=  \frac{1}{2a_I}\,\sum_{i=1}^n M_i \,,\qquad  Q_{\text{KKm}} \= \frac{1}{2a_0} \,\sum_{i=1}^n M_i.
\end{equation}
\begin{itemize}
\item[•] \underline{A unique rod: map to the four-charge non-extremal black hole \cite{Cvetic:1995kv,Chong:2004na,Chow:2014cca}.}
\end{itemize} 
We consider a unique rod, $n=1$, and the spherical coordinates $$(\rho,\,z)= \left(\sqrt{r \left(r -M_{1}\right)} \sin \theta,\, \left(r-\frac{M_{1}}{2}\right) \cos \theta +z_{1}\right). $$ The four-dimensional metric, gauge fields and scalars \eqref{eq:4dprofilechainofBH} simplify to
\begin{align}
ds_4^2 &\= -\frac{r(r-M_1)\,dt^2}{\sqrt{r_0 r_1 r_2 r_3}} \+ \sqrt{r_0 r_1 r_2 r_3} \left(\frac{dr^2}{r(r-M_1)}+d\theta^2 +\sin^2\theta d\phi^2\right),\quad \Phi_\Lambda \=0, \label{eq:4chargeBH} \\
z^I &\=  i\,\frac{a_0 a_I\,|\epsilon_{IJK}|}{2\sinh b_0 \sinh b_I}\sqrt{\frac{r_J r_K}{r_0\,r_I} } \,,\quad \bar{F}^{0}  \= \frac{M_1}{2 \sinh b_0}\sin \theta \,d\theta \wedge d\phi \,,\quad \bar{F}^{I}  \= \frac{M_1}{2 \sinh b_I\,r_I^2}dt\wedge dr\,, \nonumber
\end{align}
where we have defined $r_\Lambda \equi r+ \frac{e^{-b_\Lambda}}{2 \sinh b_\Lambda}\,M_1$.\footnote{Note that the metric no longer depend on $a_\Lambda$. In four dimensions, the dependence in $a_\Lambda$ for regular solutions is a function of $\prod_\Lambda a_\Lambda$ which is fixed in terms of $b_\Lambda$ to have asymptotically flat solutions \eqref{eq:condonasymp}.} We retrieve the static solutions of a four-charge non-extremal black hole \cite{Cvetic:1995kv,Chong:2004na,Chow:2014cca} by relating the boost parameters in these papers, $\delta_\Lambda$, to our parameters, $b_\Lambda$, such as $\cosh 2\delta_\Lambda = \coth b_\Lambda$.
\begin{itemize}
\item[•] \underline{Multiple four-charge black holes on a line.}
\end{itemize}
By considering $n>1$, we perform the multicentric generalization of the static solutions in \cite{Cvetic:1995kv,Chong:2004na,Chow:2014cca}.
In the IR, the geometry is made of a chain of four-charge static non-extremal black holes on the $z$-axis held apart by struts (see Fig.\ref{fig:ChainofBHStruts}). At each rod, the timelike Killing vector $\partial_t$ vanishes defining the locus of a horizon with area \eqref{eq:Area&SurfaceGrav}. If one wants the black holes to be in thermal equilibrium, their surface gravity must be equal \eqref{eq:Area&SurfaceGrav}, which non-trivially constrains their positions and sizes. Moreover, in between each rod the $\phi$-circle shrinks with a conical excess of order $d_i^{-1} >1$ \eqref{eq:dialphaDefcharged} defining the loci of the struts.

The struts carry an energy that counterbalances the attraction between the black holes. This energy corresponds to the interaction between the black holes in string theory. It can be derived using the method of \cite{Costa:2000kf,Bah:2021owp}. More precisely, the energy of the strut that separates the $(i-1)^\text{th}$ and $i^\text{th}$ black holes on the chain carries an energy given by
\begin{equation}
E_i \= - \frac{1-d_i}{4 G_4}\,(z_i^- - z_{i-1}^+)\,, \qquad G_4 = \frac{G_{11}}{(2\pi)^{7} \prod_{i=1}^{7} R_{y_i}}.
\end{equation}
For a configuration of two identical black holes separated by a distance $\ell$, the energy of the strut in between them is given  by
\begin{equation}
E \= - \frac{M^2 \ell}{4 G_4 (M+\ell)^2},
\end{equation}
where $M$ is the length of each rod. When the separation is large, $\ell \gg M$, the energy of the strut approximates the Newtonian potential between two particles of mass, $\frac{M}{2G_4}$, in four dimensions.  From this perspective, the strut measures the binding energy between the two black holes, or rather the potential energy needed to keep the two black holes from collapsing on each other. An important observation is that the effective ADM masses of the black holes \eqref{eq:MassBHchain}, from the Newtonian point of view, depend on $b_\Lambda$, which are associated to the four charges of the black holes \eqref{eq:4chargeBH} with the regularity condition \eqref{eq:condonasymp}.  This implies that the binding energy as measured by the strut also accounts for effects due to the electromagnetic fields of the non-extremal black holes.

For finite separation between the two black holes, the gravitational potential between them deviates significantly from that of the Newtonian limit.  In particular, the gravitational potential between two black holes vanish when $\ell \to 0$.  In this limit, the two rod sources merge and the two-body configuration becomes a single non-extremal black hole.

\subsubsection{Black holes separated by smooth bubbles}
\label{sec:chainBHbu}

Struts are singularities that do not have a consistent UV description in string theory as they correspond to cosmic strings with negative tension. Therefore, it is crucial to find a mechanism to resolve them and to obtain more relevant configurations in string theory. The non-BPS regime will require a new mechanism for this task. In the analysis of \cite{Costa:2000kf} and continued in \cite{Bah:2021owp}, it was shown that struts can be classically resolved into smooth bubbles. More precisely, one can consider the same black hole systems as before but sourcing each segment between them by a bubble rod. The resulting configuration will consist in a chain of connected rods which would be a succession of black holes and bubbles without struts. The condition for having connected rods fixes their centers in terms of their size:
\begin{equation}
z_i^+ \= z_{i+1}^- \,,\quad \forall i \qquad \Rightarrow \qquad z_i \= z_n^+ + \frac{M_i}{2} - \sum_{j=i}^n M_j\,.
\label{eq:connectedcond}
\end{equation}
One can also freely choose the origin of the $z$-axis so that $z_n^+=0$. 

For simplicity, we consider the bubble rods to be smooth KKm bubbles given by the weights \eqref{eq:bubblerodWeight2}, but one could a priori take any of the seven species of bubble rods detailed in section \ref{sec:regM4d}. This allows a faithful description of the solutions in the five-dimensional framework \eqref{eq:5dframework}. 

\begin{figure}[h]
\centering
\includegraphics[width=0.55\textwidth]{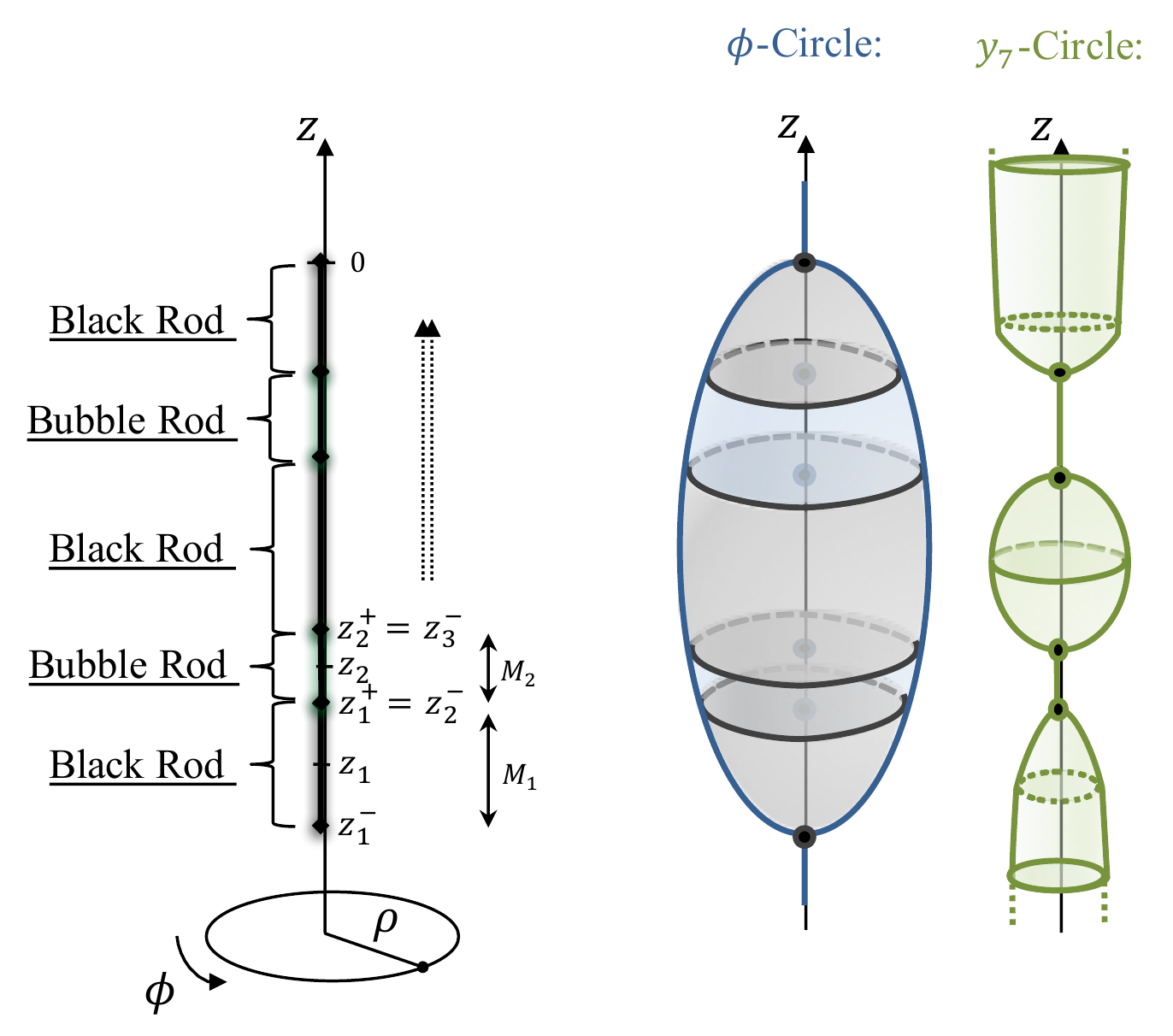}
\caption{Description of a chain of static four-charge non-extremal black holes and KKm bubbles in the five-dimensional framework \eqref{eq:5dprofilechainofBH} and the behavior of the $\phi$ and $y_7$ circles on the $z$-axis.}
\label{fig:ChainofBHBub}
\end{figure}
Therefore, we consider $n=2N+1$ rods of lengths $M_i$ where the $N+1$ odd rods correspond to black rods \eqref{eq:blackrodWeight} and the $N$ even rods correspond to KKm bubble rods \eqref{eq:bubblerodWeight2} (see Fig.\ref{fig:ChainofBHBub}). As in the previous section, we divide the warp factors $Z_\Lambda$ \eqref{eq:WarpfactorsRods} in meaningful pieces such as
\begin{equation}
\begin{split}
Z_0 &\= \frac{\cZ_0}{U_{y_7}\sqrt{U_t}}, \qquad Z_I \= \frac{\cZ_I}{\sqrt{U_t}}\,,\\
U_t &\equi \prod_{i=1}^{N+1} \left(1-\frac{2 M_{2i-1}}{R_+^{(2i-1)}} \right)\,, \qquad U_{y_7} \equi \prod_{i=1}^N \left(1-\frac{2 M_{2i}}{R_+^{(2i)}} \right)\,, \\
\cZ_0 &\equi \frac{e^{b_0}-e^{-b_0}\,U_t U_{y_7}^2}{2 a_0} \,,\qquad   \cZ_I \equi \frac{e^{b_I}-e^{-b_I}\,U_t}{2 a_I}.
\end{split}
\end{equation}

The remaining functions \eqref{eq:WarpfactorsRods} give
\begin{align}
& W_\Lambda = 1,\qquad  H_0 \= \frac{1}{2 a_0} \left[r_-^{(1)}- r_+^{(n)}+\sum_{i=1}^N \left(r_-^{(2i)}-r_+^{(2i)} \right) \right],\qquad T_I\=-\sqrt{a_I^2+\frac{U_t}{\cZ_I^2}},  \nonumber  \\
& e^{2 \nu}=\frac{E_{-+}^{(1, n)}}{\sqrt{E_{++}^{(n, n)} E_{--}^{(1,1)}}} \prod_{i=1}^{N} \prod_{j=1}^{N+1} \sqrt{\frac{E_{--}^{(2 i, 2 j-1)} E_{++}^{(2 i, 2 j-1)}}{E_{-+}^{(2 i, 2 j-1)} E_{+-}^{(2 i, 2 j-1)}}}\,,
\end{align}
where $R_\pm^{(i)}$ and $E_{\pm \pm}^{(i,j)}$ are given in \eqref{eq:RpmdefRemind}. The metric and gauge potential in M-theory are given in \eqref{eq:nonBPSfloatingRemind}, while the five-dimensional reduction along T$^6$ is given by \eqref{eq:5dframework}
\begin{align}
ds_5^2 &\= - \frac{U_t\,dt^2}{\left(\cZ_1 \cZ_2 \cZ_3 \right)^{\frac{2}{3}}} \+ \left(\cZ_1 \cZ_2 \cZ_3 \right)^{\frac{1}{3}} \left[\frac{U_{y_7}}{\cZ_0} \left(dy_7 +H_0 \,d\phi\right)^2 \+ \frac{\cZ_0}{U_t U_{y_7}} \left( e^{2\nu} \left(d\rho^2 + dz^2 \right) +\rho^2 d\phi^2\right)\right], \nonumber \\
X^I &\= \frac{|\epsilon_{IJK}|}{2} \left(\frac{\cZ_J \cZ_K}{\cZ_I^2} \right)^\frac{1}{3},\qquad \Phi_\Lambda \= 0 \,,\qquad F^{(I)}  \= dT_I \wedge dt\,.
\label{eq:5dprofilechainofBH}
\end{align}
The warp factors $U_t$ and $U_{y_7}$ force the degeneracy of the $t$ and $y_7$ fibers at the rods. In the regions where $U_{y_7}$ vanishes, that is at the even rods, the $y_7$ direction shrinks forming a smooth origin of an $\IR^2$ (with potential conical defects) if we have \eqref{eq:RegCondBu}
\be
k_{i}\,R_{y_7}  \equi \,\frac{d_{2i} M_{2i}\,e^{b_{0}}}{a_{0}}\, \prod_{j \neq i}^N\left(\frac{z_{2j}^{+}-z_{2i}^{-}}{z_{2j}^{-}-z_{2i}^{-}}\right)^{\operatorname{sign}(j-i)} \,\prod_{j =1}^{N+1}\left(\frac{z_{2j-1}^{+}-z_{2i}^{-}}{z_{2j-1}^{-}-z_{2i}^{-}}\right)^{\frac{\operatorname{sign}(j-i-1/2)}{2}}, \quad \forall i=1,...,N,
\ee
where we remind that $z_j^\pm$ are the rod endpoint coordinates on the $z$-axis \eqref{eq:RpmdefRemind} and $d_j$ are the aspect ratios \eqref{eq:dialphaDefcharged}. These $N$ ``bubble equations'' fix the lengths of the bubble rods, $M_{2i}$, according to the other parameters. Moreover, if one wants the black holes to be in thermal equilibrium, one needs in addition to impose that their surface gravities \eqref{eq:Area&SurfaceGrav} are equal, which also constrains the length of the black rods. 

The $\phi$-circle does not shrink anymore along the rod configuration and the four-charge non-extremal black holes are held apart by KKm bubbles. The struts have been then successfully replaced by smooth bubbles. This mechanism has been explored in \cite{Costa:2000kf,Bah:2021owp}. In one word, vacuum bubbles are reluctant to be squeezed as it wants to expand \cite{Witten:1981gj}, and provide the necessary pressure between two non-BPS objects \cite{Costa:2000kf}. If the electromagnetic fluxes stabilize the bubbles \cite{Stotyn:2011tv}, their reluctance to be squeezed remains \cite{Bah:2021owp}.

Finally, the solutions are asymptotic to $\IR^{1,3}\times$S$^1$ if \eqref{eq:condonasymp} is satisfied. The ADM mass is given by \eqref{eq:ADMmass4d}
\begin{equation}
\cM \= \frac{\sum_{\Lambda=0}^3 \coth b_\Lambda\,\sum_{i=1}^{N+1} M_{2i-1}+2 \coth b_0\,\sum_{i=1}^{N} M_{2i}}{8 G_4}\,.
\end{equation}
Moreover, the M2-M2-M2-KKm charges of the solutions in M-theory can be obtained from the sum of the charges carried by each rod \eqref{eq:individualMcharges}: 
\begin{equation}
Q^I_{\text{M2}} \=  \frac{1}{2a_I}\,\sum_{i=1}^{N+1} M_{2i-1}\,,\qquad  Q_{\text{KKm}} \= \frac{\sum_{i=1}^{2N+1} M_{i}+\sum_{i=1}^{N} M_{2i}}{2a_0}.
\end{equation}

\subsection{Smooth bubbling geometries}

We now seek to construct smooth non-BPS bubbling solutions without black hole sources and struts. This consists in building a chain of connected bubble rods of different species. In M-theory alone, there are 7 species of smooth bubbles that can be used, and there are several others in different duality frameworks as described in section \ref{sec:dualFrame}. This induces a wide variety of configurations possible. In this section, we will construct an explicit M2-M2-M2-KKm configuration and a D1-D5-KKm configuration in type IIB. 

\subsubsection{In M-theory}
\label{sec:BuSolMtheory}

As described in section \ref{sec:regM4d}, among the seven species of physical bubble rods in M-theory, six correspond to the smooth degeneracy of one of the T$^6$ direction and carry a unique M2 charge and a KKm charge. To construct a smooth configuration with M2-M2-M2-KKm charges, one needs at least three species of these kinds.

\begin{figure}[h]
\centering
\includegraphics[width=0.8\textwidth]{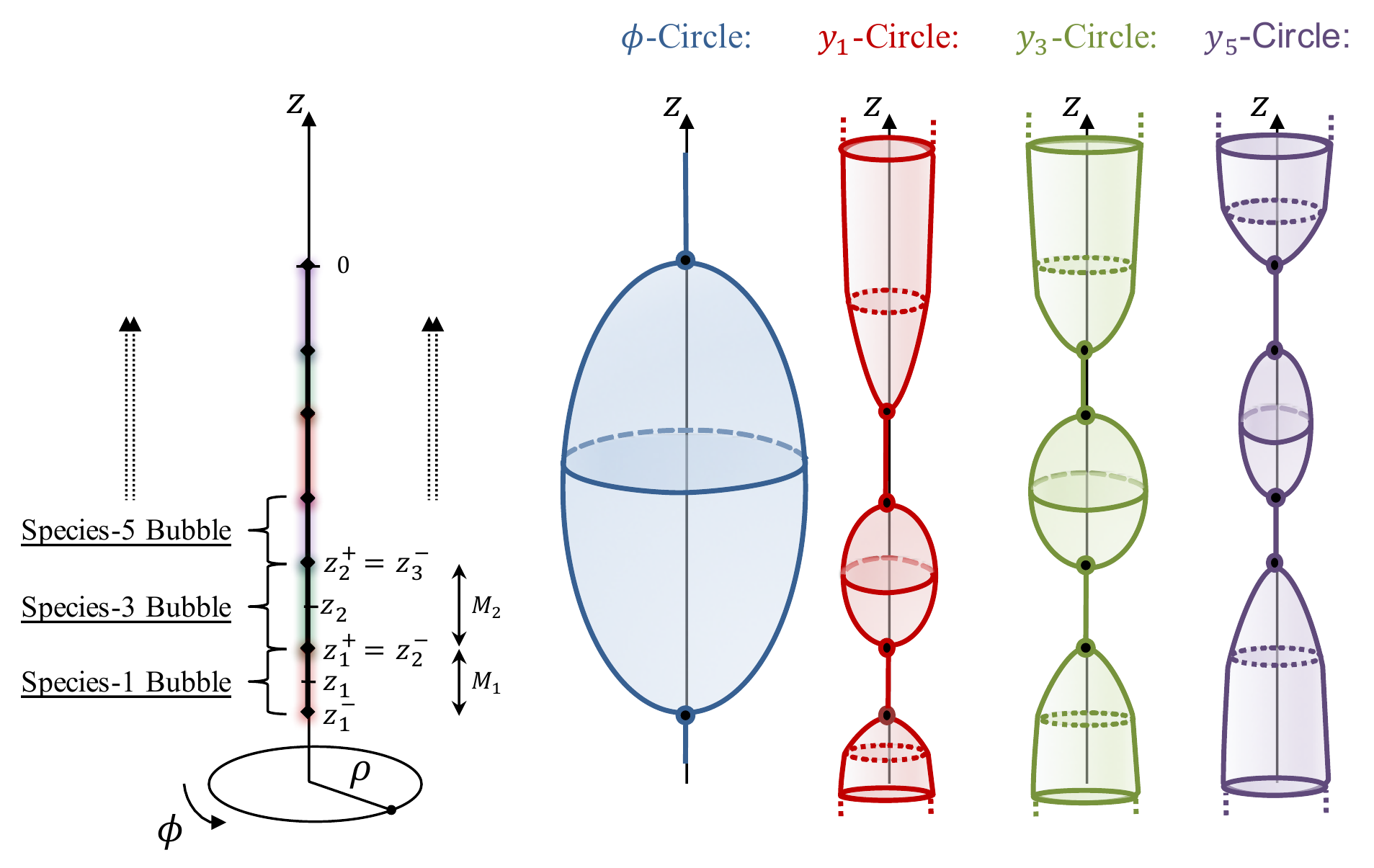}
\caption{Description of a chain of smooth M2-KKm bubbles in M-theory and the behavior of the $\phi$, $y_1$, $y_3$ and $y_5$ circles on the $z$-axis (the other fibers have finite size everywhere). The species-$k$ bubble rod corresponds to the degeneracy of the $y_k$-circle as an origin of a $\IR^2$.}
\label{fig:MtheoryBubbles}
\end{figure}

Therefore, we will consider a chain of $n=3N$ connected bubble rods of length $M_i$ which will successively make the $y_1$, $y_3$ and $y_5$ circles shrink (see Fig.\ref{fig:MtheoryBubbles}). Having connected rods constrains the rod centers in terms of their sizes as in \eqref{eq:connectedcond}.

The expression of the warp factors and gauge potentials are given by \eqref{eq:WarpfactorsRods} with the following choices of weights, obtained from \eqref{eq:bubblerodWeight1},
\begin{equation}
a_0 P_i^{(0)}=-G_i^{(0)}= \frac{1}{2}\,,\quad \begin{array}{l} G_{3i-2}^{(1)}  = a_1 P_{3i-2}^{(1)} =\frac{1}{2} , \quad G_{3i-2}^{(2)}=G_{3i-2}^{(3)}=P_{3i-2}^{(2)}=P_{3i-2}^{(3)}= 0, \\
G_{3i-1}^{(2)}  = a_2 P_{3i-1}^{(2)} =\frac{1}{2} , \quad G_{3i-1}^{(1)}=G_{3i-1}^{(3)}=P_{3i-1}^{(1)}=P_{3i-1}^{(3)}= 0, \\
G_{3i}^{(3)}  = a_3 P_{3i}^{(3)} =\frac{1}{2} , \quad G_{3i}^{(1)}=G_{3i}^{(2)}=P_{3i}^{(1)}=P_{3i}^{(2)}= 0.
\end{array}
\end{equation}
As in the previous constructions, it is convenient to split the warp factors $Z_\Lambda$ such as
\begin{align}
Z_0 &\= \frac{\cZ_0}{\sqrt{U_1 U_2 U_3}}, \qquad Z_I \= \frac{\cZ_I}{\sqrt{U_I}}\,,\label{eq:WarpfactorMtheoryBu1} \\
U_I &\equi \prod_{i=0}^{N-1} \left(1-\frac{2 M_{3i+I}}{R_+^{(3i+I)}} \right)\,, \qquad \cZ_0 \equi \frac{e^{b_0}-e^{-b_0}\,U_1 U_2 U_3}{2 a_0} \,,\qquad   \cZ_I \equi \frac{e^{b_I}-e^{-b_I}\,U_I}{2 a_I}. \nonumber
\end{align}
The functions $U_I$ are products of ``Schwarzschild factors'' that vanish at all the $(3i+I)^\text{th}$ rods while the $\cZ_\Lambda$ are non-zero and finite. The remaining functions \eqref{eq:WarpfactorsRods} give
\begin{align}
& W_0 = \left( U_1 U_2 U_3\right)^{\frac{1}{2}},\quad W_I \= U_I^{-\frac{1}{2}} ,\quad  H_0 \= \frac{1}{2 a_0} \left( r_-^{(1)}- r_+^{(n)} \right),\quad T_I\=-\sqrt{a_I^2+\frac{U_I}{\cZ_I^2}},  \nonumber  \\
& e^{2 \nu}=\frac{E_{-+}^{(1, n)}}{\sqrt{E_{++}^{(n, n)} E_{--}^{(1,1)}}} \prod_{i,j=0}^{N-1} \,\prod_{\substack{I,J=1\\I<J}}^3 \sqrt{\frac{E_{--}^{(3 i+I, 3 j+J)} E_{++}^{(3i+I, 3 j+J)}}{E_{-+}^{(3 i+I, 3 j+J)} E_{+-}^{(3 i+I, 3 j+J)}}}.
\label{eq:WarpfactorMtheoryBu2}
\end{align}
where $R_\pm^{(i)}$ and $E_{\pm \pm}^{(i,j)}$ are given in \eqref{eq:RpmdefRemind}. The metric and gauge potentials in M-theory, obtained from \eqref{eq:nonBPSfloatingRemind}, are given by
\begin{equation}
\begin{split}
ds_{11}^2 \= &- \frac{dt^2}{\left(\cZ_1 \cZ_2 \cZ_3 \right)^{\frac{2}{3}}} \+ \left(\cZ_1 \cZ_2 \cZ_3 \right)^{\frac{1}{3}}\, \left[\frac{1}{\cZ_0} \left(dy_7 +H_0 \,d\phi\right)^2 \+ \frac{\cZ_0}{U_1 U_2 U_3} \left( e^{2\nu} \left(d\rho^2 + dz^2 \right) +\rho^2 d\phi^2\right)\right]\\
&\hspace{-0.5cm} + \left(\cZ_1 \cZ_2 \cZ_3\right)^{\frac{1}{3}} \left[\frac{1}{\cZ_1} \left(U_1\,dy_1^2 \+ \,dy_2^2 \right) + \frac{1}{\cZ_2} \left(U_2 \,dy_3^2 \+ dy_4^2 \right) \+ \frac{1}{\cZ_3} \left(U_3\,dy_5^2\+ dy_6^2 \right)\right], \\
F_4 \= &d\left[ T_1 \,dt \wedge dy_1 \wedge dy_2 \+T_2 \,dt \wedge dy_3 \wedge dy_4 \+T_3\,dt \wedge dy_5 \wedge dy_6 \right]\,.
\end{split}
\end{equation}

The solutions are regular out of the rods since $U_I$ and $\cZ_\Lambda$ are finite and positive there. At the rods, the $y_1$, $y_3$ and $y_5$ fibers shrink to zero size where $U_1$, $U_2$ and $U_3$ vanish respectively (see Fig.\ref{fig:MtheoryBubbles}). These loci end the spacetime as smooth origin of $\IR^2$ (with potential conical defects) if $3N$ bubble equations are satisfied. These bubble equations arise from the regularity condition \eqref{eq:RegCondBu} at each rod. They are non-trivial multivariate polynomials that will fix all rod lengths $M_i$ in terms of the independent parameters: the extra-dimension radii $R_{y_a}$, the gauge-field parameters $(a_\Lambda,b_\Lambda)$, the number of bubbles $n=3N$ and the orbifold parameters $k_i \in \mathbb{N}$. These equations are a priori not solvable analytically if $N$ is not small, but can be solved numerically. Moreover, interesting approximations can be performed in the large $N$ limit, that is when the number of bubbles is large, and analytic solutions can be found. In this limit, the bubbles highly deform the spacetime as a ``bubble bag end'' geometry (see \cite{Bah:2021rki} for such an example with a configuration of two species of bubbles).

Each species of bubbles carries a M2 charge and a KKm charge as discussed in section \ref{sec:regM4d}. Therefore, by combining the three species, the solutions have M2-M2-M2-KKm charges. They can be derived by summing the individual rod charges \eqref{eq:individualMcharges}, and we find 
\begin{equation}
Q^I_{\text{M2}} \= \frac{1}{2a_I}\,\sum_{i=0}^N M_{3i+I} ,\qquad Q_{\text{KKm}}  \= \frac{1}{2a_0}\,\sum_{i=1}^{3N} M_i.
\label{eq:ChargeMtheoBu}
\end{equation}

In the four-dimensional frame obtained after reduction along the T$^6\times$S$^1$ \eqref{eq:4dSol}, the solutions are given by 
\begin{align}
ds_4^2 &\= -\frac{\sqrt{U_1 U_2 U_3}\,dt^2}{\sqrt{\cZ_0 \cZ_1 \cZ_2 \cZ_3}} \+ \frac{\sqrt{\cZ_0 \cZ_1 \cZ_2 \cZ_3}}{\sqrt{U_1 U_2 U_3}}\left( e^{2\nu} \left(d\rho^2 + dz^2 \right) +\rho^2 d\phi^2\right), \quad \Phi_0 \= \frac{1}{2} \log \left(U_1 U_2 U_3\right), \nonumber \\
\Phi_I &\= -\frac{1}{2}\log U_I  \,,\quad z^I \= i\,\frac{|\epsilon_{IJK}|}{2} \sqrt{\frac{U_I\,\cZ_J \cZ_K}{\cZ_0\,\cZ_I} },\quad \bar{F}^{0}  \= -dH_0 \wedge d\phi \,,\quad \bar{F}^{I}  \= dT_I \wedge dt\,.
\end{align}
Therefore, the solutions are singular at the rods in four dimensions, and these singularities are resolved in M-theory as the degeneracy of the extra dimensions. Moreover, the solutions are asymptotic to $\IR^{1,3}$ if \eqref{eq:condonasymp} is satisfied. The ADM mass is given by \eqref{eq:ADMmass4d}
\begin{equation}
\cM \= \frac{1}{8G_4}\left[\coth b_0 \,\sum_{i=1}^{3N} M_i \+ \sum_{I=1}^3 \coth b_I \sum_{i=0}^{N-1} M_{3i+I}\right].
\label{eq:MassMtheoBu}
\end{equation}
Furthermore, since the extra scalars, $\Phi_\Lambda$, are turned on, the four-dimensional solutions are not solutions of the STU Lagrangian as detailed in section \ref{sec:4dreduction}. However, they have the same conserved charges as non-extremal four-charge static black holes given by \eqref{eq:4chargeBH}, but they are horizonless and terminate the spacetime as a chain of non-trivial bubbles wrapped by fluxes. We leave the comparison of the two geometries for a later project. More precisely, it will be interesting to study, in the manner of \cite{Bah:2021rki}, the compactness of the bubble structure with respect to the size of the horizon of the corresponding black hole.

\begin{itemize}
\item[•] \underline{BPS limit:}
\end{itemize}

We have seen in \eqref{eq:BPSlimit} that generic non-BPS solutions approach the BPS regime by taking $(b_\Lambda, a_\Lambda) \to 0 $. One can derive the BPS limit of our specific non-BPS smooth bubbling solutions by considering $(b_\Lambda, a_\Lambda)$ small. 

We consider the specific flow where $b_\Lambda= a_\Lambda=\lambda$ with $\lambda \to 0$. First, from the bubble equations \eqref{eq:RegCondBu}, we have 
\begin{equation}
M_{i} \underset{\lambda\to 0}{\sim} \lambda \,q_{i},\qquad i=0,...,3N,
\end{equation}
where $q_i$ are finite constants as $\lambda \to 0$. Therefore, the full rod configuration shrinks to a point. However, it is clear that the charges \eqref{eq:ChargeMtheoBu} and mass \eqref{eq:MassMtheoBu} do not vanish such as
\begin{equation}
Q^I_{\text{M2}} \sim \frac{1}{2}\,\sum_{i=0}^N q_{3i+I} ,\qquad Q_{\text{KKm}}  \sim \frac{1}{2}\,\sum_{i=1}^{3N} q_i, \qquad \cM \sim \frac{1}{4 G_4}\left(Q_{\text{KKm}}+Q^1_{\text{M2}}+Q^2_{\text{M2}}+Q^3_{\text{M2}} \right),
\end{equation}
and the solutions converge to a non-trivial BPS solutions. More precisely, the warp factors, \eqref{eq:WarpfactorMtheoryBu1} and \eqref{eq:WarpfactorMtheoryBu2}, converge to
\begin{equation}
\begin{split}
U_I &\sim1\,, \qquad \cZ_0 \sim 1+ \frac{Q_\text{KKm}}{r}\,,\qquad  \cZ_I \sim 1+ \frac{Q_\text{M2}^I}{r}, \\
H_0 &\sim Q_\text{KKm}\,\cos \theta\,,\qquad T_I \sim - \frac{1}{\cZ_I}\,,\qquad e^{2\nu} \sim 1\,.
\end{split}
\end{equation}
where $(r,\theta)$ are the spherical coordinates $(\rho,z)=(r\sin\theta,r\cos  \theta)$. Therefore, the solutions approach a four-charge static BPS black hole, or static BMPV black hole.

At small $(b_\Lambda, a_\Lambda)$, the bubbling geometries are almost indistinguishable from the BPS black hole. They must develop an AdS$_2$ throat that caps off smoothly at $r=\cO(\lambda)$ as non-BPS bubbles. This is the first example of such a resolution in the non-BPS regime. It also shows that the geometries can be as compact as a black hole, unlike the previous ones constructed by the author \cite{Bah:2021owp,Bah:2021rki}.

It would be interesting to push the comparison further. Specifically, it would be to analyze how the BMPV black hole horizon at $r=0$ is smoothly resolved by adding a small amount of non-extremity when moving in the non-BPS regime, and how the geometries develop an AdS$_2$ throat.

\subsubsection{In type IIB}
\label{sec:BuSolIIB}

We now work in the D1-D5-P-KKm framework given by \eqref{eq:D1D5PKKm}. 
The advantage of constructing smooth non-BPS bubbling geometries in type IIB is that there are two species of bubbles that do not require to turn on the warp factors $W_\Lambda$: a KKm bubble given by the weights \eqref{eq:bubblerodWeight2} and a D1-D5-KKm bubble given by \eqref{eq:bubblerodWeight4}. Such configurations are solutions of the STU Lagrangian when reduced to four dimensions \eqref{eq:KKredAction4d}. However, the regularity of the D1-D5-KKm bubble requires that the solutions have no momentum P charge.

\begin{figure}[h]
\centering
\includegraphics[width=0.65\textwidth]{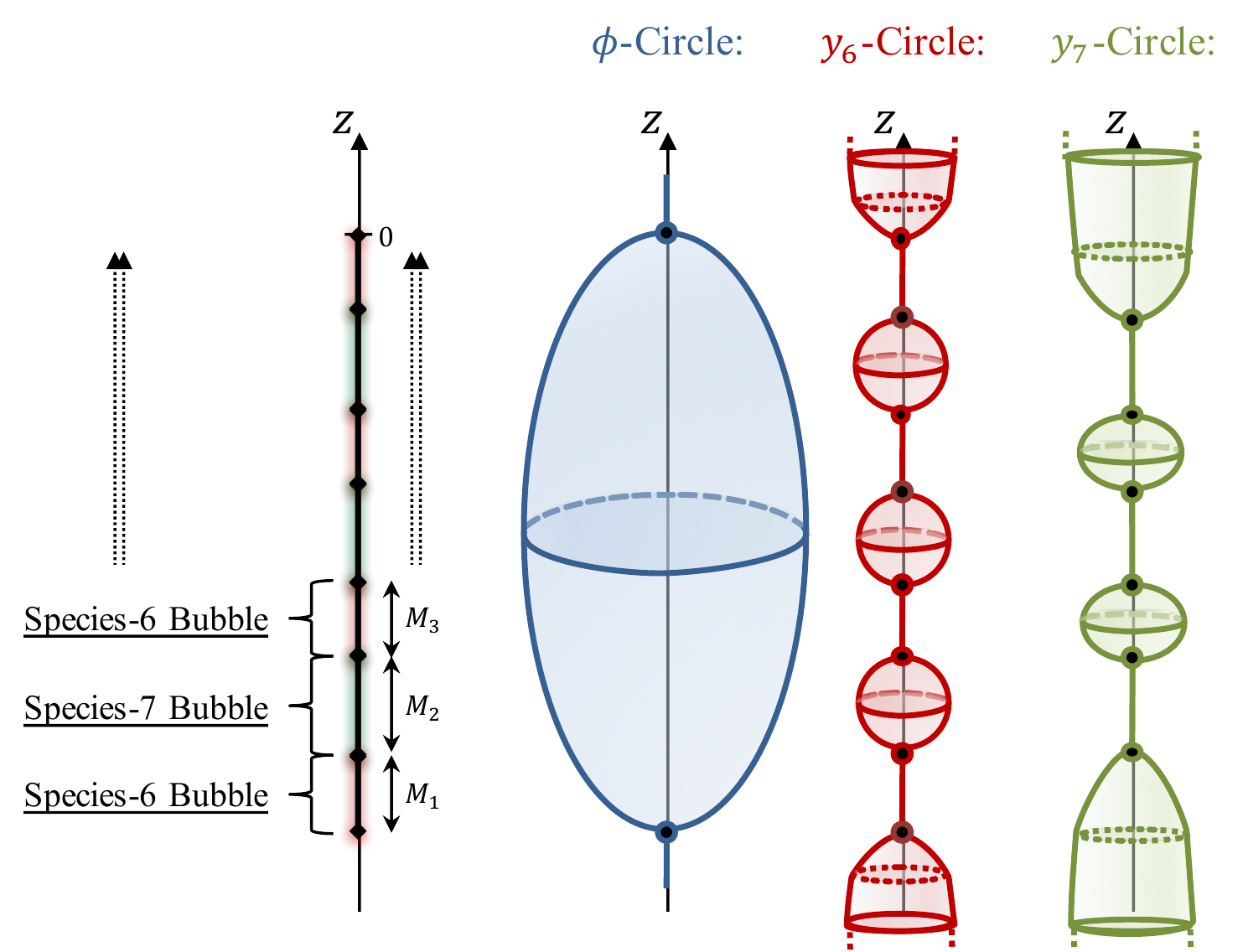}
\caption{Description of a chain of smooth D1-D5-KKm and KKm bubbles in type IIB and the behavior of the $\phi$, $y_6$, and $y_7$ circles on the $z$-axis (the other fibers have finite size everywhere). The species-$k$ bubble rod corresponds to the degeneracy of the $y_k$-circle as an origin of a $\IR^2$.}
\label{fig:IIBtheoryBubbles}
\end{figure}

We will therefore consider a chain of $n=2N+1$ connected rods of lengths $M_i$ that consists of a succession of D1-D5-KKm and KKm bubbles (see Fig.\ref{fig:IIBtheoryBubbles}).\footnote{We are considering an odd number of rods to make the link with the solutions of \cite{Bah:2021owp,Bah:2021rki}, but one could have taken any other configuration.} Having connected rods constrains the rod centers in terms of their sizes as in \eqref{eq:connectedcond}.

The warp factors and gauge potentials \eqref{eq:WarpfactorsRods} give\footnote{ As discussed in section \ref{sec:dualFrame}, one needs to consider the neutral limit \eqref{eq:neutrallimit} for the pair $(Z_3,T_3)$. That is we have considered $a_3=e^{b_3}/2=\infty$ and $a_3 P^{(3)}_{2i-1} = -1/2$ in \eqref{eq:WarpfactorsRods}.}
\begin{align}
Z_0 &\= \frac{\cZ_0}{U_{y_7}\sqrt{U_{y_6}}}, \quad Z_1 \= \frac{\cZ_1}{\sqrt{U_{y_6}}}\,, \quad Z_2 \= \frac{\cZ_2}{\sqrt{U_{y_6}}}\,, \quad Z_3 \= \sqrt{U_{y_6}}\,,\quad W_\Lambda\= 1,\\
H_0 &\= \frac{1}{2 a_0} \left[r_-^{(1)}- r_+^{(n)}+\sum_{i=1}^N \left(r_-^{(2i)}-r_+^{(2i)} \right) \right],\quad H_2 \=  \frac{1}{2 a_2} \sum_{i=1}^{N+1} \left(r_-^{(2i-1)}-r_+^{(2i-1)} \right)\,,\nonumber\\ 
 T_1&\=-\sqrt{a_1^2+\frac{U_{y_6}}{\cZ_1^2}}, \quad T_3 \= 0,\quad  e^{2 \nu}\=\frac{E_{-+}^{(1, n)}}{\sqrt{E_{++}^{(n, n)} E_{--}^{(1,1)}}} \prod_{i=1}^{N} \prod_{j=1}^{N+1} \sqrt{\frac{E_{--}^{(2 i, 2 j-1)} E_{++}^{(2 i, 2 j-1)}}{E_{-+}^{(2 i, 2 j-1)} E_{+-}^{(2 i, 2 j-1)}}}\,, \nonumber
\end{align}
where we have defined
\begin{equation}
\begin{split}
U_{y_6} &\equi \prod_{i=1}^{N+1} \left(1-\frac{2 M_{2i-1}}{R_+^{(2i-1)}} \right)\,, \qquad U_{y_7} \equi \prod_{i=1}^N \left(1-\frac{2 M_{2i}}{R_+^{(2i)}} \right)\,, \\
\cZ_0 &\equi \frac{e^{b_0}-e^{-b_0}\,U_{y_6} U_{y_7}^2}{2 a_0} \,,\qquad   \cZ_I \equi \frac{e^{b_I}-e^{-b_I}\,U_{y_6}}{2 a_I}.
\end{split}
\end{equation}
The metric, the dilaton and gauge fields in type IIB \eqref{eq:D1D5PKKm} lead to
\begin{align}
ds_{10}^2 \= &\frac{1}{\sqrt{\cZ_1 \cZ_2}}\left[-dt^2+U_{y_6}\,dy_6^2 \right] \+\sqrt{\frac{\cZ_1}{\cZ_2}}\left[ dy_1^2+dy_2^2+dy_3^2+dy_4^2 \right]  \nonumber\\
&  \+ \sqrt{ \cZ_1 \cZ_2 } \left[\frac{U_{y_7}}{\cZ_0} \left(dy_7 +H_0 \,d\phi\right)^2 \+ \frac{\cZ_0}{U_{y_6}U_{y_7}}\left( e^{2\nu} \left(d\rho^2 + dz^2 \right) +\rho^2 d\phi^2\right)\right] \label{eq:D1D5PKKm2} \\
  \Phi  \=& \frac{1}{2} \log \frac{\cZ_1}{\cZ_2}\,,\quad B_2 \= 0\,, \quad C^{(0)} \=  C^{(4)} \= 0 \,,\quad C^{(2)}\=  H_2 \,d\phi \wedge dy_7 \- T_1 \,dt \wedge dy_6\,.\nonumber 
\end{align}
The solutions carry D1, D5 and KKm charges given by 
\begin{equation}
Q_{D1} \= \frac{1}{2a_1}\,\sum_{i=1}^{N+1} M_{2i-1}\,,\quad Q_{D5} \= \frac{1}{2a_2}\,\sum_{i=1}^{N+1} M_{2i-1}\,, \quad Q_{\text{KKm}} \= \frac{\sum_{i=1}^{2N+1} M_{i}+\sum_{i=1}^{N} M_{2i}}{2a_0}.
\end{equation} 
The solutions are asymptotic to $\IR^{1,3}$ when reduced to four dimensions if \eqref{eq:condonasymp} is satisfied. Moreover, the ADM mass \eqref{eq:ADMmass4d} is given by
\begin{equation}
\cM \=  \frac{1}{8 G_4} \left(2\coth b_0 \, \sum_{i=1}^N M_{2i} + \left(-1+\sum_{I=1}^3 \coth b_I \right) \sum_{i=1}^{N+1} M_{2i-1} \right).
\end{equation}

If we assume $(a_1,b_1)=(a_2,b_2)$, we retrieve the non-BPS smooth bubbling solutions constructed in \cite{Bah:2021owp,Bah:2021rki}. Therefore, we refer the reader to \cite{Bah:2021owp,Bah:2021rki} for an exhaustive analysis of these backgrounds, their regularity and their depiction as ``bubble bag ends'' \cite{Bah:2021rki}. In short, the geometries are strongly distorted and have a S$^2$ that suddenly opens up near the bubble loci like a bag of smooth bubbles. However, in the absence of P charge, their BPS limit is not a BPS black hole but a Taub-NUT space. Thus, they cannot develop an AdS throat and cannot be as compact as a black hole.

\section{Explicit five-dimensional solutions}
\label{sec:Sol5d}

In this section, we derive explicit solutions that are asymptotic to $\IR^{1,4}$ plus extra-compact dimensions. We will be more brief than in the previous section. Indeed, one can transform any asymptotically-$\IR^{1,3}$ solution into an asymptotically-$\IR^{1,4}$ geometry while preserving the brane structure of the solution by simply taking a neutral limit \eqref{eq:neutrallimit} for the pair $(Z_0,H_0)$ and adding a semi-infinite rod in $Z_0$:\footnote{The neutral limit for the pair $(Z_0,H_0)$ has sent $a_0$ to infinity while keeping $a_0 P^{(0)}_{i} = G_i$ finite, and we find the expressions \eqref{eq:WarpfactorsRods5d} with the regularity condition \eqref{eq:condonasymp5d}.}
\begin{align}
Z_0^{(5d)} &\=  \left(r_+^{(n)}-(z-z_n^+) \right)^{-1} \,Z_0^{(4d)}\,, \\
e^{2\nu^{(5d)}} & \=  \frac{r_+^{(n)}-(z-z_n^+)}{2 r_+^{(n)}}\,\prod_{i=1}^n \left(\frac{E_{++}^{(n,i)}\,R_-^{(i)}}{E_{+-}^{(n,i)}\,R_+^{(i)}} \right)^{a_0 P^{(0)}_i } \, e^{2\nu^{(4d)}} \,. \nonumber
\end{align}
All other warp factors and gauge potentials can be left unchanged. Five-dimensional solutions can therefore be easily extracted from the four-dimensional solutions constructed before. However, the regularity constraints at the sources are modified by the presence of the semi-infinite rod as discussed in section \ref{sec:regM5d}. 

We will first derive solutions consisting of a chain of three-charge non-extremal black holes before constructing smooth bubbling geometries. The generic solutions are given in section \ref{sec:5dNonBPSGen}, and the warp factors and gauge potentials are given in \eqref{eq:WarpfactorsRods5d}.

\subsection{Chain of static non-extremal black holes}

We consider that the $n$ finite rods of length $M_i$ on the $z$-axis correspond to black rods given by the weights \eqref{eq:blackrodWeight5d}. As for four-dimensional solutions, it is more convenient to divide the warp factors into meaningful parts: 
\begin{align}
Z_I &\= \frac{\cZ_I}{\sqrt{U_t}}\,, \qquad Z_0 \= \frac{1}{\left(r_+^{(n)}-(z-z_n^+) \right)\, \sqrt{U_t} }\\
U_t &\equi \prod_{i=1}^n
\left(1-\frac{2 M_i}{R_+^{(i)}} \right)\,,\qquad \cZ_\Lambda \equi \frac{e^{b_\Lambda}-e^{-b_\Lambda}\,U_t}{2 a_\Lambda}. \nonumber
\end{align}
The remaining functions \eqref{eq:WarpfactorsRods5d} give
\begin{align}
& W_\Lambda = 1,\qquad e^{2\nu} \= \frac{r_+^{(n)}-(z-z_n^+)}{2r_+^{(n)}} \,\prod_{i=1}^n \sqrt{\frac{E_{++}^{(n,i)}\,R_-^{(i)}}{E_{+-}^{(n,i)}\,R_+^{(i)}}  }\,\prod_{i,j=1}^n\, \sqrt{  \frac{E_{+-}^{(i,j)}E_{-+}^{(i,j)}}{E_{++}^{(i,j)}E_{--}^{(i,j)}}} ,\\
&  H_0 \=0 ,\qquad T_I\=-\sqrt{a_I^2+\frac{U_t}{\cZ_I^2}}, \nonumber 
\end{align}
where $R_\pm^{(i)}$ and $E_{\pm \pm}^{(i,j)}$ are given in \eqref{eq:RpmdefRemind}. The metric and gauge potentials in M-theory are given in \eqref{eq:nonBPSfloatingRemind}, while the five-dimensional reduction along T$^6$ is given by \eqref{eq:5dframework}:
\begin{align}
ds_5^2 &\= -\frac{U_t\,dt^2}{(\cZ_1 \cZ_2 \cZ_3)^\frac{2}{3}} \+( \cZ_1 \cZ_2 \cZ_3)^\frac{1}{3} \Biggl[\left(r_+^{(n)}-(z-z_n^+) \right)\,dy_7^2  \nonumber \\
& \hspace{5.5cm}+ \frac{1}{\left(r_+^{(n)}-(z-z_n^+) \right) \,U_t} \left( e^{2\nu} \left(d\rho^2 + dz^2 \right) +\rho^2 d\phi^2\right) \Biggr], \nonumber \\
X^I &\= \frac{|\epsilon_{IJK}|}{2} \left(\frac{\cZ_J \cZ_K}{\cZ_I^2} \right)^\frac{1}{3},\qquad \Phi_\Lambda \=0 \,,\qquad F^{(I)} \= dT_I \wedge dt\,.
\end{align}
The warp factors $U_t$ vanishes at each rod as a product of ``Schwarzschild factors,'' thereby inducing the horizons. The $\cZ_I$ depend only on the gauge field parameters, and are always positive and finite. The condition for having an asymptotically flat five-dimensional space requires \eqref{eq:condonasymp5d}, 
and one can read the ADM mass from \eqref{eq:ADMmass5d}:
\begin{equation}
\cM \= \frac{\pi\,\sum_{I=1}^3 \coth b_I}{4 G_5}\,\sum_{i=1}^n M_i\,.
\label{eq:MassBHchain5d}
\end{equation}
Moreover, the solutions carry three charges that are M2-M2-M2 charges in M-theory. They can be obtained from the sum of the rod charges \eqref{eq:individualMcharges}: 
\begin{equation}
Q^I_{\text{M2}} \=  \frac{1}{2a_I}\,\sum_{i=1}^n M_i.
\end{equation}

If we consider a unique rod, $n=1$, one retrieves the static limit of the single three-charge black hole constructed in \cite{Cvetic:1996xz,Cvetic:1997uw,Cvetic:1998xh}. The map can be done by changing coordinates 
$$(\rho,\,z)= \frac{1}{2}\left(\sqrt{r^2 \left(r^2 -2M_{1}\right)} \sin 2\theta,\, \left(r^2-M_1\right) \cos 2\theta +2z_{1}\right), $$
and identifying the boost parameters $\delta_I$ used in \cite{Cvetic:1996xz,Cvetic:1997uw,Cvetic:1998xh} to the gauge field parameter $b_I$ such as $\cosh 2\delta_I =\coth b_I$.

Taking $n>1$ consists therefore in a chain of static three-charge black holes in five dimensions. The black holes are separated by struts, that are segments where the $\phi$-circle degenerates with a conical excess of order $d_i^{-1} >1$ \eqref{eq:dialphaDefcharged}. The struts account for the lack of repulsion between the non-extremal black holes and encodes the binding energy of the system.

The singular struts can be replaced by regular bubbles in a manner similar to that used for the chain of four-dimensional black holes in the section \ref{sec:chainBHbu}. Specifically, one can source the M-theory solutions with bubble rods at each segment between the black holes. The bubble rods can either be vacuum bubbles that make the $y_7$ circle degenerate smoothly \eqref{eq:bubblerodWeight25d}, or M2 bubbles that correspond to the degeneracy of one of the T$^6$ directions \eqref{eq:bubblerodWeight15d}.

\subsection{Smooth bubbling geometries}

We now construct smooth non-BPS bubbling solutions that consist in a chain of connected bubble rods of different species. We will restrict to the five-dimensional analogs of the M-theory solutions constructed in section \ref{sec:BuSolMtheory}. One could have also built asymptotically-$\IR^{1,4}$ D1-D5 non-BPS bubbling solutions by taking the equivalent of the four-dimensional solutions of section \ref{sec:BuSolIIB}.

We consider a chain of $n=3N$ connected bubble rods of length $M_i$ where the circles $y_1$, $y_3$ and $y_5$ successively degenerate. Having connected rods constrains the rod centers in terms of their sizes as in \eqref{eq:connectedcond}, and we also assume that the origin of the $z$-axis is at $z_n^+=0$. The warp factors and gauge potential are given by \eqref{eq:WarpfactorsRods5d} with the following choices of weights, obtained from \eqref{eq:bubblerodWeight15d},
\begin{equation}
G_i =-G_i^{(0)}= \frac{1}{2}\,,\quad \begin{array}{l} G_{3i-2}^{(1)}  = a_1 P_{3i-2}^{(1)} =\frac{1}{2} , \quad G_{3i-2}^{(2)}=G_{3i-2}^{(3)}=P_{3i-2}^{(2)}=P_{3i-2}^{(3)}= 0, \\
G_{3i-1}^{(2)}  = a_2 P_{3i-1}^{(2)} =\frac{1}{2} , \quad G_{3i-1}^{(1)}=G_{3i-1}^{(3)}=P_{3i-1}^{(1)}=P_{3i-1}^{(3)}= 0, \\
G_{3i}^{(3)}  = a_3 P_{3i}^{(3)} =\frac{1}{2} , \quad G_{3i}^{(1)}=G_{3i}^{(2)}=P_{3i}^{(1)}=P_{3i}^{(2)}= 0.
\end{array}
\end{equation}
It is also convenient to divide the warp factors $Z_\Lambda$ such as
\begin{align}
Z_0 &\= \frac{1}{\left(r_+^{(n)}-z\right)\,\sqrt{U_1 U_2 U_3}}, \qquad Z_I \= \frac{\cZ_I}{\sqrt{U_I}}\,,\label{eq:WarpfactorMtheoryBu15d} \\
U_I &\equi \prod_{i=0}^{N-1} \left(1-\frac{2 M_{3i+I}}{R_+^{(3i+I)}} \right)\,, \qquad  \cZ_I \equi \frac{e^{b_I}-e^{-b_I}\,U_I}{2 a_I}\,. \nonumber
\end{align}
The remaining functions \eqref{eq:WarpfactorsRods5d} give
\begin{align}
& W_0 = \left( U_1 U_2 U_3\right)^{\frac{1}{2}},\qquad W_I \= U_I^{-\frac{1}{2}} ,\qquad  H_0 \= 0 ,\qquad T_I\=-\sqrt{a_I^2+\frac{U_I}{\cZ_I^2}},  \nonumber  \\
& e^{2 \nu}=\frac{1}{Z_0}\,\sqrt{\frac{E_{-+}^{(1, n)}}{2E_{++}^{(n, n)} E_{--}^{(1,1)}}} \prod_{i,j=0}^{N-1} \,\prod_{\substack{I,J=1\\I<J}}^3 \sqrt{\frac{E_{--}^{(3 i+I, 3 j+J)} E_{++}^{(3i+I, 3 j+J)}}{E_{-+}^{(3 i+I, 3 j+J)} E_{+-}^{(3 i+I, 3 j+J)}}}.
\label{eq:WarpfactorMtheoryBu25d}
\end{align}
where $R_\pm^{(i)}$ and $E_{\pm \pm}^{(i,j)}$ are given in \eqref{eq:RpmdefRemind}. The metric and gauge potentials in M-theory, obtained from \eqref{eq:nonBPSfloatingRemind}, are given by
\begin{align}
ds_{11}^2 \= &- \frac{dt^2}{\left(\cZ_1 \cZ_2 \cZ_3 \right)^{\frac{2}{3}}} \+ \left(\cZ_1 \cZ_2 \cZ_3 \right)^{\frac{1}{3}}\, \Biggl[\left(r_+^{(n)}-z \right) \,dy_7^2 \nonumber \\
& \hspace{5.1cm}\+ \frac{1}{\left(r_+^{(n)}-z \right)U_1 U_2 U_3} \left( e^{2\nu} \left(d\rho^2 + dz^2 \right) +\rho^2 d\phi^2\right)\Biggr]  \nonumber\\
& + \left(\cZ_1 \cZ_2 \cZ_3\right)^{\frac{1}{3}} \left[\frac{1}{\cZ_1} \left(U_1\,dy_1^2 \+ \,dy_2^2 \right) + \frac{1}{\cZ_2} \left(U_2 \,dy_3^2 \+ dy_4^2 \right) \+ \frac{1}{\cZ_3} \left(U_3\,dy_5^2\+ dy_6^2 \right)\right], \nonumber\\
F_4 \= &d\left[ T_1 \,dt \wedge dy_1 \wedge dy_2 \+T_2 \,dt \wedge dy_3 \wedge dy_4 \+T_3\,dt \wedge dy_5 \wedge dy_6 \right]\,.
\end{align}

The solutions are regular out of the finite and semi-infinite rods since $U_I$, $\cZ_I$ and $r_+^{(n)}-z$ are finite and positive there. 

The semi-infinite rod, $\rho=0$ and $z>z_n^+=0$, corresponds to a coordinate degeneracy since $r_+^{(n)}-z=0$. The $y_7$-circle reduces to zero size as a smooth origin of $\IR^2$ (see appendix \ref{App:AttheRod2} for more details).

At the finite rods, the $y_1$, $y_3$ and $y_5$ fibers shrink successively to zero size since $U_1$, $U_2$ and $U_3$ vanish alternatively. These loci correspond to smooth origin of $\IR^2$ with potential conical defects if $3N$ bubble equations are satisfied (see Appendix \ref{App:AttheRod2} for more details). These bubble equations arise from the regularity condition \eqref{eq:RegCondBu5d} at each rod. They are non-trivial multivariate polynomials that fix all rod lengths, $M_i$. 

Each species of bubble rod carries a M2 charge as discussed in section \ref{sec:regM5d}. Therefore, by combining the three species, the solutions have M2-M2-M2 charges. They are given by
\begin{equation}
Q^I_{\text{M2}} \= \frac{1}{2a_I}\,\sum_{i=0}^N M_{3i+I}.
\label{eq:ChargeMtheoBu5d}
\end{equation}
In the five-dimensional frame after reduction on the T$^6$ \eqref{eq:5dframework}, the solutions are 
\begin{align}
ds_5^2 &\= \left(U_1 U_2 U_3 \right)^{\frac{1}{3}} \left[- \frac{dt^2}{\left(\cZ_1 \cZ_2 \cZ_3 \right)^{\frac{2}{3}}} \+ \left(\cZ_1 \cZ_2 \cZ_3 \right)^{\frac{1}{3}}\, \Biggl[\left(r_+^{(n)}-z \right) \,dy_7^2  \right.  \\
& \left. \hspace{5.5cm}\+ \frac{1}{\left(r_+^{(n)}-z \right)U_1 U_2 U_3} \left( e^{2\nu} \left(d\rho^2 + dz^2 \right) +\rho^2 d\phi^2\right)\Biggr] \right] \nonumber  \\
\Phi_0 &= \frac{1}{2}\log (U_1 U_2 U_3)  \,,\quad \Phi_I = -\frac{1}{2}\log U_I  \,,\quad X^I =\frac{|\epsilon_{IJK}|}{2} \left(\frac{U_I\,\cZ_J \cZ_K}{\sqrt{U_J U_K}\,\cZ_I} \right)^\frac{1}{3},\quad F^{I}  = dT_I \wedge dt\,. \nonumber
\end{align}
Therefore, they are singular at the rods in five dimensions, and these singularities are resolved in M-theory as the degeneracy of the extra dimensions. Moreover, the solutions are asymptotic to $\IR^{1,4}$ if \eqref{eq:condonasymp5d} is satisfied. The ADM mass is given by \eqref{eq:ADMmass5d}
\begin{equation}
\cM \= \frac{\pi}{4G_5} \sum_{I=1}^3 \left( \coth b_I \sum_{i=0}^{N-1} M_{3i+I}\right)\,.
\label{eq:MassMtheoBu5d}
\end{equation}
Furthermore, since the extra scalars, $\Phi_\Lambda$, are turned on, the five-dimensional solutions are not solutions of the STU Lagrangian as detailed in section \ref{sec:5dreduction}. Nevertheless, they have the same conserved charges as a non-extremal three-charge static black hole in five dimensions, but they are horizonless and terminate the spacetime as a chain of non-BPS bubbles in M-theory.

\begin{itemize}
\item[•] \underline{BPS limit:}
\end{itemize}

As for the four-dimensional bubbling geometries constructed in section \ref{sec:BuSolMtheory}, one can derive the BPS limit of the present five-dimensional solutions. This consists of taking the gauge field parameters to zero, $(b_I, a_I) \to 0 $. For simplicity, we consider $b_I= a_I=\lambda$ with $\lambda \to 0$. First, from the bubble equations \eqref{eq:RegCondBu5d}, we have 
\begin{equation}
M_{i} \underset{\lambda\to 0}{\sim} \lambda \,q_{i},\qquad i=0,...,3N,
\end{equation}
where $q_i$ are finite constants as $\lambda \to 0$, and the whole structure reduces to a point. However, it is clear that the charges \eqref{eq:ChargeMtheoBu5d} and the mass \eqref{eq:MassMtheoBu5d} do not vanish:
\begin{equation}
Q^I_{\text{M2}} \sim \frac{1}{2}\,\sum_{i=0}^N q_{3i+I} , \qquad \cM \sim \frac{\pi}{4 G_5}\left(Q^1_{\text{M2}}+Q^2_{\text{M2}}+Q^3_{\text{M2}} \right),
\end{equation}
The warp factors and gauge potentials, \eqref{eq:WarpfactorMtheoryBu15d} and \eqref{eq:WarpfactorMtheoryBu25d}, behave as
\begin{equation}
\begin{split}
U_I &\sim1\,, \qquad \cZ_I \sim 1+ \frac{Q_\text{M2}^I}{r^2}, \qquad r_+^{(n)}-z  \sim r^2 \sin^2 \theta \,, \qquad  T_I \sim - \frac{1}{\cZ_I}\,,\qquad e^{2\nu} \sim 1\,.
\end{split}
\end{equation}
where $(r,\theta)$ are the five-dimensional spherical coordinates $(\rho,z)=\frac{1}{2}(r^2\sin2\theta,r^2\cos 2 \theta)$. Therefore, the solutions approach a three-charge static BPS black hole in five dimensions.

The BPS black hole horizon is thus smoothly resolved into smooth non-BPS bubbles wrapped by flux by adding a small amount of non-extremity. At small $(a_I,b_I)$, the solutions are almost indistinguishable from the BPS black hole. They must develop an AdS$_2$ throat that does not end in a horizon but as non-BPS bubbles.

\section{Discussion}

In this paper, we have shown that non-BPS linear ansatz that allow non-trivial matter fields and topology can be directly derived from Maxwell-Einstein equations in string theory. In M-theory on T$^6\times$S$^1$, our ansatz enables four gauge potentials corresponding to three stacks of M2 branes and a KKm vector, and it can be dualized to other string frames such as the D1-D5-P-KKm frame. Focusing on solutions that are asymptotic to four- and five-dimensional flat space, we derived families of four-charge non-extreme black holes on a line and non-BPS bubbling geometries. The latter can have the same charges and mass as non-extremal black holes but are horizonless and smooth. We have highlighted examples that are almost indistinguishable from the BPS black hole when they are slightly non-BPS but terminate the spacetime smoothly as a chain of non-BPS bubbles.

While the present non-BPS floating brane ansatz opens a new door to the study of non-BPS solitons in string theory, several questions need to be explored in future work. First, although the smooth charged bubble is known to be a meta-stable vacuum \cite{Stotyn:2011tv,StabilityPaper}, the stability of chains of such objects remains to be studied. Second, one can think of constructing new families of solutions with different boundaries than the flat asymptotic.  Third, it will also be interesting to add brane degrees of freedom to the ansatz in M-theory, such as M5 or P brane charges. This will enable Chern-Simons interactions to be turned on as in the BPS floating brane ansatz for multicenter solutions. In addition, to construct more astrophysically-interesting solutions, a rotational degree of freedom will need to be added. 

Furthermore, understanding the origin of the solutions as bound states of strings and branes will require careful analysis of the geometric transition that occurs. For BPS solutions, for example, the geometric transition that gives rise to smooth geometries is well understood \cite{Bena:2013dka}. The present one uses very different mechanisms because our solutions are far from being the non-BPS extensions of known BPS smooth bubbling solutions as their BPS limit suggests, and they are based on non-trivial topology changes of compact tori. They should therefore be very illuminating on constructions of string and brane boundary states in the non-supersymmetric regime, but one should not expect these bound states to use transitions similar to those of BPS bound states.

Finally, it would be interesting to have a better geometrical and physical understanding of the smooth bubbling geometries. One can compare them with the non-extremal black hole with the same mass and charges as in \cite{Bah:2021rki}. Moreover, one can derive their multipole moments, probe them by light geodesics or compute their quasi-normal modes using technologies developed for similar geometries \cite{Bena:2020uup,Bianchi:2020miz,Bah:2021jno,Mayerson:2020tpn,Bacchini:2021fig,Ikeda:2021uvc,Bena:2020yii}. One could also being interested in performing M2-brane probe computation in the smooth bubbling solutions in M-theory. If we do not expect the branes to feel an entirely flat potential that allows them to ``float'' everywhere in space, our non-linear ansatz suggest special flat regions where probe can be trapped.

\section*{Acknowledgments}
I am grateful to Ibrahima Bah for his rich advice, and to Iosif Bena, Nejc Ceplak,  Anthony Houppe and Nick Warner for interesting discussions. This work is supported in part by NSF grant PHY-1820784.

\vspace{1cm}

\newpage

\appendix
\leftline{\LARGE \bf Appendices}

\section{Equations for $\nu$}
\label{App:Weyl}

The equations for the base warp factor, $\nu$, obtained from \eqref{eq:EOMMtheory} with the ansatz of metric and gauge field given by \eqref{eq:nonBPSfloating}, are
\begin{align}
\frac{2}{\rho} \partial_z \nu \= &\sum_{I=1}^3 \left[\partial_\rho \log W_I\, \partial_z \log W_I+ \partial_\rho \log Z_I\, \partial_z \log Z_I - Z_I^2 \partial_\rho T_I\,\partial_z T_I \right]\nonumber \\
& +  \partial_\rho \log W_0\, \partial_z \log W_0+ \partial_\rho \log Z_0\, \partial_z \log Z_0 +\frac{1}{\rho^2 Z_0^2} \partial_\rho H_0\,\partial_z H_0\,, \label{App:nueq}\\
\frac{4}{\rho} \partial_\rho \nu \= &\sum_{I=1}^3 \left[\left(\partial_\rho \log W_I \right)^2 - \left( \partial_z \log W_I \right)^2 \right] +  \left(\partial_\rho \log W_0 \right)^2 - \left( \partial_z \log W_0 \right)^2\nonumber \\
& +\sum_{I=1}^3  \left[\left(\partial_\rho \log Z_I \right)^2 - \left( \partial_z \log Z_I \right)^2- Z_I^2 \left( \left(\partial_\rho T_I \right)^2-\left(\partial_z T_I \right)^2\right)\right]\nonumber \\
& + \left(\partial_\rho \log Z_0 \right)^2 - \left( \partial_z \log Z_0 \right)^2 +\frac{1}{\rho^2 Z_0^2} \left( \left(\partial_\rho H_0\right)^2- \left( \partial_z H_0 \right)^2 \right). \nonumber
\end{align}

\section{The non-BPS floating-brane ansatz in other frames}
\label{App:AllFrames}

In this section, we derive several frames dual to our M-theory ansatz. We first detail the four- and five-dimensional reduction, before discussing the ansatz in different type IIB and IIA frameworks.

\subsection{Profile in five and four dimensions}
\label{App:5d&4dframe}

We reduce the ansatz \eqref{eq:nonBPSfloating} to five and four dimensions after compactification along the T$^6$ and the S$^1$. We apply a series of KK reductions from M-theory using the generic rules of \cite{Lu:1995yn,Cremmer:1997ct}. We briefly summarize them here by truncating the degrees of freedom that are not present in our ansatz. 

We aim to describe the reduction of the eleven-dimensional action \eqref{eq:Action11D} to $D=4,5$ dimensions for solutions of the following type
\begin{equation}
d s_{11}^{2} \=e^{\frac{1}{3} \vec{g} \cdot \vec{\phi}}  \, d s_{D}^{2} +\sum_{i=1}^{11-D} e^{2 \vec{\gamma}_{i} \cdot \vec{\phi}} \,\left(dy_i+ \cA^{(i)}\right)^{2}\,,\quad F_4  \= d\left[\,\sum_{I=1}^{\lfloor \frac{11-D}{2}\rfloor} A^{(I)} \wedge dy_{2I-1} \wedge dy_{2I} \right],
\label{eq:KKredGenApp}
\end{equation}
where $\vec{\phi}$ is a vector of $11-D$ dilatons, $\cA^{(i)}$ and $A^{(I)}$ are one-form gauge fields that have components only along the $D$-dimensional directions and $\vec{g}$ and $\vec{\gamma}_i$ are constant vectors that we will make precise. We assume that all $y_i$ directions are U(1) isometries. 

After the series of reduction along the $y_i$, we obtain an Einstein-Maxwell-dilaton theory with the following action in $D=4$ or $5$ dimensions
\begin{align}
(16 \pi G_D) S_D \= &\int d^D x  \sqrt{-\det g} \,R  \- \int \cL_D +\cL_{C\text{-}S},\label{eq:KKredActionGenApp}\\
\cL \= & \frac{1}{2} \star d \vec{\phi} \wedge d \vec{\phi} + \frac{1}{2} \sum_{I=1}^{\lfloor \frac{11-D}{2}\rfloor} e^{\vec{a}_I.\vec{\phi}} \, \star d F^{(I)} \wedge d F^{(I)} + \frac{1}{2}\sum_{i=1}^{11-D}  e^{\vec{b}_i.\vec{\phi}} \, \star d \cF^{(i)} \wedge d \cF^{(i)} ,\nonumber\\
\cL_{C\text{-}S} \= &\begin{cases} \frac{1}{6} \sum_{IJK} |\epsilon_{IJK}|\, F^{(I)}\wedge F^{(J)} \wedge A^{(K)} \,, \quad \text{if }D=5,\\
0 \,,\quad \text{if }D=4, \end{cases} \nonumber
\end{align}
where $F^{(I)}=dA^{(I)}$ and $\cF^{(i)}=d\cA^{(i)}$, $\vec{a}_I$ and $\vec{b}_i$ are also constant vectors, and the Newton constant is given according to the eleven-dimensional one by $$G_D = \frac{G_{11}}{(2\pi)^{11-D} \prod_{i=1}^{11-D} R_{y_i}}.$$ The constant vectors are all obtained from
\begin{equation}
\begin{split}
\vec{g} & \equi3\left(s_{1}, s_{2}, \ldots, s_{11-D}\right) ,\\
\vec{f}_{i} &\equi(\underbrace{0,0, \ldots, 0}_{i-1},(10-i) s_{i}, s_{i+1}, s_{i+2}, \ldots, s_{11-D}),
\end{split}
\end{equation} 
where $s_{i}\equiv \sqrt{2 /((10-i)(9-i))}$, such as
\begin{equation}
\vec{a}_I \equi \vec{f}_{2I-1} + \vec{f}_{2I}- \vec{g}, \qquad \vec{b}_i \equi- \vec{f}_{i}, \quad \vec{\gamma}_i \equi \frac{1}{6}\vec{g} - \frac{1}{2} \vec{f}_i.
\end{equation}
In $D$ dimensions, our solutions are defined by a metric, given by $ds_D^2$, $11-D$ scalars and KK gauge fields and $\lfloor \frac{11-D}{2}\rfloor$ one-form gauge fields that arise from $F_4$.

\subsubsection{Reduction to five dimensions}
\label{App:5dframe}

We first perform a series of KK reductions to $D=5$ dimensions by compactifying along $\{y_i\}_{i=1,..,6}$. For the ansatz \eqref{eq:nonBPSfloating}, there are no KK gauge fields along those fibers, and therefore we take $\cA^{(i)}=\cF^{(i)} = 0$. Moreover, it is convenient to change the basis of six scalars $\vec{\phi}=(\phi_i)_{i=1,..,6}$ to a more suitable one that allows a simpler final Lagrangian and a better comparison with the STU model in five dimensions:
\begin{equation}
\begin{split}
\phi_1 &= -\frac{3\left( \log X^1- \Phi_1 \right) + \Phi_0 }{4} , \quad \phi_2 = -\frac{9 \log X^1  +3\Phi_0+7 \Phi_1}{4\sqrt{7}}, \quad \phi_3 = \frac{\log \frac{{X^3}^2}{{X^2}^5} -3 \Phi_0 + 7\Phi_2}{2\sqrt{21}}, \\
\phi_4 &=  \frac{\log \frac{{X^3}^2}{{X^2}^5} -3 \Phi_0 - 5\Phi_2}{2\sqrt{15}}, \quad \phi_6 = -\frac{3 \log X^3 + 3 \Phi_0 - 5 \Phi_3}{2\sqrt{10}},\quad \phi_5 = -\frac{\sqrt{3}( \log X^3 +  \Phi_0+ \Phi_3)}{2\sqrt{2}},
\end{split}
\end{equation}
where we have defined seven scalars with the constraint $X^1 X^2 X^3=1$. With such a definition, the metric and gauge field \eqref{eq:KKredGenApp} for $D=5$ are given by
\begin{equation}
d s_{11}^{2} =e^{-\frac{2}{3} \Phi_0}  \, d s_{5}^{2} \+ e^{\frac{1}{3}\Phi_0}\, \sum_{I=1}^{3} X^I \left( e^{-\Phi_I} dy_{2I-1}^2+ e^{\Phi_I} dy_{2I}^2 \right)\,,\quad F_4  = d\left[\,\sum_{I=1}^{3} A^{(I)} \wedge dy_{2I-1} \wedge dy_{2I} \right],
\label{eq:KKredGenApp2}
\end{equation}
We apply \eqref{eq:KKredActionGenApp}, and the reduced five-dimensional action is then
\begin{align}
(16 \pi G_5) S_5 \= &\int \left( \cL_{STU}^{5D} \- \frac{1}{2}\,\sum_{\Lambda=0}^3 \star d\Phi_\Lambda \wedge \Phi_\Lambda \right),
\label{eq:KKredAction5dApp}
\end{align}
where $\cL_{STU}^{5D}$ is the STU Lagrangian of five-dimensional $\cN=2$ supergravity:
\begin{equation}
\cL_{STU}^{5D} = R\,\star \mathbb{I} - Q_{IJ} \left( \star F^{(I)} \wedge F^{(J)} + \star d X^I \wedge d X^J \right) + \frac{|\epsilon_{IJK}|}{2} \,A^{(I)} \wedge F^{(J)} \wedge F^{(K)}\,,
\end{equation}
where repeated indices are summed over, and we have defined the scalar kinetic term $Q_{IJ}=\frac{1}{2}\, \text{diag} \left((X^1)^{-2} ,(X^2)^{-2} ,(X^3)^{-2} \right)$. Note that the Cherns-Simons term can be dropped because it is trivially zero for our ansatz. Nevertheless, we have kept it for a clearer identification with the STU Lagrangian.

In this framework, the ansatz \eqref{eq:nonBPSfloating} defines solutions of the five-dimensional action such as
\begin{align}
ds_5^2 &\= - \frac{dt^2}{\left(Z_1 Z_2 Z_3 \right)^{\frac{2}{3}}} \+ \left(Z_1 Z_2 Z_3 \right)^{\frac{1}{3}} \left[\frac{1}{Z_0} \left(dy_7 +H_0 \,d\phi\right)^2 \+ Z_0 \left( e^{2\nu} \left(d\rho^2 + dz^2 \right) +\rho^2 d\phi^2\right)\right], \nonumber \\
X^I &\= \frac{|\epsilon_{IJK}|}{2} \left(\frac{Z_J Z_K}{Z_I^2} \right)^\frac{1}{3},\qquad \Phi_\Lambda \= \log W_\Lambda \,,\qquad F^{(I)} \= dA^{(I)} \= dT_I \wedge dt\,.
\label{eq:5dframeworkApp}
\end{align}

\subsubsection{Reduction to four dimensions}
\label{App:4dframe}

We further reduce to four dimensions by compactifying along $y_7$ for solutions where $y_7$ corresponds to a compact direction. One could reapply the generic reduction rules summarized before, but, thanks to the identification \eqref{eq:KKredAction5d}, it is simpler to directly use known results about the reduction of the STU model from five to four dimensions \cite{Goldstein:2008fq,DallAgata:2010srl,Bah:2021jno}. 

The reduced four-dimensional action is given by 
\begin{align}
(16 \pi G_4) S_4 \= &\int \left( \cL_{STU}^{4D}\- \frac{1}{2}\,\sum_{\Lambda=0}^3 \star d\Phi_\Lambda \wedge \Phi_\Lambda \right),
\label{eq:KKredAction4dApp}
\end{align}
where $\cL_{STU}^{4D}$ is the STU Lagrangian of four-dimensional $\cN=2$ supergravity and $G_4 \= \frac{G_5}{2\pi R_{y_7}}$. This Lagrangian is better written in terms of three independent complex scalars $z^I$ and the field strengths of four one-form gauge fields $\bar{F}^\Lambda= \bar{A}^\Lambda$, and is given by
\begin{equation}
\cL_{STU}^{4D} \= R \,\star \mathbb{I} - 2 g_{IJ} \,\star dz^I \wedge d\bar{z}^J \- \frac{1}{2} \mathcal{I}_{\Lambda \Sigma} \,\star \bar{F}^\Lambda \wedge \bar{F}^\Sigma + \frac{1}{2} \cR_{\Lambda \Sigma}\, \bar{F}^\Lambda \wedge \bar{F}^\Sigma.
\end{equation}
Relabelling the scalar fields as $z^{I}=\{S=\sigma-i s, T=$
$\tau-i t, U=v-i u\},$ the metric of the scalar $\sigma$-model $g_{I J}$ follows from the Kähler potential
\be
\mathcal{K} =-\log (8 s t u),\qquad g_{IJ} = \frac{\partial^2 \mathcal{K}}{\partial z^I \partial \bar{z}^J}, \label{eq:gscalcouplingApp}
\ee
the gauge kinetic couplings are
\be \label{eq:IscalcouplingApp}
\mathcal{I}= s t u\left(\begin{array}{cccc}
1+\frac{\sigma^{2}}{s^{2}}+\frac{\tau^{2}}{t^{2}}+\frac{v^{2}}{u^{2}} & -\frac{\sigma}{s^{2}} & -\frac{\tau}{l^{2}} & -\frac{v}{u^{2}} \\
-\frac{\sigma}{s^{2}} & \frac{1}{s^{2}} & 0 & 0 \\
-\frac{\tau}{l^{2}} & 0 & \frac{1}{t^{2}} & 0 \\
-\frac{v}{u^{2}} & 0 & 0 & \frac{1}{u^{2}}
\end{array}\right),
\ee
and the axionic couplings are
\be \label{eq:RscalcouplingApp}
\mathcal{R}=\left(\begin{array}{cccc}
-2 \sigma \tau v & \tau v & \sigma v & \sigma \tau \\
\tau v & 0 & -v & -\tau \\
\sigma v & -v & 0 & -\sigma \\
\sigma \tau & -\tau & -\sigma & 0
\end{array}\right).
\ee
The four-dimensional metric, scalars and gauge fields arise from the five dimensional ones such as
\begin{equation}
ds_5^2 = e^{2\Phi} \,ds_4^2 + e^{-4\Phi} (dy_7-\bar{A}^0)^2,\quad
A^{(I)} = \bar{A}^I \+ C^I (dy_7 - \bar{A}^0),\quad z^I = C^I + i \,e^{-2\Phi} X^I\,.
\end{equation}
In this context, our ansatz is composed of a metric, three purely-imaginary complex scalars, four real scalars and four one-form gauge fields given by
\begin{align}
ds_4^2 &\= -\frac{dt^2}{\sqrt{Z_0 Z_1 Z_2 Z_3}} \+ \sqrt{Z_0 Z_1 Z_2 Z_3} \left( e^{2\nu} \left(d\rho^2 + dz^2 \right) +\rho^2 d\phi^2\right), \nonumber \\
z^I &\= i\,\frac{|\epsilon_{IJK}|}{2} \sqrt{\frac{Z_J Z_K}{Z_0\,Z_I} },\quad \Phi_\Lambda \= \log W_\Lambda  \,,\quad \bar{F}^{0}  \= -dH_0 \wedge d\phi \,,\quad \bar{F}^{I}  \= dT_I \wedge dt\,.
\label{eq:4dSolApp}
\end{align}

\subsection{Other string theory frames}
\label{App:dualFrame}

In this section, we present the non-BPS floating brane ansatz \eqref{eq:nonBPSfloating} in different duality frames in type IIA and type IIB theories. We will make a great use of T-duality rules and compactification from M-theory, which we summarize here following \cite{Bena:2008dw}.

\noindent The set of bosonic fields in M-theory is given by the metric $G_{\mu \nu} $ and the three-form gauge field $A_{3\,\,\mu \nu \rho}$. After the compactification along $y_i$ we are left with type IIA supergravity with the fields
$$
g_{\mu \nu}, \qquad C_{\mu \nu \rho}^{(3)}, \qquad B_{\mu \nu}, \qquad C_{\mu}^{(1)}, \qquad \Phi\,.
$$
which are related to the eleven-dimensional fields as follows (note that we are working in string frame):
\begin{align}
g_{\mu \nu}=\sqrt{G_{y_i y_i}}\left(G_{\mu \nu}+\frac{G_{\mu y_i} G_{\nu y_i}}{G_{y_iy_i}}\right), \qquad C_{\mu}^{(1)}=\frac{G_{\mu y_i}}{G_{y_iy_i}},\nonumber \\
C_{\mu \nu}^{(3)}=A_{\mu \nu \rho}, \qquad B_{\mu \nu}=A_{\mu \nu y_i}, \qquad \Phi=\frac{3}{4} \log \left(G_{y_iy_i}\right).
\end{align}
Solutions in Einstein frame are obtained by transforming ${g_\text{E}}_{\mu \nu} \= e^{\frac{\Phi}{2}}\,g_{\mu\nu}$.

\noindent T-duality transformations act on the supergravity fields mixing them according to Buscher's rules \cite{Buscher:1987sk}. We assume that $y_j$ is the direction along which one performs the T-duality transformation, and we decompose the string metric, B-fields and RR gauge fields such as
\begin{equation}
\begin{array}{l}
d s_{10}^{2}\=G_{y_j y_j}\left(d y_j+A_{\mu} d x^{\mu}\right)^{2} \+\widehat{g}_{\mu \nu} d x^{\mu} d x^{\nu}, \\
B^{(2)}\=B_{\mu y_j} \,d x^{\mu} \wedge\left(d y_j+A_{\mu} d x^{\mu}\right) \+\widehat{B}^{(2)} \\
C^{(p)} \=C_{y_j}^{(p-1)} \wedge\left(dy_j+A_{\mu} d x^{\mu}\right)\+\widehat{C}^{(p)} \,.
\end{array}
\end{equation}
where the forms, $\widehat{B}^{(2)}, C_{y}^{(p-1)}$ and $\widehat{C}^{(p)}$, do not have legs along $y_j$ and are functions only of the $x^{\mu}$ coordinates, the dualized fields are
\begin{align}
d \widetilde{s}_{10}^{2}& \=G_{y_j y_j}^{-1}\left(d y_j-B_{\mu y_j} d x^{\mu}\right)^{2} \+ \widehat{g}_{\mu \nu} d x^{\mu} d x^{\nu}, \qquad e^{2 \widetilde{\Phi}} \=\frac{e^{2 \Phi}}{G_{y_j y_j}} , \nonumber\\
\widetilde{B}^{(2)} &\=-A_{\mu} \, d x^{\mu} \wedge \left(d y_j-B_{\mu y_j} d x^{\mu}\right) \+\widehat{B}^{(2)}, \\
\widetilde{C}^{(p)}& \=\widehat{C}^{(p-1)} \wedge\left(d y_j-B_{\mu y_j} d x^{\mu}\right)\+C_{y_j}^{(p)} .\nonumber
\end{align}

\subsubsection{The D2-D2-D2-D6 frame}
\label{App:D2D2D2D6}

By reducing the M-theory ansatz \eqref{eq:nonBPSfloating} along $y_7$, the solutions correspond to D2-D2-D2-D6 solutions given by (in string frame)
\begin{align}
ds_{10}^2 \= &- \frac{dt^2}{W_0\sqrt{Z_0 Z_1 Z_2 Z_3}} \+ \frac{\sqrt{Z_0 Z_1 Z_2 Z_3}}{W_0} \left( e^{2\nu} \left(d\rho^2 + dz^2 \right) +\rho^2 d\phi^2\right) \nonumber\\
& +\frac{\sqrt{ Z_1 Z_2 Z_3}}{\sqrt{Z_0}} \left[\frac{1}{Z_1} \left(\frac{dy_1^2}{W_1} + W_1 \,dy_2^2 \right) + \frac{1}{Z_2} \left(\frac{dy_3^2}{W_2} + W_2 \,dy_4^2 \right) + \frac{1}{Z_3} \left(\frac{dy_5^2}{W_3} + W_3 \,dy_6^2 \right)\right],\nonumber\\
\Phi \=& \frac{1}{4} \log \frac{Z_1 Z_2 Z_3}{W_0^2\,Z_0^3}\,,\qquad B_2 \= 0\,, \label{eq:D2D2D2D6}\\
C^{(1)} \= &H_0\, d\phi\,,\qquad C^{(3)} \= T_1 \,dt \wedge dy_1 \wedge dy_2 \+T_2 \,dt \wedge dy_3 \wedge dy_4 \+T_3\,dt \wedge dy_5 \wedge dy_6\,.
\nonumber
\end{align}
This framework requires that the $y_7$ direction in M-theory is compact and is ill-defined for the family of asymptotically-$\IR^{1,4}$ solutions constructed in the sections \ref{sec:Sol5d} and \ref{sec:5dNonBPSGen}.

\subsubsection{The D2-D2-F1-KKm frame}
\label{App:D2D2F1KKm}

By reducing the M-theory ansatz \eqref{eq:nonBPSfloating} along $y_5$, the solutions correspond to  D2-D2-F1-KKm solutions given by (in string frame)
\begin{align}
ds_{10}^2 \= &- \frac{dt^2}{Z_3 \sqrt{W_3 W_0 Z_1 Z_2}} \+ \sqrt{\frac{ Z_1 Z_2}{W_3 W_0}}  \left[\frac{1}{Z_0} \left(dy_7 +H_0 \,d\phi\right)^2 \+ Z_0\left( e^{2\nu} \left(d\rho^2 + dz^2 \right) +\rho^2 d\phi^2\right)\right] \nonumber\\
& +\sqrt{\frac{W_0}{ W_3}} \left[\sqrt{\frac{Z_2}{Z_1}} \left(\frac{dy_1^2}{W_1} + W_1 \,dy_2^2 \right) + \sqrt{\frac{Z_1}{Z_2}}\left(\frac{dy_3^2}{W_2} + W_2 \,dy_4^2 \right) \right] +  \frac{\sqrt{W_3 W_0 Z_1 Z_2}}{Z_3}\,dy_6^2\,, \nonumber \\
\Phi \=& \frac{1}{4} \log \frac{W_0 Z_1 Z_2}{W_3^3 Z_3^2}\,,\qquad B_2 \= -T_3 \,dt\wedge dy_6\,, \label{eq:D2D2F1KKm}\\
C^{(1)} \= &0\,,\qquad C^{(3)} \= T_1 \,dt \wedge dy_1 \wedge dy_2 \+T_2 \,dt \wedge dy_3 \wedge dy_4 \,. \nonumber
\end{align}

\subsubsection{The D1-D3-F1-KKm frame}
\label{App:D1D3F1KKm}

By applying a T-duality along $y_1$, the solutions correspond to D1-D3-F1-KKm solutions given by
\begin{align}
ds_{10}^2 \= &- \frac{dt^2}{Z_3 \sqrt{W_3W_0 Z_1 Z_2}} \+ \sqrt{\frac{ Z_1 Z_2}{W_3 W_0}}  \left[\frac{1}{Z_0} \left(dy_7 +H_0 \,d\phi\right)^2 \+ Z_0\left( e^{2\nu} \left(d\rho^2 + dz^2 \right) +\rho^2 d\phi^2\right)\right] \nonumber \\
& \+\sqrt{\frac{W_0}{W_3}} \left[\sqrt{\frac{Z_2}{Z_1}} \,W_1\,dy_2^2+ \sqrt{\frac{Z_1}{Z_2}}\left(\frac{W_1 W_3 \,dy_1^2}{W_0}\+\frac{dy_3^2}{W_2} \+ W_2 \,dy_4^2 \right) \right] \+ \frac{\sqrt{W_3 W_0 Z_1 Z_2}}{Z_3}\,dy_6^2\,, \nonumber \\
\Phi \=& \frac{1}{2} \log \frac{W_1  Z_1 }{W_3  Z_3}\,,\qquad B_2 \= - T_3 \,dt\wedge dy_6\,,\label{eq:D1D3F1KKm}\\
C^{(0)} \= &0\,,\qquad C^{(2)} \= -T_1 \,dt \wedge dy_2\,,\qquad C^{(4)}\=T_2 \,dt \wedge dy_1\wedge dy_3 \wedge dy_4 \+ \ldots \,.\nonumber
\end{align}
The ``$\ldots$'' in $C^{(4)}$ corresponds to the term arising from $C^{(5)}$ in type IIA. It is appropriate to use the self-duality of $\widetilde{F}^{(5)}$ to derive this term with
\begin{equation}
\widetilde{F}^{(5)} \= dC^{(4)} \+ dB_2 \wedge C^{(2)} \= dC^{(4)}\,.
\end{equation}
We find 
\begin{equation}
C^{(4)} \= T_2 \,dt \wedge dy_1\wedge dy_3 \wedge dy_4 \- H_2 \,d\phi \wedge dy_2 \wedge dy_6 \wedge dy_7\,,
\end{equation}
where $H_2$ has the same form as $H_0$ in \eqref{eq:WarpfactorsRods} but with $P_i^{(0)} \to P_i^{(2)}$, that is 
\begin{equation}
H_2 \equi \sum_{i=1}^n P^{(2)}_i \left(r_-^{(i)}-r_+^{(i)} \right)\,,
\end{equation} 
Note that the electromagnetic dual of $T_2 \,dt \wedge dy_1\wedge dy_3 \wedge dy_4$ only couples with $Z_2$ since we have $\star_{10} \left( dx_a \wedge dt \wedge dy_1 \wedge dy_3 \wedge dy_4 \right) \= - \rho Z_2^2\, \epsilon_a^{\,\,b} dx_b \wedge d\phi \wedge dy_2 \wedge dy_6 \wedge dy_7$ where $x_a=(\rho,z)$.

\subsubsection{The D0-D4-F1-KKm frame}
\label{App:D0D4F1KKm}

By applying a T-duality along $y_2$ to the ansatz above, the solutions correspond to D0-D4-F1-KKm solutions given by
\begin{equation}
\begin{split}
ds_{10}^2 \= &- \frac{dt^2}{Z_3 \sqrt{W_3W_0 Z_1 Z_2}} \+ \sqrt{\frac{ Z_1 Z_2}{W_3 W_0}}  \left[\frac{1}{Z_0} \left(dy_7 +H_0 \,d\phi\right)^2 \+ Z_0\left( e^{2\nu} \left(d\rho^2 + dz^2 \right) +\rho^2 d\phi^2\right)\right]\\
& +\sqrt{\frac{Z_1}{Z_2}}\left[\sqrt{\frac{W_3}{W_0}} \,\left(W_1\,dy_1^2+ \frac{dy_2^2}{W_1}\right) +\sqrt{\frac{W_0}{W_3}} \left(\frac{dy_3^2}{W_2} + W_2 \,dy_4^2 \right) \right] \+  \frac{\sqrt{W_3 W_0 Z_1 Z_2}}{Z_3}\,dy_6^2\,, \\
\Phi \=& \frac{1}{4} \log \frac{Z_1^3}{W_0 W_3 Z_2 Z_3^2},\quad B_2 = -T_3 \,dt\wedge dy_6,\quad C^{(1)} =  -T_1 \,dt ,\quad C^{(3)}=-H_2 \,d\phi \wedge dy_6 \wedge dy_7 .
\end{split}
\label{eq:D0D4F1KKm}
\end{equation}

\subsubsection{The D1-D5-P-KKm frame}
\label{App:D1D5PKKm}

After T-duality along $y_6$, the solutions correspond to D1-D5-P-KKm solutions given by
\begin{align}
ds_{10}^2 \= &- \frac{dt^2}{Z_3 \sqrt{W_3W_0 Z_1 Z_2}} \+ \sqrt{\frac{ Z_1 Z_2}{W_3 W_0}}  \left[\frac{1}{Z_0} \left(dy_7 +H_0 \,d\phi\right)^2 \+ Z_0\left( e^{2\nu} \left(d\rho^2 + dz^2 \right) +\rho^2 d\phi^2\right)\right] \nonumber\\
& +\sqrt{\frac{Z_1}{Z_2}}\left[\sqrt{\frac{W_3}{W_0}} \,\left(W_1\,dy_1^2+ \frac{dy_2^2}{W_1}\right) +\sqrt{\frac{W_0}{W_3}} \left(\frac{dy_3^2}{W_2} + W_2 \,dy_4^2 \right) \right] \label{eq:D1D5PKKmApp} \\
& + \frac{Z_3}{\sqrt{W_3 W_0 Z_1 Z_2}}\,\left(dy_6+T_3\, dt \right)^2\,, \qquad \Phi \= \frac{1}{2} \log \frac{Z_1}{W_0 W_3 Z_2}\,,\qquad B_2 \= 0\,, \nonumber \\
C^{(0)} \=&  0 \,,\qquad C^{(2)}\=  H_2 \,d\phi \wedge dy_7 \- T_1 \,dt \wedge dy_6\,,\qquad C^{(4)} \= 0\,.\nonumber 
\end{align}

\subsubsection{The P-M5-M2-KKm frame}
\label{App:PM5M2KKm}

If we uplift the D0-D4-F1-KKm solutions to M-theory, we obtain P-M5-M2-KKm solutions given by
\begin{equation}
\begin{split}
ds_{11}^2 \= &- \frac{1}{\left(W_3 W_0 Z_2 Z_3^2 \right)^\frac{1}{3}}\frac{dt^2}{Z_1} + \left(\frac{Z_2^2 Z_3}{W_0 W_3}\right)^{\frac{1}{3}} \left[\frac{1}{Z_0} \left(dy_7 +H_0 \,d\phi\right)^2 + Z_0\left( e^{2\nu} \left(d\rho^2 + dz^2 \right) +\rho^2 d\phi^2\right)\right]\\
& \+\left(\frac{Z_3}{W_0 W_3 Z_2} \right)^{\frac{1}{3}} \left[W_3 \left(W_1 \,dy_1^2\+ \frac{dy_2^2}{W_1}  \right) +W_0 \left(\frac{dy_3^2}{W_2} \+ W_2 \,dy_4^2 \right) \right]\\
& \+ \left( \frac{W_0 W_3 Z_2}{Z_3}\right)^\frac{2}{3}\,dy_6^2 \+  \frac{Z_1}{\left(W_0 W_3 Z_2 Z_3^2 \right)^\frac{1}{3}} \left( dy_5 -T_1 dt\right)^2 \,,\\
F_4 \= &d\left[ -H_2 \,d\phi \wedge dy_6 \wedge dy_7 \+T_3\,dt \wedge dy_5 \wedge dy_6 \right]\,.
\end{split}
\label{eq:PM5M2KKm}
\end{equation}

\section{Regularity analysis}
\label{App:Reg}

In this section, we detail the regularity of the non-BPS M-theory solutions constructed in section \ref{sec:GenWeylSol}. We first discuss solutions that are asymptotic to $\IR^{1,3}\times$S$^1\times$T$^6$ before the asymptotically-$\IR^{1,4}\times$T$^6$ solutions.

\subsection{Four-dimensional solutions}
\label{App:Reg4d}

Generic non-BPS Weyl solutions in M-theory, that are asymptotic to $\IR^{1,3}$ plus compactification circles, are sourced by $n$ rods of length $M_i$, and are obtained from the ansatz \eqref{eq:nonBPSfloating}, with the warp factors and gauge potentials given in \eqref{eq:WarpfactorsRods}.

\subsubsection{Regularity out of the $z$-axis}

At large distance, $(\rho,z)=(r\sin \theta ,r\cos \theta)$ and $r$ large, the warp factors and gauge potentials \eqref{eq:WarpfactorsRods} behave as
$$
Z_\Lambda \sim \frac{\sinh  b_\Lambda}{a_\Lambda} \,,\quad W_\Lambda \sim 1 \,,\quad H_0 \sim \sum_{i=1}^n M_i P^{(0)}_i \,\cos \theta\,,\quad T_I\=-a_I \coth b_I,\quad e^{2\nu} \sim1 \,.
$$
Therefore, the four-dimensional reduction of the solution \eqref{eq:4dSol}, is asymptotically flat if and only if 
\begin{equation}
\prod_{\Lambda=0}^3  \frac{\sinh b_\Lambda}{a_\Lambda} \= 1\,.
\end{equation}

At finite distance but out of the $z$-axis, the warp factors and gauge potentials are finite. However, the warp factors $Z_\Lambda$ can change sign due to their $\sinh$ form. This will necessarily induce closed timelike curves. By noting that $R_-^{(i)} \to 0$ at the $i^\text{th}$ rod and $R_-^{(i)} < R_+^{(i)}$, we see that imposing 
\begin{equation}
b_\Lambda\geq 0 \,,\quad a_\Lambda P_i^{(\Lambda)} \geq 0\,, \qquad \forall \Lambda,i\,,
\end{equation}
is enough to guarantee that all $Z_\Lambda$ are positive and that the solutions are regular out of the $z$-axis.

\subsubsection{Regularity on the $z$-axis}

We first discuss the regularity at the rods before the regularity elsewhere on the $z$ axis.

\subsubsection{At the $i^\text{th}$ rod}
\label{App:AttheRod}

The local spherical coordinates around the i$^{\text{th}}$ rod are given by $r_i \rightarrow 0$ for $0\leq \theta_i \leq \pi$ with
\begin{equation}
\rho = \sqrt{r_i(r_i+M_i)}\,\sin\theta_i \,,\qquad z = \left(r_i+\frac{M_i}{2} \right) \cos\theta_i +z_i\,.
\label{eq:localcoorApp}
\end{equation}
The two-dimensional base behaves as
\begin{equation}
d\rho^2 + dz^2 \sim \frac{M_i \sin^2 \theta_i}{4} \left(\frac{dr_i^2}{r_i}+M_i\, d\theta_i^2 \right)\,.
\end{equation}
Moreover,
\begin{equation}
\frac{R_+^{(i)}}{R_-^{(i)}} \sim \frac{M_i}{r_i} \,,\qquad \frac{R_+^{(j)}}{R_-^{(j)}} \sim \left(\frac{z_j^+-\left(z_i + \frac{M_i}{2} \cos \theta_i \right)}{z_j^--\left(z_i + \frac{M_i}{2}  \cos \theta_i \right)}\right)^{\text{sign}(j-i)} \equi A_{ij}(\theta_i)\,,\qquad j\neq i\,.
\label{eq:RpRmexp}
\end{equation}
Thus,
\begin{equation}
\begin{split}
W_\Lambda &\sim\left( \frac{M_i}{r_i} \right)^{ G^{(\Lambda)}_i}\,\prod_{j\neq i} A_{ij}(\theta_i)^{G^{(\Lambda)}_j}\,,\\
Z_\Lambda &\sim \begin{cases}&\dfrac{e^{b_\Lambda}}{2 a_\Lambda}\,\left( \dfrac{M_i}{r_i} \right)^{ a_\Lambda P^{(\Lambda)}_i}\,\prod_{j\neq i} A_{ij}(\theta_i)^{a_\Lambda P^{(\Lambda)}_j}\,,\qquad \text{if }a_\Lambda P^{(\Lambda)}_i \neq 0, \\
&\dfrac{1}{a_\Lambda}\,\sinh \left[ \log \left(\prod_{j\neq i} A_{ij}(\theta_i)^{a_\Lambda P^{(\Lambda)}_j} \right)+ b_\Lambda \right]\,,\qquad \text{if }a_\Lambda P^{(\Lambda)}_i = 0,
\end{cases}
\end{split}
\label{eq:expansionZIApp}
\end{equation}
As for $e^{2\nu}$, we have
\begin{equation}
\begin{split}
e^{\nu_{ii}}&\sim \frac{16 r_i^2}{M_i^2\sin^4 \theta_i}\,,\qquad \quad e^{\nu_{jk}}\ \sim \begin{cases}
1 \qquad &\text{if } i<j\leq k  \text{ or } j\leq k<i  \\
\dfrac{(z_k^- - z_j^+)^2(z_k^+ - z_j^-)^2}{(z_k^+ - z_j^+)^2(z_k^- - z_j^-)^2} \qquad &\text{if } j<i<k 
\end{cases}\,,
\end{split}
\label{eq:EEEEatrod}
\end{equation}
\begin{equation}
\begin{split}
e^{\nu_{ij}} &\sim \begin{cases} 
\dfrac{(z_j^+ - z_i^-)^2}{(z_j^- - z_i^-)^2} \,A_{ij}(\theta_i)^{-2} \qquad &\text{if } j>i,  \\
\dfrac{(z_j^- - z_i^+)^2}{(z_j^+ - z_i^+)^2} \,A_{ij}(\theta_i)^{-2}\qquad &\text{if } j<i\,. 
\end{cases} \hspace{10cm}
\end{split}
\nonumber
\end{equation}
Therefore,
\begin{equation}
\begin{split}
e^{2 \nu}\, \sim \, d_i^2 \,\left(\frac{4 r_i}{M_i \sin^2 \theta_i} \right)^{\alpha_{ii}} \,\prod_{j\neq i}A_{ij}(\theta_i)^{-2\alpha_{ij}} \,  \left(\frac{z_j^+ - z_i^-}{z_j^- -z_i^-} \right)^{2\,\text{sign}(j-i) \,\alpha_{ij}} \,,
\end{split}
\end{equation}
where we have defined the constants
\begin{equation}
d_1 \equi 1\,,\qquad d_i \equi  \prod_{j=1}^{i-1} \prod_{k=i}^n \left(\dfrac{(z_k^- - z_j^+)(z_k^+ - z_j^-)}{(z_k^+ - z_j^+)(z_k^- - z_j^-)}  \right)^{\alpha_{jk}}\quad \text{when } i=2,\ldots n\,,
\label{eq:diDef}
\end{equation}
and considered that the product ``$\prod_{j=1}^{i-1}$'' is equal to $1$ for the first rod, $i=1$.
Finally, the metric components and gauge potentials behave around the $i^\text{th}$ rod as\footnote{We have denoted $\widetilde{g}_{\phi \phi}$ the metric component of $\phi$ without the connection term in the $y_7$ fiber, that is $\widetilde{g}_{\phi \phi}= \rho^2 \frac{Z_0\,\left(Z_1 Z_2 Z_3 \right)^{1/3}}{W_0^{2/3}} $. Moreover, by anticipating what is coming next, we have assumed that $P_i^{(0)}$ is necessarily non-zero and the expansion of $Z_0$ is given by the first line in \eqref{eq:expansionZIApp}. As for the other $Z_I$, we kept the possibility of having $P_i^{(I)}=0$ by keeping the expressions of $Z_I$ arbitrary.}

\begin{align}
H_0  &\sim -M_i P_i^{(0)}\,\cos \theta_i \+ \sum_{j\neq i} \text{sign}(j-i) M_j P_j^{(0)} \,,\quad T_I \sim -a_I \left[ 1 + \cO \left( r_i^{2 a_I P_i^{(I)}} \right)\right] , \nonumber\\
g_{tt} &\sim- \frac{1}{\left( Z_1 Z_2 Z_3\right)^{\frac{2}{3}}}\left( \frac{M_i}{r_i} \right)^{ -\frac{2}{3}G^{(0)}_i}\,\prod_{j\neq i} A_{ij}(\theta_i)^{ -\frac{2}{3}G^{(0)}_j}\,, \nonumber\\
 g_{x_\pm^{(I)}x_\pm^{(I)}} &\sim \left(\frac{|\epsilon_{IJK}|}{2}\, \frac{Z_J Z_K}{Z_I^2}\right)^{\frac{1}{3}}\left( \frac{M_i}{r_i} \right)^{ \frac{1}{3}G^{(0)}_i \mp G_i^{(I)} }\,  \prod_{j\neq i} A_{ij}(\theta_i)^{ \frac{1}{3}G^{(0)}_j \mp G_j^{(I)} }\,,\nonumber
 \end{align}
 \begin{align}
 g_{y_7 y_7} &\sim \frac{2 a_0}{e^{b_0}} \left( Z_1 Z_2 Z_3\right)^{\frac{1}{3}}\left( \frac{M_i}{r_i} \right)^{ -\frac{2}{3}G^{(0)}_i-a_0 P^{(0)}_i}\,\prod_{j\neq i} A_{ij}(\theta_i)^{-\frac{2}{3}G^{(0)}_j-a_0 P^{(0)}_j}\,, \label{eq:ytphibehaviorBubblerod}\\
\widetilde{g}_{\phi\phi} &\sim  \frac{M_i^2\, e^{b_0}\,\sin^2\theta_i}{2 a_0}\, \left( Z_1 Z_2 Z_3\right)^{\frac{1}{3}}\left( \frac{M_i}{r_i} \right)^{ -1-\frac{2}{3}G^{(0)}_i+a_0 P^{(0)}_i}\,\prod_{j\neq i} A_{ij}(\theta_i)^{-\frac{2}{3}G^{(0)}_j+a_0 P^{(0)}_j}\,,\nonumber\\
 g_{r_i r_i}& \sim  \frac{d_i^2 e^{b_0}}{2 a_0} \left( Z_1 Z_2 Z_3\right)^{\frac{1}{3}}\,\left(\frac{4 r_i}{M_i \sin^2 \theta_i} \right)^{\alpha_{ii}-1} \left( \frac{M_i}{r_i} \right)^{ -\frac{2}{3}G^{(0)}_i+a_0 P^{(0)}_i}\nonumber\\
& \hspace{0.5cm} \times \prod_{j\neq i}A_{ij}(\theta_i)^{-2\alpha_{ij}-\frac{2}{3}G^{(0)}_j+a_0 P^{(0)}_j} \,  \left(\frac{z_j^+ - z_i^-}{z_j^- -z_i^-} \right)^{2\,\text{sign}(j-i) \,\alpha_{ij}} \,, \nonumber \\
g_{\theta_i \theta_i}& \sim r_i  M_i \,g_{r_i r_i}\,,  \nonumber
\end{align}
where the expansion of $Z_I$ can take two forms if $P_i^{(i)}$ is zero or not given by \eqref{eq:expansionZIApp}, $\epsilon_{IJK}$ is the rank-3 Levi-Civita tensor and we have used the convenient notation 
\begin{equation}
(y_1,y_2,y_3,y_4,y_5,y_6)=(x^{(1)}_+,x^{(1)}_-,x^{(2)}_+,x^{(2)}_-,x^{(3)}_+,x^{(3)}_-).
\end{equation}

Therefore, the Killing vector along the extra dimensions, $\partial_{y_a}$, or the timelike Killing vector, $\partial_t$, can vanish at the rod while all other have a finite norm by fixing the 8 weights $(G_i^{(\Lambda)},P_i^{(\Lambda)})$. We then have 8 possible choices of regular rods. Because all T$^6$ directions are similar by permutation, we will treat three categories: a black rod where $\partial_t$ vanishes defining the horizon of a M2-M2-M2-KKm non-extremal black hole, a bubble rod where $\partial_{y_1}$ or $\partial_{y_2}$ vanishes defining the locus of a smooth M2-KKm bubble and another bubble rod where $\partial_{y_7}$ vanishes defining the locus of a smooth KKm bubble.

We first assume that the rod is disconnected from the others.
\begin{itemize}
\item[•] \underline{A four-charge non-extremal static black hole.}

We consider
\begin{equation}
G^{(\Lambda)}_i =0,\qquad  P^{(\Lambda)}_i = \frac{1}{2 a_\Lambda}.
\end{equation}
The metric components along the compact dimensions and the $\phi$-circle are finite while the time component vanishes as $g_{tt} = \cO(r_i)$. More concretely, the $i^\text{th}$ rod corresponds to a horizon where the timelike Killing vector $\partial_t$ shrinks.  Indeed, we have $$ \alpha_{ii} \= 1 \,,\qquad \alpha_{ij} \=  \frac{1}{2} \sum_{\Lambda=0}^3 a_\Lambda P^{(\Lambda)}_j \,,$$
which implies that the $\theta_i$-dependent factors in $g_{tt}$ and $g_{r_i r_i}$ are remarkably the same. The local metric around the i$^\text{th}$ rod is then
\begin{align}
ds^2 \bigl|_{r_i = 0} \= &\bar{g}_{r_i r_i}(\theta_i) \left(d\rho_i^2 - \kappa_i^2\,\rho_i^2\,dt^2 \right)+ g_{\theta_i\theta_i}(\theta_i)  \,\left( d\theta_i^2 + \bar{g}_{\phi\phi}(\theta_i) \,\sin^2 \theta_i\, \,d\phi^2\right) \label{eq:metricatrodBS}\\
&+ \sum_{I=1}^3 \left(g_{x_+^{(I)}x_+^{(I)}}(\theta_i) \,d{x_+^{(I)}}^2+g_{x_-^{(I)}x_-^{(I)}}(\theta_i) \,d{x_-^{(I)}}^2\right) + g_{y_7 y_7}(\theta_i) \left( dy_7+H_0 d\phi\right)^2, \nonumber
\end{align}
where $\rho_i^2 \equiv 4 r_i$, the $g_{xx}(\theta_i)$ are all finite and non-zero for $0\leq \theta_i \leq \pi$  \eqref{eq:ytphibehaviorBubblerod} and the surface gravity, $\kappa_i$, is given by
\begin{equation}
\kappa_i \equi \frac{2}{d_i\,M_i}\, \sqrt{\frac{a_0 a_1 a_2 a_3}{e^{b_0+b_1+b_2+b_3}}}\prod_{j\neq i} \left(\frac{z_j^+ - z_i^-}{z_j^- -z_i^-} \right)^{\text{sign}(i-j)\, \alpha_{ij}}\,.
\label{eq:surfaceGravVac}
\end{equation}

The metric corresponds to the horizon of a black hole with a S$^2\times$T$^7$ topology. One can relate the surface gravity to the temperature of the black hole by requiring smoothness of the Euclideanized solution. We find
\begin{equation}
\cT_i \= \frac{\kappa_i}{2\pi}\,.
\end{equation}
Note that if we study axisymmetric solutions with multiple black holes in thermal  equilibrium the temperature associated to each black rod must be fixed to be equal.

Moreover, as $g_{y_7y_7}(\theta_i)g_{\theta_i \theta_i}(\theta_i)^2\bar{g}_{\phi\phi}(\theta_i)\prod_{I=1}^3 g_{x_+^{(I)}x_+^{(I)}}(\theta_i)g_{x_-^{(I)}x_-^{(I)}}(\theta_i)$ is remarkably independent of $\theta_i$, the area of the horizon is simple to derive. We find
\begin{equation}
A_i \= \int_{S^2\times T^7} \sqrt{\det g\bigl|_{S^2\times T^7}} \= \frac{(2\pi)^8\,M_i}{\kappa_i}\,\prod_{a=1}^7 R_{y_a}\,.
\end{equation}
Because all $P_i^{(\Lambda)}$ are finite, the black hole carries generically three M2 charges and a KKm charge given by \eqref{eq:individualMcharges}. Each charge can be taken to be zero by considering the neutral limit \eqref{eq:neutrallimit}, which corresponds to $a_\Lambda =  \sinh b_\Lambda\to \infty $. Moreover, since we have $T_I \sim -a_I \left(1+\cO(\rho_i^2)\right)$ \eqref{eq:ytphibehaviorBubblerod}, the field strength is vanishing at the rod which guarantees its regularity.

\item[•] \underline{A M2-KKm bubble rod.}

We consider that one of the directions of the first T$^2$ shrinks at the rod while all other directions have finite size, that is 
\begin{equation}
\pm G_i^{(1)} = -G_i^{(0)} = a_0 P_i^{(0)} = a_1 P_i^{(1)} =\frac{1}{2} , \quad G_i^{(2)}=G_i^{(3)}=P_i^{(2)}=P_i^{(3)}= 0,
\end{equation}
where the ``$\pm$'' imposes the degeneracy of the $x_\pm^{(1)}$ fiber. One can obtain any other directions of the T$^6$ by permuting the $I=1,2,3$ indexes.

For these weights, the spacelike Killing vector $\partial_{x_\pm^{(1)}}$ shrinks on the $i^\text{th}$ rod corresponding to a coordinate singularity of an origin of $\IR^2$ space. We have $$ \alpha_{ii} \= 1 \,,\qquad \alpha_{ij} \=  \frac{1}{2} \left(-G_j^{(0)}\pm G_j^{(1)}+a_0 P_j^{(0)}+a_1 P_j^{(1)} \right)\,,$$
which implies that the $\theta_i$-dependent factors in $g_{x_\pm^{(1)}x_\pm^{(1)} }$ and $g_{r_i r_i}$ are the same. The local metric around the i$^\text{th}$ rod is then
\begin{align}
ds^2 \bigl|_{r_i = 0} \= &\bar{g}_{r_i r_i}(\theta_i) \left(d\rho_i^2 +\frac{\rho_i^2}{C_i^2}\,d{x_\pm^{(1)}}^2 \right)+ g_{\theta_i\theta_i}(\theta_i)  \,\left( d\theta_i^2 + \bar{g}_{\phi\phi}(\theta_i) \,\sin^2 \theta_i\, \,d\phi^2\right) \nonumber\\
&+ \sum_{I=2}^3 \left(g_{x_+^{(I)}x_+^{(I)}}(\theta_i) \,d{x_+^{(I)}}^2+g_{x_-^{(I)}x_-^{(I)}}(\theta_i) \,d{x_-^{(I)}}^2\right) + g_{y_7 y_7}(\theta_i) \left( dy_7+H_0 d\phi\right)^2 \nonumber \\
& +g_{x_\mp^{(1)}x_\mp^{(1)}}(\theta_i) \,d{x_\mp^{(1)}}^2 -g_{tt}(\theta_i) dt^2\,,\label{eq:metricatrodBu}
\end{align}
where all the $g_{xx}(\theta_i)$ can be obtained from \eqref{eq:ytphibehaviorBubblerod} and are finite and non-zero for $0\leq  \theta_i \leq \pi$. Moreover, we have defined $\rho_i^2 \equiv 4 r_i$ and the constant, $C_i$, is given by 
\begin{equation}
C_i \equi  d_{i} M_{i}\,\sqrt{\frac{e^{b_{1}+b_0}}{a_{1}a_{0}}}\, \prod_{j \neq i}\left(\frac{z_{j}^{+}-z_{i}^{-}}{z_{j}^{-}-z_{i}^{-}}\right)^{\operatorname{sign}(j-i) \alpha_{i j}}\,.
\end{equation}
The two-dimensional subspace $(\rho_i,x_\pm^{(1)})$ describes a smooth origin of $\IR^2$ or a smooth discrete quotient $\IR^2/\mathbb{Z}_{k_i}$ if the parameters are fixed according to the radius of the $x_\pm^{(1)}$-circle as
\begin{equation}
\begin{split}
R_{x_\pm^{(1)}} = \frac{C_i}{k_i}\,, \qquad k_i \in \mathbb{N}.
\end{split}
\label{eq:condRymultibubble}
\end{equation}
To conclude, the time slices of the eleven-dimensional space at the i$^\text{th}$ rod is a bolt described by a warped S$^2\times$T$^6$ fibration over an origin of a $\IR^2/\mathbb{Z}_{k_i}$ space. 

Moreover, one can check that metric determinant of the S$^2\times$T$^6$ is remarkably independent of $\theta_i$ if there are no black rods in the configuration. Therefore, one can easily derive the area of the S$^2\times$T$^6$ bubble for such configurations and we find
\begin{equation}
A_{\text{B}i} \= \int_{S^2\times T^6} \sqrt{\det g\bigl|_{S^2\times T^6}} \= (2\pi)^7\,k_i\,M_i\, \left(\frac{e^{b_1+b_2+b_3}}{8a_1 a_2 a_3} \right)^\frac{1}{3}\,\prod_{a=1}^7 R_{y_a}\,.
\label{eq:AreaBubbleApp}
\end{equation}
Finally, since only $P_i^{(1)}$ and $P_i^{(0)}$ are finite, the bubble carries a unique M2 charge and a KKm charge given by \eqref{eq:individualMcharges}. Each charge can be taken to be zero by considering the neutral limit \eqref{eq:neutrallimit}, which corresponds to $a_\Lambda = \sinh b_\Lambda \to \infty$. Moreover, since we have $T_1 \sim -a_1 \left(1+\cO(\rho_i^2)\right)$ \eqref{eq:ytphibehaviorBubblerod}, the component of the field strength along $x_\pm^{(1)}$ is vanishing at the rod which guarantees its regularity.

\item[•] \underline{A KKm bubble.}

The $y_7$ fiber shrinks to zero size at the rod while all others are finite if the rod has the following weights
\begin{equation}
a_0\,P_i^{(0)}\= 1 , \qquad G_i^{(\Lambda)} \= P_i^{(I)} \= 0\,.
\end{equation}
The analysis is similar to the M2-KKm bubble. We have
\begin{equation}
\alpha_{ii} \= 1 \,,\qquad \alpha_{ij} \=  a_0 P_i^{(0)}\,,
\end{equation}
and the local metric takes the same form as in \eqref{eq:metricatrodBu} by replacing $x_\pm^{(1)}$ by $y_7$ and $C_i$ is now given by
\begin{equation}
C_i \equi  \frac{d_{i} M_{i}\,e^{b_0}}{a_{0}}\, \prod_{j \neq i}\left(\frac{z_{j}^{+}-z_{i}^{-}}{z_{j}^{-}-z_{i}^{-}}\right)^{\operatorname{sign}(j-i) \alpha_{i j}}\,.
\end{equation}
The two-dimensional subspace $(\rho_i,y_7)$ describes a smooth origin of $\IR^2$ or a discrete quotient $\IR^2/\mathbb{Z}_{k_i}$ if the parameters are fixed according to the radius of the $y_7$-circle
\begin{equation}
\begin{split}
R_{y_7} = \frac{C_i}{k_i}\,, \qquad k_i \in \mathbb{N}.
\end{split}
\label{eq:condRymultibubble2}
\end{equation}
Therefore, the time slices of the eleven-dimensional space at the i$^\text{th}$ rod is a bolt described by a warped S$^2\times$T$^6$ fibration over an origin of a $\IR^2/\mathbb{Z}_{k_i}$ space. Because all $P^{(\Lambda)}_i$ are zero except $P^{(0)}_i$, the bubble carries only a KKm charge given by \eqref{eq:individualMcharges}.

Finally, the area of the S$^2\times$T$^6$ KKm bubble is also derivable when there are no black rods and give the same formula as the M2-KKm bubbles \eqref{eq:AreaBubbleApp}.

\end{itemize}

If we assume now that the rod is connected, we still have the eight same choices of weights but the local topology might change. For instance, we consider that the $i^\text{th}$ rod is a black rod. If the rod is not connected from below and connected from above to a M2-KKm bubble rod where $x_+^{(1)}=y_1$ shrinks, then the local metric at the horizon is still given \eqref{eq:metricatrodBS}. However, $g_{x_+^{(1)}x_+^{(1)}}(\theta_i)$ and $\bar{g}_{\phi \phi} (\theta_i)$ are not finite for $0 \leq \theta_i \leq \pi$ anymore. More precisely, we have $g_{x_+^{(1)}x_+^{(1)}}(\theta_i) \sim 0$ and $\bar{g}_{\phi \phi} (\theta_i) \sin^2 \theta_i $ finite around $\theta_i \to 0$. Therefore, the $y_1$-circle pinches off at the north pole and the horizon has a S$^3\times$T$^6$ topology. Note that the surface gravity computed in \eqref{eq:surfaceGravVac} is still the same and still well-defined. If the rod is now connected from above and below to two M2-KKm bubble rods where $y_1$ shrinks, we have a S$^2\times$T$^7$ horizon again but the S$^2$ is now described by $(\theta_i,y_1)$. Similar scenarios happen if the $i^\text{th}$ rod is a bubble rod: we can have either an S$^2\times$T$^6$ bubble or a S$^3\times$T$^5$ bubble depending on what is surrounding the rod.

To conclude, we have eight types of rods that can source physically our solutions on the axis. For each type of rods the $\phi$-circle has a finite size. The different rods and their physics has been depicted in Fig.\ref{fig:RodCategories} and Fig.\ref{fig:TouchingRods}.

Moreover, note that the exponents $\alpha_{ij}$ drastically simplify for the physical rods:
\begin{equation}
\begin{split}
\alpha_{ij}& \= \begin{cases} 
1 \qquad &\text{if the }i^\text{th}\text{ and }j^\text{th}\text{ rods are of the same nature,}  \\
\frac{1}{2} \qquad &\text{otherwise.}
\end{cases} 
\end{split}
\label{eq:alphasimApp}
\end{equation}
Therefore, the three-dimensional base, determined by the warp factor $e^{2\nu}$ \eqref{eq:WarpfactorsRods}, does not depend on the specific nature of the rods.

\subsubsection{On the $z$-axis and out of the rods}
\label{App:OutoftheRod}

We now study the behavior of the solutions on the $z$-axis, $\rho \to 0$, and out of the rods where the $\phi$-circle can shrink to zero size. On these segments, each $R^{(i)}_\pm$ \eqref{eq:Rpmdef} is non-zero and finite
\begin{equation}
R^{(i)}_\pm = 2 |z-z_i| \pm M_i\,.
\end{equation}
Thus, the warp factors $(W_\Lambda,Z_\Lambda)$ are also non-zero and finite there \eqref{eq:WarpfactorsRods}. The regularity reduces to the study of the three-dimensional subspace $(\rho,z,\phi)$,
\begin{equation}
ds_3^2 \= e^{2 \nu} \left(d\rho^2 +dz^2 \right) +\rho^2 d\phi^2\,.
\end{equation}
At $\rho=0$ and out of the rods, we want this space to correspond to the cylindrical coordinate degeneracy. First we have
\begin{equation}
e^{\nu_{jk}} \sim \begin{cases}
1 \qquad &\text{if } j\leq k  \text{ and } z \not\in [z_j + \frac{M_j}{2} , z_k - \frac{M_k}{2} ]  \\
\dfrac{(z_k^- - z_j^+)^2(z_k^+ - z_j^-)^2}{(z_k^+ - z_j^+)^2(z_k^- - z_j^-)^2} \qquad &\text{if } j<k   \text{ and } z \in ] z_j + \frac{M_j}{2} , z_k - \frac{M_k}{2} [
\end{cases}\,.
\end{equation}
Therefore, we get
\begin{equation}
e^{2 \nu} \sim \begin{cases}
~~ 1  &\text{if } z < a_1 - \frac{M_1}{2}  \text{ and } z > a_n + \frac{M_n}{2} \\
~~ d_i^2\qquad  &\text{if } z \in \, ]z_{i-1} + \frac{M_{i-1}}{2} , z_i - \frac{M_i}{2} [\,,\quad \forall i 
\end{cases}
\end{equation}
where $d_i$ is given in \eqref{eq:diDef}. First we notice that, asymptotically, $z > z_n^+$ and $z<z_1^-$, the base space is directly flat $\IR^3$ without conical singularity. However, in between two rods, we have two possibilities if they are connected or disconnected.

\begin{itemize}
\item[•] \underline{Disconnected rods:}
\end{itemize}

We consider the segment in between the $(i-1)^\text{th}$ rod and $i^\text{th}$ rods with $z_{i-1}^+ <  z_i^- $. The three-dimensional base is then given by the metric
 \begin{equation}
ds_3^2 \sim d_i^2 \left(d\rho^2 +dz^2 +\frac{\rho^2}{d_i^2} d\phi^2 \right) \,.
\end{equation}
The segment corresponds to a $\IR^3$ base with the local cylindrical angle $\phi_i \equi \frac{\phi}{d_i}$. Moreover, we have
\begin{equation}
\dfrac{(z_k^- - z_j^+)(z_k^+ - z_j^-)}{(z_k^+ - z_j^+)(z_k^- - z_j^-)}  = \frac{(z_k-z_j)^2 - \frac{1}{4} (M_k+M_j)^2}{(z_k-z_j)^2 - \frac{1}{4} (M_k-M_j)^2} < 1 \,,\qquad \alpha_{jk} \= 1 \text{ or } \frac{1}{2}>0\,.
\label{eq:simpleform}
\end{equation}
Thus, we necessarily have $d_i <1$. Thus, the segment has a \emph{conical excess} and the period of the local angle, $\phi_i= \frac{\phi}{d_i}$, is $\frac{2\pi}{d_i}>2\pi$. This manifests itself as a string with negative tension, or strut, between the two rods. The strut exerts the necessary repulsion so that the whole structure does not collapse. We can calculate the stress tensor and the energy of the strut using the method described in \cite{Costa:2000kf,Bah:2021owp}, and we will similarly find that the energy is given by 
\begin{equation}
E_i \= - \frac{1-d_i}{4 G_4}\,(z_i^- - z_{i-1}^+)\,, \qquad G_4 = \frac{G_{11}}{(2\pi)^{7} \prod_{i=1}^{7} R_{y_i}}.
\end{equation}

\begin{itemize}
\item[•] \underline{Connected rods:}
\end{itemize}

We consider the intersection between the $(i-1)^\text{th}$ rod and $i^\text{th}$ rods with  $z_{i-1}^+ =  z_i^- $. The intersection then consists of a point with coordinates $(\rho,z)\=(0,z_{i-1}^+)\=(0,z_{i}^-)\,.$
We first define local spherical coordinates as follows
\begin{align}
\mathfrak{r}_i &\equi  \sqrt{\rho^2 +\left(z-z_{i}^-\right)^2}\,,\qquad \cos \tau_i \equi \frac{z -z_{i}^- }{\mathfrak{r}_i }\,, \\
\mbox{that is} \qquad
\rho &= \mathfrak{r}_i \,\sin \tau_i \,,\qquad z =  \mathfrak{r}_i \cos \tau_i + z_i^-\,. \nonumber
\end{align}
The two-dimensional base transforms to
\begin{equation}
d\rho^2+dz^2 = d {\mathfrak{r}_i^+}^2+  {\mathfrak{r}_i^+}^2\,d{\tau_i^+}^2\,.
\end{equation}
At $\mathfrak{r}_i \rightarrow 0$ we have
\begin{align}
\frac{R_+^{(i-1)}}{R_-^{(i-1)}} &\sim \frac{2 M_{i-1}}{\left(1+\cos \tau_i\right)  \mathfrak{r}_i}\,,\quad \frac{R_+^{(i)}}{R_-^{(i)}} \sim \frac{2 M_{i}}{\left(1-\cos \tau_i\right)  \mathfrak{r}_i}\,,\quad \frac{R_+^{(j)}}{R_-^{(j)}} \sim  \left(\frac{z_j^+ -z_i^+}{z_j^--z_i^+} \right)^{\text{sign}(j-i)}\,,\quad j \neq i-1,i\,, \nonumber \\
e^{\nu_{i-1\,i-1}}& \sim \frac{(1+\cos \tau_i)^2}{4}\,,\qquad e^{\nu_{ii}} \sim \frac{(1-\cos \tau_i)^2}{4}\,,\qquad e^{\nu_{i -1\,i}} \sim \frac{(z^+_{i}-z^-_{i-1})^2}{(z^+_{i}-z^+_{i-1})^2(z^-_{i}-z^-_{i-1})^2}\,{\mathfrak{r}_i}^2,  \nonumber\\
e^{\nu_{jk}} &\sim \begin{cases}
\dfrac{(z_k^- - z_j^+)^2(z_k^+ - z_j^-)^2}{(z_k^+ - z_j^+)^2(z_k^- - z_j^-)^2} \qquad &\text{if } j<i-1 \text{ and } i <k  \\
1 \qquad &\text{otherwise } 
\end{cases}\,, 
\end{align}
\begin{align}
e^{\nu_{i-1\,j}} &\sim \begin{cases} 
\dfrac{(z_j^+ - z_{i-1}^-)^2\left(z_j^- - z_{i-1}^+\right)^2}{(z_j^- - z_{i-1}^-)^2\left(z_j^+ - z_{i-1}^+\right)^2} \qquad &\text{if } j>i,  \\
1 \qquad &\text{if } j<i-1, 
\end{cases}\,, \hspace{7cm} \nonumber \\
 e^{\nu_{ij}} &\sim \begin{cases} 
1 \qquad &\text{if } j>i,  \\
\dfrac{(z_{i}^+ - z_j^-)^2\left(z_{i}^- - z_j^+\right)^2}{(z_{i}^- - z_j^-)^2\left(z_{i}^+ - z_j^+\right)^2} \qquad &\text{if } j<i-1, 
\end{cases}\,, \nonumber
\end{align}
Thus, the $\phi$-circle keeps a finite size unlike the disconnected case. This means that the intersection is protected from the conical excess associated to the degeneracy of the $\phi$-circle in between two rods.

In order to determine the local topology, a distinction must be made between different types of rods. Being connected, the $i^\text{th}$ and $(i-1)^\text{th}$ rods are necessarily of a different nature.\footnote{Otherwise, they form a single rod.}  Consequently, we have two possible scenarios: an intersection between a black rod and a bubble rod and between two different bubble rods. 

For the first scenario we consider that the $(i-1)^\text{th}$ rod is a M2-KKm bubble where $y_1$ shrinks, but all other choices would have led to the same results. From the above expressions, we find that the local metric is given by
\begin{equation}
ds^2 \bigl|_{\mathfrak{r}_i=0} \propto \alpha_i \,d\phi^2 + \sum_{a\neq 1}\beta^{(a)}_i\,dy_a^2+  \frac{d\mathfrak{r}_i^2}{\mathfrak{r}_i} + \mathfrak{r}_i \left( d\tau_i^2 - 2 \kappa_i^2\,(1-\cos \tau_i)\,dt^2+\frac{ 2 (1+\cos \tau_i)}{R_{y_1}^2\,{k_{i-1}}^2}dy_1^2\right)\,,
\nonumber
\end{equation}
where $\alpha_i$ and $\beta^{(a)}_i$ are irrelevant finite constants and the surface gravity of the black rod $\kappa_i$ and the orbifold parameter of the M2-KKm bubble rod $k_{i-1}$ are given in \eqref{eq:surfaceGravVac} and \eqref{eq:condRymultibubble}. This corresponds to the usual metric at the north pole of a horizon. At this type of loci, a circle composing the surface of the horizon is degenerating which is here $y_1$. In the present case, it degenerates with a conical defect, parametrized by $k_{i-1}$, related to the M2-KKm bubble connected to the black hole. 

We now analyze the scenario of two connected bubble rods. For simplicity, we assume that they are two M2-KKm bubbles where the $y_1$ and $y_2$ circles shrink respectively. The time slices of the metric are locally given by
\begin{equation}\begin{split}
ds^2 \bigl|_{\mathfrak{r}_i=0} \propto  \sum_{a\neq 1,2}\beta^{(a)}_i\,dy_a^2+ \alpha_i \,d\phi^2+  \frac{d\mathfrak{r}_i^2}{\mathfrak{r}_i} + \mathfrak{r}_i \left( d\tau_i^2 + \frac{2 (1-\cos \tau_i)}{R_{y_2}^2\,{k_{i}}^2}dy_2^2+ \frac{ 2 (1+\cos \tau_i)}{R_{y_1}^2\,{k_{i-1}}^2}dy_1^2\right)\,,
\nonumber
\end{split}
\end{equation}
where $\alpha_i$ and $\beta^{(a)}_i$ are irrelevant finite constants and $(k_{i-1},k_{i})$ are the orbifold parameters of the connected bubble rods. This corresponds to the metric of the origin of an orbifolded $\IR^{4}$ parametrized by $(\mathfrak{r}_i,\tau_i,y_1,y_2)$. The two angles have the same conical defects as the connected bubbles but the local topology is free from struts and conical excess. Moreover, if $ k_{i-1} = k_{i}=1$, the time slices of the metric corresponds to the origin of a $\IR^{4}$ with a T$^5$ fibration and the local spacetime is entirely smooth.

\subsection{Five-dimensional solutions}
\label{App:Reg5d}

To construct five-dimensional solutions, one needs to consider the four-dimensional spatial part of the eleven-dimensional spacetime \eqref{eq:nonBPSfloating},
\be
ds_4^2 \= \frac{1}{Z_0} \left(dy_7 +H_0 \,d\phi\right)^2 \+ Z_0 \left( e^{2\nu} \left(d\rho^2 + dz^2 \right) +\rho^2 d\phi^2\right), 
\label{eq:4dbaseApp}
\ee
as a four-dimensional base. First, we consider $y_7$ to be an angle with $2\pi$ periodicity. Moreover, as suggested in \cite{Emparan:2001wk}, one also needs  $Z_0$ to be sourced by a semi-infinite rod and to consider the neutral limit for the pair $(Z_0,H_0)$, which implies $H_0=0$. 

Generic five-dimensional Weyl solutions in M-theory are therefore given by the ansatz \eqref{eq:nonBPSfloating}, sourced by $n$ finite rods of length $M_i$ and a $(n+1)^\text{th}$ semi-infinite rod above the rod configuration. We will consider that the $n^\text{th}$ and $(n+1)^\text{th}$ rods are connected to avoid a conical excess between them, $z_n^+=z_{n+1}^-$. The warp factors and gauge potentials are  given by \eqref{eq:WarpfactorsRods5d}.

\subsubsection{Regularity out of the $z$-axis}

We first discuss the condition on the asymptotics by considering $(\rho,z)=\frac{1}{2}(r^2\sin 2\theta ,r^2\cos 2\theta)$ and $r$ large. The warp factors and gauge potentials \eqref{eq:WarpfactorsRods5d} behave as
$$
Z_I \sim \frac{\sinh  b_I}{a_I} \,,\quad W_\Lambda \sim 1 \,,\quad Z_0 \sim \frac{1}{r^2\sin^2\theta}\,,\quad T_I\=-a_I \coth b_I,\quad e^{2\nu} \sim\sin^2\theta \,.
$$
Therefore, the solutions, when reduced to five dimensions after KK reduction on the T$^6$ \eqref{eq:5dframework}, are asymptotically:
\begin{equation}
ds_5 \sim - \left(\prod_{I=1}^3  \frac{\sinh b_I}{a_I} \right)^{-\frac{2}{3}} \,dt^2 \+  \left(\prod_{I=1}^3  \frac{\sinh b_I}{a_I} \right)^{\frac{1}{3}} \left[dr^2+r^2\left( d\theta^2 +\sin^2\theta dy_7^2 + \cos^2 \theta d\phi^2 \right) \right].
\end{equation}
They are asymptotically flat if
\begin{equation}
\prod_{I=1}^3  \frac{\sinh b_I}{a_I} \= 1\,.
\end{equation}

As for four-dimensional solutions, avoiding the changes of sign in the warp factors $Z_I$ requires to impose
\begin{equation}
b_I\geq 0 \,,\quad a_I P_i^{(I)} \geq 0\,, \qquad \forall i\,, \,\,I=1,2,3,
\end{equation}
which makes the solutions regular out of the $z$-axis.

\subsubsection{Regularity on the $z$-axis}

We will first discuss the regularity at the rods before the regularity elsewhere on the $z$ axis.

\subsubsection{At the rod}
\label{App:AttheRod2}

The regularity at the $n$ finite rods is very similar to the one performed for four-dimensional solutions in section \ref{App:AttheRod}. However, the regularity at the semi-infinite rod will require a specific analysis.

\begin{itemize}
\item[•] \underline{At the semi-infinite rod: $\rho=0$ and $z>z_n^+$.}
\end{itemize}

At $\rho=0$ and $z>z_n^+$, we have
\begin{equation}
 \frac{R_+^{(i)}}{R_-^{(i)}} \sim \frac{z-z_i^-}{z-z_i^+},\qquad i=1,...,n,
\end{equation}
which is finite and non-zero. Therefore, the warp factors $(Z_I,W_\Lambda)$ are finite and non-zero \eqref{eq:WarpfactorsRods5d}, and the regularity at the rod reduces to the regularity of the four-dimensional metric \eqref{eq:4dbaseApp}. For that purpose, it is convenient to define the coordinates $(\xi,\eta)$ such that
\begin{equation}
\rho \equi \eta \xi \,,\qquad z \equi z_n^+ \+ \frac{\eta^2-\xi^2}{2}\quad \Rightarrow\quad d\rho^2 +dz^2 \= (\eta^2+\xi^2)(d\xi^2+d\eta^2)\,.
\end{equation}
The rod is described by $\xi=0$ and $\eta>0$  in this coordinate system. In the $\xi\to 0$ limit, one has
\begin{equation}
Z_0\, \sim\, \xi^{-2 G_{n+1}} \,\prod_{i=1}^n \left(\frac{z_n^{+}-z_i^++\frac{\eta^2}{2}}{z_n^{+}-z_i^-+\frac{\eta^2}{2}} \right)^{-G_i},\quad e^{2\nu}\, \sim\,  \frac{1}{2} \left(\frac{2\xi^2}{\eta^2} \right)^{{G_{n+1}}^2}\prod_{i=1}^n \left(\frac{z_n^{+}-z_i^++\frac{\eta^2}{2}}{z_n^{+}-z_i^-+\frac{\eta^2}{2}} \right)^{2 G_{n+1}G_i} \nonumber
\end{equation}
Therefore, if we impose
\begin{equation}
G_{n+1} \= 1\,,
\end{equation} 
the local four-dimensional metric \eqref{eq:4dbaseApp} behaves as
\begin{equation}
ds_4^2 \,\sim \, \prod_{i=1}^n \left(\frac{z_n^{+}-z_i^++\frac{\eta^2}{2}}{z_n^{+}-z_i^-+\frac{\eta^2}{2}} \right)^{G_i} \,\left[ d\eta^2 + d\xi^2 + \xi^2 dy_7^2 \right] \+  \prod_{i=1}^n \left(\frac{z_n^{+}-z_i^++\frac{\eta^2}{2}}{z_n^{+}-z_i^-+\frac{\eta^2}{2}} \right)^{-G_i} \eta^2 d\phi^2.
\end{equation}
The rod corresponds to a regular coordinate singularity where the $y_7$-circle pinches off defining a smooth origin of a $\IR^2$.

\begin{itemize}
\item[•] \underline{At the $i^\text{th}$ rod: $\rho=0$ and $z_i^-<z<z_i^+$.}
\end{itemize}

The analysis is very close to the one led for four-dimensional solutions in section \ref{App:AttheRod}, and we refer the reader to this section for most of the details of the expansion. The only difference is that there are new terms arising from the semi-infinite rod in $Z_0$ and $e^{2\nu}$ that must be carefully expanded. 

We also use the local spherical coordinate $(r_i,\theta_i)$ \eqref{eq:localcoorApp}, and derive the local geometry at $r_i\to 0$. The new terms behave such that
\begin{align}
&r_+^{(n)} -(z-z_n^+) \,\sim\, 2\left(z_n^+-z_i-\frac{M_i}{2} \cos \theta_i \right),\qquad \frac{r_+^{(n)}-(z-z_n^+)}{2r_+^{(n)}} \, \sim\, 1\,, \\
&\frac{E_{++}^{(n,j)}R_-^{(j)}}{E_{+-}^{(n,j)}R_+^{(j)}} \sim \begin{cases}
1 \qquad &\text{if } i<j \\
\dfrac{(z_n^+ - z_j^+)^2}{(z_n^+ - z_j^-)^2} \qquad &\text{if } j<i
\end{cases}\,, \qquad \frac{E_{++}^{(n,i)}R_-^{(i)}}{E_{+-}^{(n,i)}R_+^{(i)}} \sim \frac{\left(z_n^+-z_i-\frac{M_i}{2} \cos \theta_i \right)^2}{(z_n^+-z_i^-)^2}. \nonumber
\end{align}
From \eqref{eq:RpRmexp} and \eqref{eq:EEEEatrod}, we find that the main functions behave as
\begin{align}
W_\Lambda &\sim\left( \frac{M_i}{r_i} \right)^{ G^{(\Lambda)}_i}\,\prod_{j\neq i} A_{ij}(\theta_i)^{G^{(\Lambda)}_j}\,, \qquad Z_0 \sim \frac{1}{2\left(z_n^+-z_i-\frac{M_i}{2} \cos \theta_i \right)}\left( \frac{M_i}{r_i} \right)^{ G_i}\,\prod_{j\neq i} A_{ij}(\theta_i)^{G_j} ,\nonumber\\
 Z_I &\sim \begin{cases}&\dfrac{e^{b_I}}{2 a_I}\,\left( \dfrac{M_i}{r_i} \right)^{ a_I P^{(I)}_i}\,\prod_{j\neq i} A_{ij}(\theta_i)^{a_I P^{(I)}_j}\,,\qquad \text{if }a_I P^{(I)}_i \neq 0, \\
&\dfrac{1}{a_I}\,\sinh \left[ \log \left(\prod_{j\neq i} A_{ij}(\theta_i)^{a_I P^{(I)}_j} \right)+ b_I \right]\,,\qquad \text{if }a_I P^{(I)}_i = 0,
\end{cases} \label{eq:ExpZI5dApp}\\
 e^{2 \nu}\, &\sim \, d_i^2\,\left(\frac{4 r_i}{M_i \sin^2 \theta_i} \right)^{\alpha_{ii}}\, \left(\frac{z_n^+-z_i-\frac{M_i}{2} \cos \theta_i }{z_n^+-z_i^-}\right)^{2G_i}\, \prod_{j=1}^i  \left( \frac{z_n^+ - z_j^+}{z_n^+ - z_j^-}\right)^{2 G_j}\, \nonumber\\
 &\hspace{0.5cm}\times\prod_{j\neq i}A_{ij}(\theta_i)^{-2\alpha_{ij}} \,  \left(\frac{z_j^+ - z_i^-}{z_j^- -z_i^-} \right)^{2\,\text{sign}(j-i) \,\alpha_{ij}} \,. \nonumber
\end{align}
where the constants $d_i$ are defined in \eqref{eq:diDef}, and we have considered that the product ``$\prod_{j=1}^{i-1}$'' is equal to $1$ for the first rod, $i=1$.
Gathering all these expressions, the metric components and gauge potentials behave around the $i^\text{th}$ rod as\footnote{We have denoted $\widetilde{g}_{\phi \phi}$ the metric component of $\phi$ without the connection term in the $y_7$ fiber, that is $\widetilde{g}_{\phi \phi}= \rho^2 \frac{Z_0\,\left(Z_1 Z_2 Z_3 \right)^{\frac{1}{3}}}{W_0^2} $.}
\begin{align}
T_I &\sim -a_I \left[ 1 + \cO \left( r_i^{2 a_I P_i^{(I)}} \right)\right] ,\nonumber  \\
g_{tt} &\sim- \frac{1}{\left( Z_1 Z_2 Z_3\right)^{\frac{2}{3}}}\left( \frac{M_i}{r_i} \right)^{ -\frac{2}{3}G^{(0)}_i}\,\prod_{j\neq i} A_{ij}(\theta_i)^{ -\frac{2}{3}G^{(0)}_j}\,, \nonumber \\
 g_{x_\pm^{(I)}x_\pm^{(I)}} &\sim \left(\frac{|\epsilon_{IJK}|}{2}\, \frac{Z_J Z_K}{Z_I^2}\right)^{\frac{1}{3}}\left( \frac{M_i}{r_i} \right)^{ \frac{1}{3}G^{(0)}_i \mp G_i^{(I)} }\,  \prod_{j\neq i} A_{ij}(\theta_i)^{ \frac{1}{3} G^{(0)}_j \mp G_j^{(I)}}\,, \label{eq:ytphibehaviorBubblerod5d} \\
 g_{y_7 y_7} &\sim 2\left(z_n^+-z_i-\frac{M_i}{2} \cos \theta_i \right)\left(Z_1 Z_2 Z_3\right)^{\frac{1}{3}}\left( \frac{M_i}{r_i} \right)^{ -\frac{2}{3}G^{(0)}_i-G_i}\,\prod_{j\neq i} A_{ij}(\theta_i)^{-\frac{2}{3}G^{(0)}_j- G_j}\,,\nonumber  \\
\widetilde{g}_{\phi\phi} &\sim  \frac{M_i^2\, \sin^2\theta_i}{2\left(z_n^+-z_i-\frac{M_i}{2} \cos \theta_i \right)}\, \left( Z_1 Z_2 Z_3\right)^{\frac{1}{3}}\left( \frac{M_i}{r_i} \right)^{ -1-\frac{2}{3}G^{(0)}_i+ G_i}\, \prod_{j\neq i} A_{ij}(\theta_i)^{-\frac{2}{3}G^{(0)}_j+ G_j}\,, \nonumber \\
g_{r_i r_i}& \sim  \frac{d_i^2}{2 \left( z_n^+ -z_i^+\right)^{2 G_i}} \left(Z_1 Z_2 Z_3\right)^{\frac{1}{3}}\,\left(\frac{4 r_i}{M_i \sin^2 \theta_i} \right)^{\alpha_{ii}-1} \left( \frac{M_i}{r_i} \right)^{ -\frac{2}{3}G^{(0)}_i+ G_i} \nonumber \\
& \hspace{0.5cm} \times \left(z_n^+-z_i-\frac{M_i}{2} \cos \theta_i \right)^{2G_i-1}\,\,\prod_{j=1}^i  \left( \frac{z_n^+ - z_j^+}{z_n^+ - z_j^-}\right)^{2 G_j} \nonumber\\
& \hspace{0.5cm} \times \prod_{j\neq i}A_{ij}(\theta_i)^{-2\alpha_{ij}-\frac{2}{3}G^{(0)}_j+G_j} \,  \left(\frac{z_j^+ - z_i^-}{z_j^- -z_i^-} \right)^{2\,\text{sign}(j-i) \,\alpha_{ij}} \,, \nonumber \\
g_{\theta_i \theta_i}& \sim r_i  M_i \,g_{r_i r_i}\,, \nonumber
\end{align}
where the two possible developments of $Z_I$ are given in \eqref{eq:ExpZI5dApp}, $\epsilon_{IJK}$ is the rank-3 Levi-Civita tensor and we have also used the convenient notation 
\begin{equation}
(y_1,y_2,y_3,y_4,y_5,y_6)=(x^{(1)}_+,x^{(1)}_-,x^{(2)}_+,x^{(2)}_-,x^{(3)}_+,x^{(3)}_-).
\end{equation}

Therefore, one can make one of the Killing vector along the extra dimensions, $\partial_{y_a}$, or the timelike Killing vector, $\partial_t$, vanish at the rod while all other have a finite norm by fixing the 8 weights $(G_i^{(\Lambda)},G_i,P_i^{(I)})$. Moreover, the physical rods are similar to the ones for four-dimensional solutions, and they are given by the same weights knowing that $G_i = a_0 P_i^{(0)}$. We refer then to the analysis led in section \ref{App:AttheRod} for the description of the local geometry and topology, and to the summary in section \ref{sec:regM5d}.

In a word, we have eight types of rods that can source physically our solutions on the axis. Moreover, as for four-dimensional solutions, the exponents $\alpha_{ij}$ \eqref{eq:alpha5d} drastically simplify for the physical rods and give \eqref{eq:alphasimApp}.

Finally, the regularity analysis on the $z$-axis but out of the rods, that is where the $\phi$-circle pinches off, and at the intersections of connected rods are identical to the analysis led for the four-dimensional solutions. We refer the reader to the appendix \ref{App:OutoftheRod}.



\bibliographystyle{utphys}      

\bibliography{microstates}       


\end{document}